\label{key}
\pdfoutput=1
\documentclass[twocolumn]{aastex631}

\shorttitle{MWISP Clumps}
\shortauthors{Jiang et al.}

\graphicspath{{./}{figures/}}
\usepackage{amsmath}
\usepackage{appendix}
\usepackage{array}
\usepackage{url}
\usepackage{soul} 
\usepackage{color, xcolor} 
\usepackage{makecell}
\usepackage{booktabs}
\usepackage{threeparttable}
\usepackage{tabularx}
\usepackage{graphicx}
\usepackage{enumitem}
\usepackage{rotating}

\shortauthors{Jiang et al.}

\begin{document}

\title{Investigations of MWISP Clumps: $^{13}$CO Clump Source Catalogs and Physical Properties}

\author[0000-0002-3549-5029]{Yu Jiang}
\affiliation{Purple Mountain Observatory and Key Laboratory of Radio Astronomy, Chinese Academy of Sciences, 10 Yuanhua Road, Nanjing 210034, People’s Republic of China }
\affiliation{School of Astronomy and Space Science, University of Science and Technology of China, 96 Jinzhai Road, Hefei 230026, People’s Republic of China}

\author[0000-0003-4586-7751]{Qing-Zeng Yan}
\affiliation{Purple Mountain Observatory and Key Laboratory of Radio Astronomy, Chinese Academy of Sciences, 10 Yuanhua Road, Nanjing 210034, People’s Republic of China }

\author[0000-0001-7768-7320]{Ji Yang}
\affiliation{Purple Mountain Observatory and Key Laboratory of Radio Astronomy, Chinese Academy of Sciences, 10 Yuanhua Road, Nanjing 210034, People’s Republic of China }

\author{Sheng Zheng}
\affiliation{Center for Astronomy and Space Sciences, China Three Gorges University, 8 University Road, Yichang 443002,  People’s Republic of China}
\affiliation{College of Science, China Three Gorges University, 8 University Road, Yichang 443002, People’s Republic of China}

\author[0000-0003-3151-8964]{Xuepeng Chen}
\affiliation{Purple Mountain Observatory and Key Laboratory of Radio Astronomy, Chinese Academy of Sciences, 10 Yuanhua Road, Nanjing 210034, People’s Republic of China }
\affiliation{School of Astronomy and Space Science, University of Science and Technology of China, 96 Jinzhai Road, Hefei 230026, People’s Republic of China}

\author[0000-0002-0197-470X]{Yang Su}
\affiliation{Purple Mountain Observatory and Key Laboratory of Radio Astronomy, Chinese Academy of Sciences, 10 Yuanhua Road, Nanjing 210034, People’s Republic of China }

\author{Zhibo Jiang}
\affiliation{Purple Mountain Observatory and Key Laboratory of Radio Astronomy, Chinese Academy of Sciences, 10 Yuanhua Road, Nanjing 210034, People’s Republic of China }

\author[0000-0003-0849-0692]{Zhiwei Chen}
\affiliation{Purple Mountain Observatory and Key Laboratory of Radio Astronomy, Chinese Academy of Sciences, 10 Yuanhua Road, Nanjing 210034, People’s Republic of China }

\author[0000-0003-2418-3350]{Xin Zhou}
\affiliation{Purple Mountain Observatory and Key Laboratory of Radio Astronomy, Chinese Academy of Sciences, 10 Yuanhua Road, Nanjing 210034, People’s Republic of China }

\author{Yao Huang}
\affiliation{Center for Astronomy and Space Sciences, China Three Gorges University, 8 University Road, Yichang 443002, People’s Republic of China}
\affiliation{College of Science, China Three Gorges University, 8 University Road, Yichang 443002, People’s Republic of China}

\author[0000-0003-0592-3042]{Xiaoyu Luo}
\affiliation{Center for Astronomy and Space Sciences, China Three Gorges University, 8 University Road, Yichang 443002, People’s Republic of China}
\affiliation{College of Science, China Three Gorges University, 8 University Road, Yichang 443002, People’s Republic of China}

\author[0000-0003-1714-0600]{Haoran Feng}
\affiliation{Purple Mountain Observatory and Key Laboratory of Radio Astronomy, Chinese Academy of Sciences, 10 Yuanhua Road, Nanjing 210034, People’s Republic of China }

\author{De-Jian Liu}
\affiliation{Center for Astronomy and Space Sciences, China Three Gorges University, 8 University Road, Yichang 443002, People’s Republic of China}
\affiliation{College of Science, China Three Gorges University, 8 University Road, Yichang 443002, People’s Republic of China}



\correspondingauthor{Ji Yang}
\email{jiyang@pmo.ac.cn}

\correspondingauthor{Sheng Zheng}
\email{zsh@ctgu.edu.cn}





\begin{abstract}
We present the first comprehensive catalogs of $^{13}$CO clumps from the Milky Way Imaging Scroll Painting (MWISP) project. By developing an equivalent global detection scheme integrated with the FacetClumps algorithm, we successfully extract 71,661 molecular clumps across a high-resolution $^{13}$CO data cube spanning 2310 deg$^2$ from the MWISP Phase I survey. To determine accurate distances, we design an automatic hierarchical distance decision method using signal regions as fundamental objects, effectively resolving the kinematic distance ambiguity problem and obtaining reliable measurements for 97.94\% of the sample. Statistical analysis reveals that 65.3\% of clumps are gravitationally bound, accounting for approximately 96.3\% of the statistical total mass. Scaling relation analysis across multiple surveys reveals universal power-law behaviors in clump populations. Maser-associated clumps exhibit modified parameter distributions and scaling relations, revealing how active star formation alters clump dynamics and structure. These extensive catalogs establish a foundation for investigating molecular clump properties, star formation processes, and Galactic evolution. 
\end{abstract} 


\section{Introduction}
Molecular clouds in the interstellar medium exhibit rich and complex structures across multiple spatial scales, serving as critical components in galactic evolution. Within this hierarchical organization, molecular clumps represent intermediate-scale condensations that mark a crucial transition from diffuse gas to star formation \citep{Review_15,FIVe,Review_3,Review_12}. These clumps form through complex interactions between gravity, turbulence, magnetic fields, and stellar feedback, often occurring within the filamentary structures of larger molecular cloud complexes \citep[e.g.,][]{Review_8,Review_16,Review_11,Review_4,Review_2}. The diverse formation pathways—whether through gravitational instability, turbulent compression, or filamentary fragmentation—concentrate gas into high-density regions capable of gravitational collapse. Characterizing clump properties thus provides direct insights into both the initial conditions of star formation and the critical balance between collapsing forces and supportive mechanisms like turbulent pressure or magnetic tension \citep[e.g.][]{Gravitation_1,Hierarchical_Collapse_1,Review_4}, ultimately revealing the physical processes that govern molecular cloud evolution across galactic environments \citep[e.g.][]{Review_6,Review_3}.

While decades of surveys have mapped molecular cloud distributions, comprehensive studies of clump properties across the Galaxy remain limited \citep[e.g.,][]{CO_Survey_1,CO_Survey_2,CO_Survey_3,Review_13}. The Boston University-Five College Radio Astronomy Observatory Galactic Ring Survey (\citealt{GRS_1}) contributed important $^{13}$CO observations but with limited spatial coverage, while dust continuum surveys like ATLASGAL \citep{ATLASGAL_2009} and Hi-GAL \citep{HiGal_0} offered comprehensive coverage of high-density regions but lacked velocity information necessary for distance determination and dynamical analysis. More recent efforts, including ThrUMMS \citep{ThrUMMS}, CHIMPS \citep{CHIMPS_1}, FUGIN \citep{FUGIN_1}, SEDIGISM \citep{SEDIGISM_1}, FQS \citep{FQS}, OGHReS \citep{OGHReS}, and PAMS \citep{PAMS} have advanced our understanding of molecular cloud structures but still faced limitations in spatial coverage, sensitivity, or ability to trace gas across different density regimes.

Currently, there exists no comprehensive catalog of $^{13}$CO clumps that spans the diverse environments of the Galactic plane while maintaining uniform detection methodologies, distance determination techniques, and physical parameter derivations \citep{CHIMPS_2,MWISP_Analysis_Luo_2024,MWISP_Analysis_Yang_2025}. This absence limits our ability to develop a coherent picture of molecular cloud evolution across the Galaxy and understand how environmental factors influence star formation processes in different Galactic regions. 

The Milky Way Imaging Scroll Painting (MWISP; \citealt{CO_Survey_MWISP_1}) project offers a transformative approach to addressing these limitations through its comprehensive, unbiased survey of $^{12}$CO/$^{13}$CO/C$^{18}$O ($J=1\text{--}0$) emission across the Galactic plane. Spanning an wide-ranging region in its Phase I operations, MWISP represents one of the most spatially complete molecular line surveys of the Milky Way to date. The simultaneous acquisition of multiple CO isotopologues enables a multi-tracer perspective that captures molecular gas across diverse density regimes and environmental conditions. This coordinated multi-isotopologue strategy allows for precise determination of critical physical parameters, such as excitation temperatures, optical depths, and column densities, through local thermodynamic equilibrium (LTE) analysis \citep{CHIMPS_2,MWISP_Analysis_Dong_2023}. 

The exceptional scale of MWISP data presents unique computational challenges requiring innovative methodological solutions. Traditional clump detection strategies often struggle with the sheer volume of data in modern surveys, while subcube partitioning approaches can introduce artifacts at boundaries \citep{CHIMPS_2,SEDIGISM_2,SN_Map_2,MWISP_Analysis_Luo_2024}. Additionally, resolving the kinematic distance ambiguity (KDA) problem—where a single observed velocity can correspond to two possible distances in the inner Galaxy—remains a persistent challenge for Galactic structure studies \citep{GRS_Dist_1,ATLASGAL_2018,HiGal_Dist}. 

In this paper, we present the first comprehensive catalogs of $^{13}$CO clumps extracted from the MWISP survey. We introduce an equivalent global detection scheme that maintains detection integrity while substantially reducing computational resource requirements. We attempt to resolve the challenge of KDA through an automatic hierarchical decision method and derive physical properties for an extensive sample of molecular clumps. Through statistical analysis of their physical properties and scaling relations, we explore deviations from classical scaling laws and examine distinct characteristics in maser-associated clumps to explore the mechanisms potentially governing molecular clump structure and dynamics. 

The paper is organized as follows:  Section \ref{Data} introduces the MWISP survey and details the observational data utilized in this study. Section \ref{Method} presents the molecular clump detection methodology, including the FacetClumps algorithm and the equivalent global detection scheme. Section \ref{Distance} details the hierarchical approach to distance determination and resolution of KDAs. Section \ref{PProperties} analyzes the physical properties and scaling relations of molecular clumps, comparing our results with other major surveys and examining differences in maser-associated clump populations. Finally, Section \ref{Summary} presents the conclusions and outlines future research directions enabled by these clump catalogs. 

\section{Data} \label{Data}
\subsection{The MWISP Survey and $^{13}$CO Emission}\label{Data_CO}
The MWISP\footnote{\href{http://english.dlh.pmo.cas.cn/ic/in/}{http://english.dlh.pmo.cas.cn/ic/in/}} project is an unbiased $^{12}$CO/$^{13}$CO/C$^{18}$O ($J=1\text{--}0$) molecular line survey of the Galactic plane, conducted using the Purple Mountain Observatory (PMO) 13.7-m millimeter-wave telescope at Delingha, China \citep{CO_Survey_MWISP_1}. This multi-line survey simultaneously observes all three CO isotopologues through the 3$\times$3-beam Superconducting Spectroscopic Array Receiver \citep{CO_Survey_MWISP_2}, operating in sideband-separating mode with on-the-fly mapping. 

The Phase I survey (2011--2021) covers Galactic longitude $l = 9\fdg75$ to $229\fdg75$ with latitude range $|b| \leq 5\fdg25$ and a velocity coverage of $\sim$2600 km~s$^{-1}$. The system achieves a half-power beamwidth of $\sim$52$^{\prime\prime}$ at 115.271 GHz ($^{12}$CO) and $\sim$55$^{\prime\prime}$ at 110.201/109.782 GHz ($^{13}$CO/C$^{18}$O) with pointing accuracy better than 5$^{\prime\prime}$. Spectral data are acquired through 18 fast Fourier transform spectrometers providing 1 GHz bandwidth over 16,384 channels, yielding velocity resolutions of 0.16 km~s$^{-1}$ for $^{12}$CO and 0.17 km~s$^{-1}$ for $^{13}$CO/C$^{18}$O. The surveyed region is divided into 30$^{\prime}$ $\times$ 30$^{\prime}$ cells with 30$^{\prime\prime}$ grid spacing. After first-order baseline subtraction and quality control, the final data cubes reach typical RMS noise levels of 0.45 K per channel for $^{12}$CO and 0.25 K per channel for $^{13}$CO/C$^{18}$O. This work focuses on the extraction of clumps from Phase I whole-sky $^{13}$CO emission data spanning a velocity range of $\pm$200 km~s$^{-1}$. 

\begin{figure*}
\centering
\centerline{\includegraphics[width=7in]{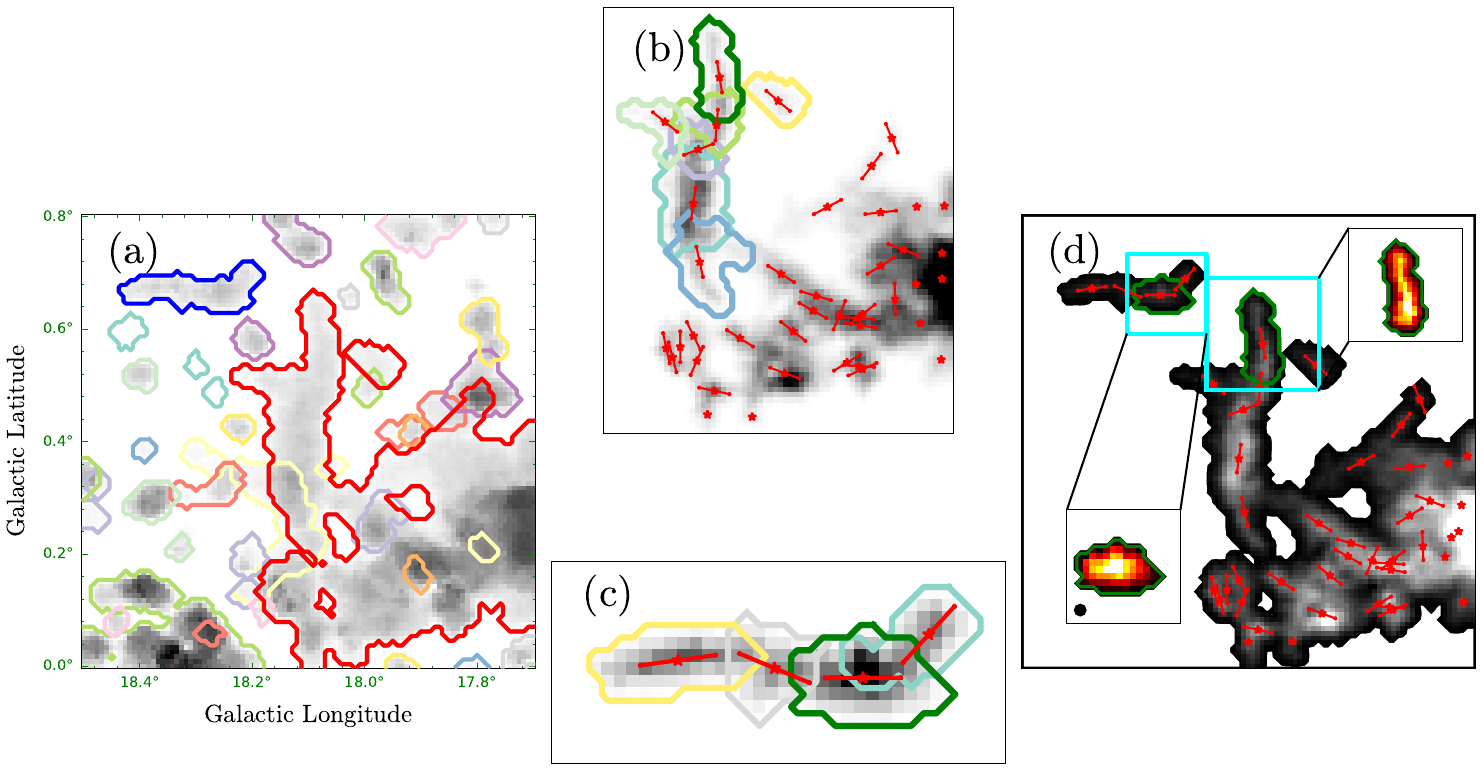}}
\caption{Schematic diagram illustrating the equivalent global detection scheme. (a) The SRs extracted from the original data (see Figure 1 of \citealt{DPConCFil}) by FacetClumps. Curves of different colors represent the boundaries of different SRs. (b) The reconstructed SR data cube, corresponding to the red contours in (a). The asterisks denote the centers of the clumps, and the direction of the red straight lines indicates the orientation of the clumps at the untouched edges. Contours in different colors represent the boundaries of the clumps. To avoid confusion, only a subset of the clump boundaries is shown. (c) The reconstructed SR data cube, corresponding to the blue contours in (a). (d) Detection results storage data cube. Transfer the detection masks from (b) and (c) to a storage data cube of the same size as the original data. The inset view shows two clumps in detail, and the black circle in the bottom-left corner marks the spatial resolution of MWISP. }
\label{DSRegion_Flow}
\end{figure*}

\section{Molecular Clump Detection} \label{Method}
\subsection{The Clump Detection Algorithm}\label{Data_Clump}
Molecular clumps, as dense conglomerations of gas and dust within molecular clouds, are defined by a central region with conspicuously higher density than the surrounding environment \citep{Review_15,FacetClumps}. This characteristic makes them critical potential sites for star formation. Morphologically, these clumps exhibit irregular geometries \citep{MWISP_Analysis_Long_2024,MWISP_Analysis_Luo_2024_2}, and their average intensity spectrum along any direction exhibits a Gaussian-like profile. 

Accurate identification of such clumps is essential to understanding star formation processes. The FacetClumps algorithm \citep{FacetClumps} has been demonstrated to exhibit robust performance in localization accuracy, detection completeness, and regional segmentation precision \citep{FellWalker,ConBased,LDC}. Initially, FacetClumps extracts signal regions (SRs) using morphological methods. By combining the facet model with the extremum determination theorem of multivariate functions, it autonomously locates clump centers within the preprocessed SRs. The algorithm then applies a connectivity-based minimum distance clustering method to merge locally gradient-segmented regions, effectively delineating the boundaries of individual clumps. Given its technical advantages and demonstrated reliability, this study utilizes FacetClumps for clump detection. Detailed parameter configurations are presented in Table \ref{InputPar} of Appendix \ref{Parameters_FacetClumps}. 

\subsection{The Equivalent Global Detection Scheme}\label{Equivalent_Dec}
Our study seeks to achieve direct global detection across entire 2310 deg$^2$ observational datasets. However, such global detection requires prohibitively high computational resources that become increasingly challenging with growing data volumes. Traditional strategies using subdivided data cubes -- even with overlapping region partitioning -- suffer from two critical limitations: (1) mask stitching ambiguities at subcube boundaries, and (2) selection ambiguities for clumps in overlapping areas. These issues become particularly severe for extended structures that span multiple subcubes, where discrepancies between subcube-level and global-level detection are inevitable. In such cases, even a single misidentified clump can trigger cascading mask stitching errors. 

To overcome these challenges, we propose an equivalent global detection scheme outlined in Figure \ref{DSRegion_Flow}. The original illustration data and direct detection results have been presented in Figures 1 and 2 of \cite{DPConCFil}. The implementation consists of three stages: (1) initial SR extraction using the first detection phase of FacetClumps (Figure~\ref{DSRegion_Flow}a), ensuring the extraction of the same objects across different-scale detections; (2) independent clump detection for each SR (Figure \ref{DSRegion_Flow}b–c), with simultaneous mapping of generated masks to a global storage array matching the size of the original data (Figure \ref{DSRegion_Flow}d); (3) final catalog calculation using original data and integrated global masks. The magnified inset demonstrates the characteristic morphologies of clumps and confirms they exceed the telescope's beam-scale threshold required for physically meaningful source extraction. 

This scheme achieves strict equivalence to whole-sky detection through two key innovations: distributed processing preserves large-scale structural relationships by avoiding subcube segmentation biases, while mask inheritance ensures spatial integrity via real-time global coordinate mapping. By effectively decoupling the computational load, the approach substantially reduces resource requirements and improves efficiency compared to direct global detection, enabling identical outputs across datasets of various scales. The implementation has been integrated as a subfunction of FacetClumps and is publicly available\footnote{\href{https://github.com/JiangYuTS/FacetClumps}{https://github.com/JiangYuTS/FacetClumps}}. 

\begin{figure*}
\centering
\centerline{\includegraphics[width=7.2in]{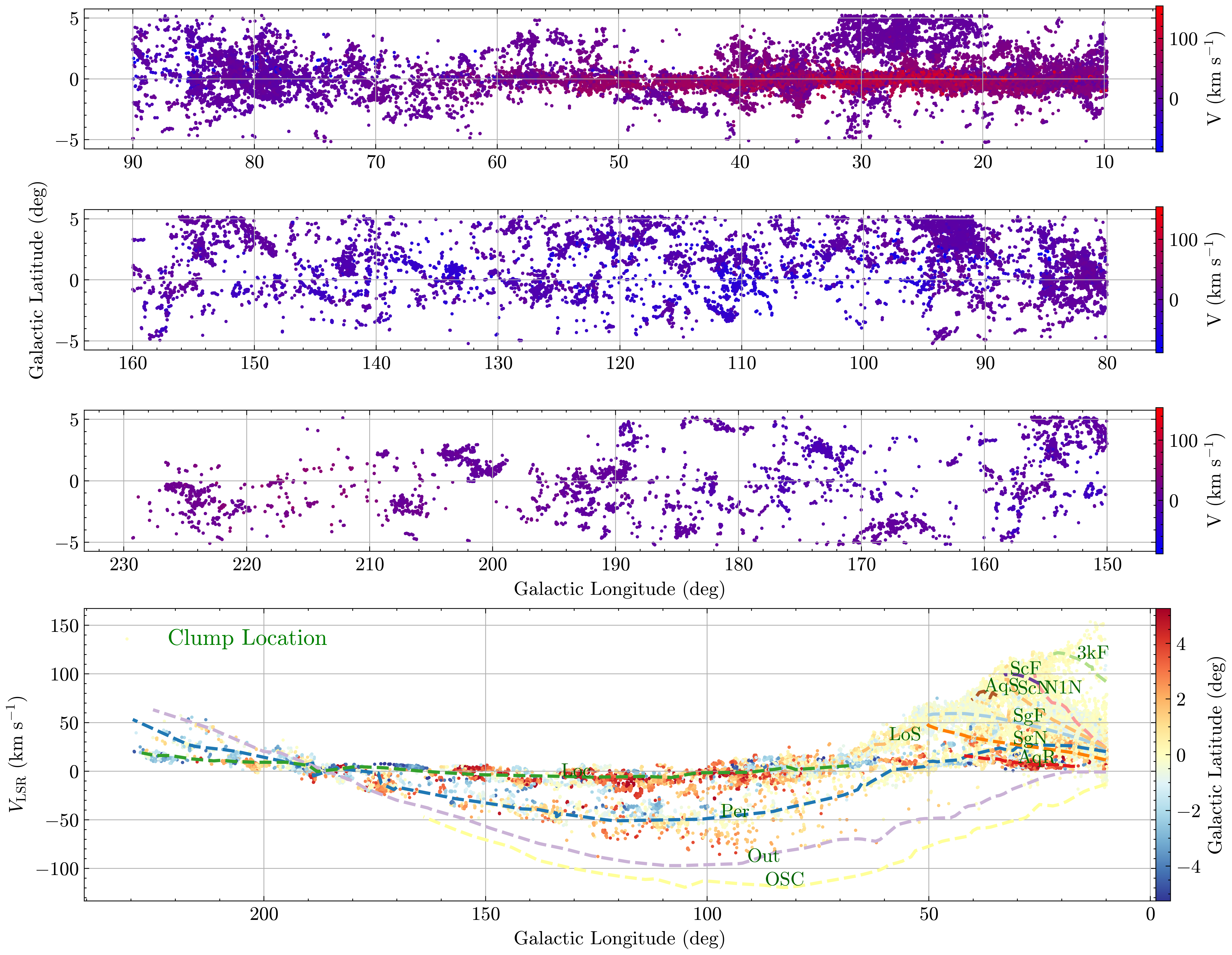}}
\caption{Distribution of all clumps in the Galactic longitude–latitude and Galactic longitude–velocity planes. Different colored dashed lines illustrate various spiral arms from the model by \cite{Distance_1}, with their names labeled.}
\label{MWISP_Clumps_LBV}
\end{figure*}

\begin{figure*}
\centering
\centerline{\includegraphics[width=7in]{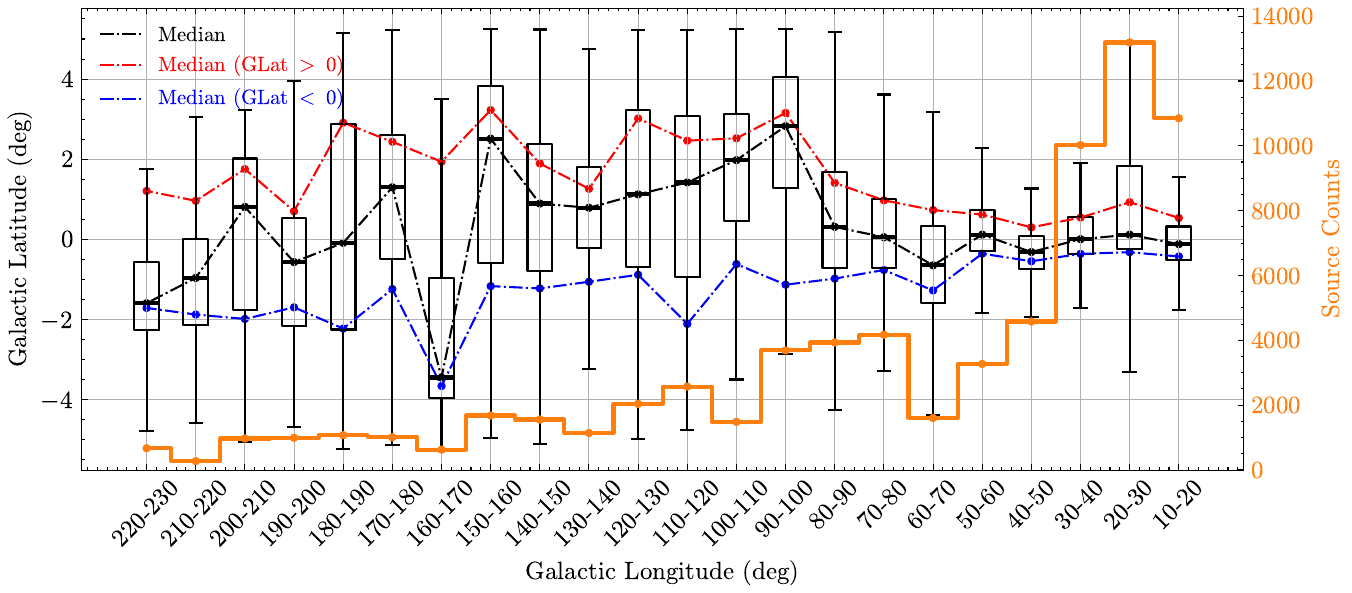}}
\caption{Distribution of clumps across the Galactic longitude--latitude plane. Box plots illustrate the statistical distribution of Galactic latitude within discrete longitude intervals. Each box displays the interquartile range (IQR), with upper and lower boundaries marking the third and first quartiles, respectively, and the horizontal line indicating the median. The colored dashed-dotted lines represent median values for specific populations: red for positive Galactic latitudes (GLat $>$ 0), blue for negative latitudes (GLat $<$ 0), and black for the overall sample. Whiskers extend to the most extreme data points within 1.5 $\times$ IQR, revealing the dispersion across different Galactic longitudes. The orange line (right axis) quantifies the source count within each longitude bin.}
\label{MWISP_Clumps_LV}
\end{figure*}

\subsection{Clump Information and Detection Catalog}\label{Data_Products}
The MWISP Phase I, spanning a decade of observations, maintains generally uniform sensitivity across its datasets, yet localized variations emerged due to temporal changes in observational parameters (e.g., atmospheric transparency fluctuations and receiver configuration upgrades). To mitigate these variations, we perform spectral-axis-normalized noise calibration through pixel-wise division of the raw data cube by its corresponding noise map, generating a unified signal-to-noise ratio (S/R) cube (where the noise RMS = 1). This S/R-based processing approach optimally preserves high-S/R sources in low-background regions while suppressing false detections in high-noise areas \citep[e.g.,][]{SN_Map_1,CHIMPS_2,SN_Map_2}. 

We implement the equivalent global detection scheme in conjunction with  FacetClumps on the S/N cube, where the S/N cube serves exclusively as the basis for mask extraction. Clump property calculations utilize the original brightness temperature data with the corresponding global mask. We have extracted 71,661 molecular clumps within approximately 6 hours of computational time. Representative clump images are displayed in Figure~\ref{Imgs_ExClumps} of Appendix~\ref{Clump_Examples}, and the detection catalog is detailed in Table \ref{Catalogue_Table_Detection} of Appendix~\ref{Tables}. 


Figure~\ref{MWISP_Clumps_LBV} shows the spatial distribution of clumps in both Galactic longitude-latitude and longitude-velocity projections, with the latter overlaid with major Galactic spiral arm tracers. Figure~\ref{MWISP_Clumps_LV} reveals the distribution of clumps across Galactic coordinates, highlighting a notable asymmetry in their vertical distribution. The overall median Galactic latitude (black line) oscillates between approximately $-2^\circ$ and $+3^\circ$ throughout the observed longitude range. The positive latitude population (red line) maintains higher median values (typically $+1^\circ$ to $+3^\circ$) compared to their negative latitude counterparts (blue line, ranging from $-2^\circ$ to $-0.5^\circ$), indicating a north-south asymmetry in molecular gas distribution. Source counts exhibit a significant concentration within the $10^\circ$–$40^\circ$ longitude range, reflecting the molecular-rich regions in the inner Galaxy. This distribution peaks in the $20^\circ$–$30^\circ$ interval, where substantial star-forming complexes such as Aquila Rift are located \citep{MWISP_Analysis_Su_2020}. Similarly, a notable concentration is observed within the $70^{\circ}$--$100^{\circ}$ longitude range, associated with the prominent star-forming region in Cygnus \citep{MWISP_Analysis_Zhang_2024}.

\section{Distance Determination}\label{Distance}
Accurate distance determination to clumps forms the cornerstone for deriving their physical properties. While distances to a number of nearby large-scale molecular clouds have been established using reddened background stars, these measurements continue to undergo updates and refinements \citep{Emission_D6,Emission_D5,Emission_D1,Emission_D2,MWISP_Analysis_SunL_2024,MWISP_Analysis_Zhang_2024,MWISP_Analysis_Zhuang_2024}. To overcome the challenge of distance determination for broad sky-survey samples such as MWISP, we introduce a hierarchical distance decision method that prioritizes maser parallax distances when available and incorporates kinematic distance measurements \citep{ATLASGAL_2018}, ensuring maximum completeness in distance determination across the entire sample. 

KDA is a major issue in the first Galactic quadrant (FGQ, $l=0^\circ-90^\circ$), where a single velocity can correspond to both near and far distances. Clumps in the second (SGQ, $l=90^\circ-180^\circ$) and third (TGQ, $l=180^\circ-270^\circ$) Galactic quadrants are free from KDA due to their positions outside the solar circle, where the rotation curve gives clear distance solutions. Traditional clump-centric KDA resolution indicates that clumps within larger-scale structures, such as filaments \citep[e.g.][]{MST,MST_Apply_1} and clouds \citep[e.g.][]{SEDIGISM_2}, often exhibit both near and far kinematic distance solutions. Cross-matching with higher-spatial-resolution surveys (e.g., Hi-GAL, \citealt{HiGal_Dist}; ATLASGAL, \citealt{ATLASGAL_2018}) further shows that a single MWISP clump may correspond to multiple distance-assigned clumps in these surveys. Additionally, when using automatic HI self-absorption (HISA) determination as outlined in Section \ref{KDA_SR}, faint clumps may lack corresponding HI absorption dips and are more susceptible to interference from noise in the HI spectral lines. 

To address these inconsistencies, we adopt SRs (extended structures encompassing both clumps and filaments) as the fundamental objects for KDA resolution. Furthermore, using SRs as determination units enables more reliable distance classification for clumps located near boundaries, such as those around the solar circle, tangent range, and quadrants ($l=90^\circ$). While we calculate distances for individual clumps within each SR, a limitation emerges when SRs contain multiple velocity components, as assigning uniform near or far distances to all contained clumps may introduce inaccuracies, particularly at low Galactic longitudes (\(\sim 40^\circ\)). If in the future we can effectively separate different multiple velocity components, the precision of distance classification will be further improved. Despite this constraint, the SR-centric approach holds considerable advantages over clump-centric determination. 

\begin{figure*}
\centering
\centerline{\includegraphics[width=7in]{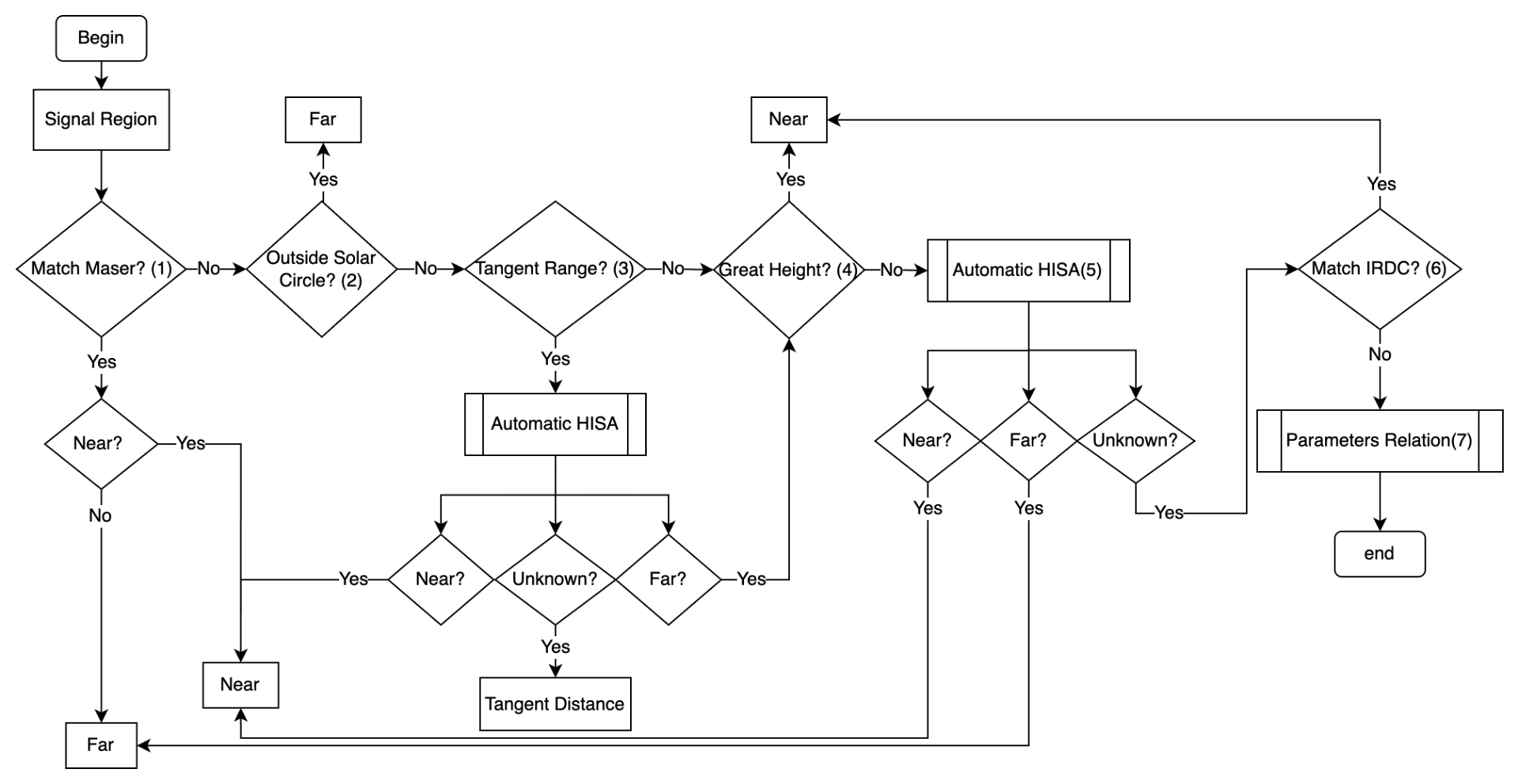}}
\caption{Flow chart showing the KDA solutions procedure adopted for the MWISP SRs in the FGQ.}
\label{KDA_Flow}
\end{figure*}

\subsection{Resolving the KDA of SRs}\label{KDA_SR}
The KDA problem in the FGQ SR is resolved using the process outlined in Figure \ref{KDA_Flow}. This process consists of seven main subprocesses, with the reliability of each step gradually decreasing. Before making determinations, the kinematic near, far, and tangent point distances are calculated for each SR and clump using the Galactic rotation curve model ($R_0 = 8.15$ kpc and $\theta_0 = 240$ \text{km}~\text{s}$^{-1}$; \citealt{Distance_2,Distance_1}). 

\begin{itemize}
\item Matching masers (1): The maser sources with accurate parallaxes are matched with the SRs (considered valid if located within the mask region). A SR may match multiple masers. Then, the classification is based on the distance relationship between all matched masers and the tangent point distance at that longitude. If half of the masers have a distance less than the tangent point distance, the SR is classified as near distance; otherwise, it is classified as far distance. 

\item Outside solar circle (2): The $V_{\text{LSR}}$ of the SR is checked to determine whether it is less than 0. If it is, the SR is located in the outer solar circle, and the distance is classified as far distance. 

\item Tangent point zone (3): The tangent point range is defined as the speed range where \( V_{\text{LSR}} - V_{\text{Tangent}}\) $>$ -5 km s$^{-1}$ \citep{SEDIGISM_2}. If the SR’s $V_{\text{LSR}}$ falls within this range, the distance is first judged using the automatic HISA method described in (5). If classified as far distance, further analysis using the Galactic disk height constraint method described in (4) is applied. If the HISA method does not obtain a clear distance solution, the distance at the tangent point is adopted. 

\item Height constraint (4): When a far-distance solution results in a height \( Z > 120~\mathrm{pc} \) \citep[the galactic disk height threshold of clumps;][]{KDA_Method_Height_1,ATLASGAL_2018,KDA_Method_Height_2}, the SR is classified as near distance. 

\item Automatic HISA method (5): Molecular clumps at nearby distances typically exhibit absorption features in the hydrogen spectral line corresponding to their source velocity, as cold clumps absorb radiation from warmer hydrogen gas situated behind them. Conversely, distant clumps show no such absorption features at their source velocity when warmer hydrogen gas is distributed throughout the galactic plane. This phenomenon has been extensively employed in numerous studies tackling distance ambiguity \citep{GRS_Dist_1,ATLASGAL_2018,SEDIGISM_2,CHIMPS_2,HiGal_Dist}. 

We have compiled HI surveys overlapping with the MWISP survey in the FGQ, prioritizing data sources based on resolution: the Galactic Arecibo $L$-Band Feed Array Hi survey \citep[GALFA-Hi;][]{GalFa}, the VLA Galactic Plane Survey \citep[VGPS;][]{VGPS}, the Southern Galactic Plane Survey \citep[SGPS;][]{SGPS}, the Canadian Galactic Plane Survey \citep[CGPS;][]{CGPS}, and HI4PI \citep{HI4PI}. Figure \ref{HI_Surveys_Range} illustrates the distribution of SRs determined by HISA from these various projects.

To systematically and objectively evaluate HISA features across such a broad sky area, we develope an automated classification method illustrated in Figure \ref{KDA_HI_Spectral}. When self-absorption features are detected near the SR, the distance is classified as near. Conversely, the presence of HI emission peaks suggests a far classification. If neither feature is apparent, the distance remains undetermined. 

\begin{figure*}
\centering
\centerline{\includegraphics[width=7in]{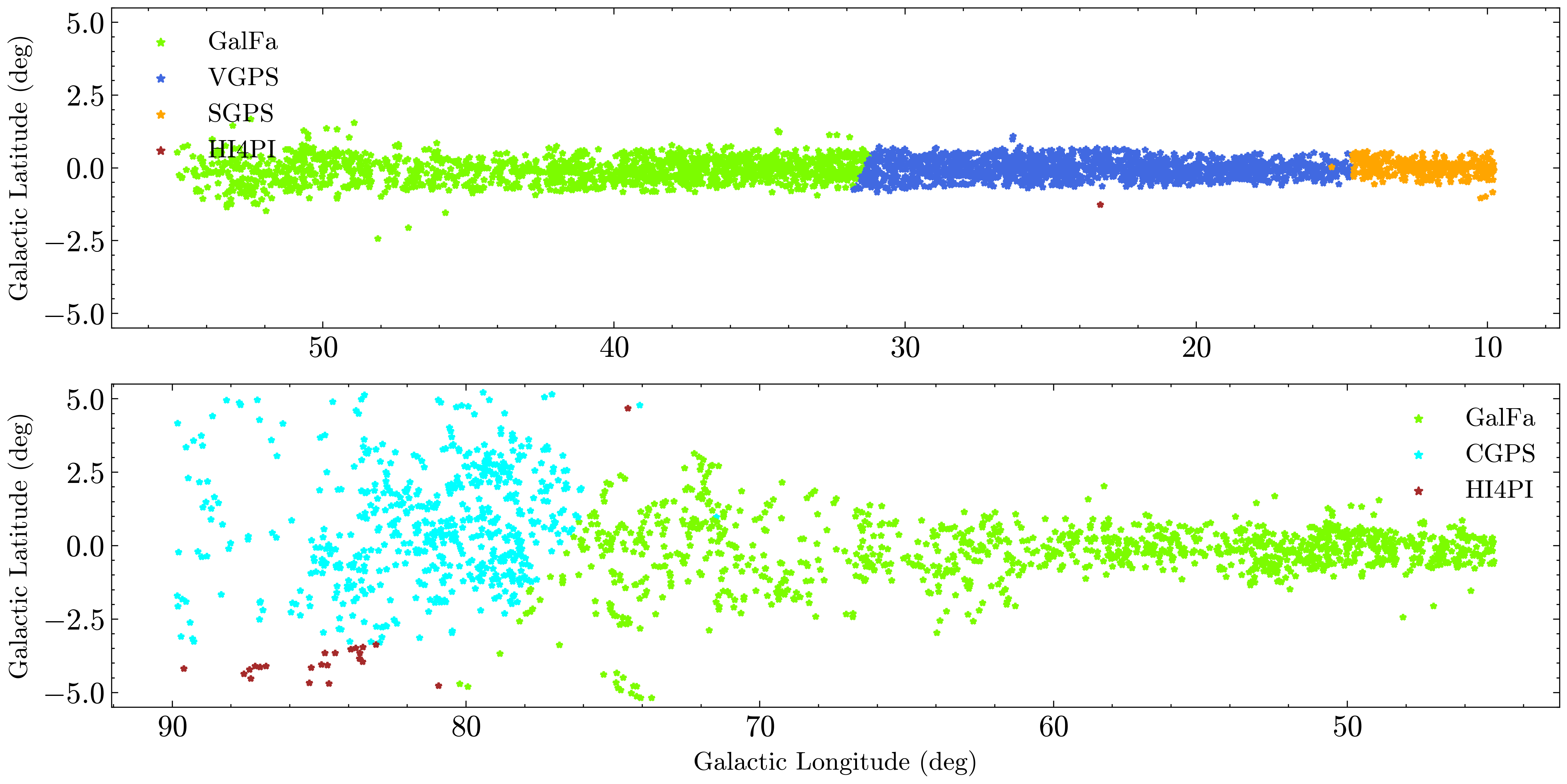}}
\caption{The coverage of HI data from different surveys utilized in the automatic HISA method. The priority order of HI data is arranged in the sequence of GALFA, VGPS, CGPS, SCPS, and HI4PI. Asterisks denote the positions of SRs.}
\label{HI_Surveys_Range}
\end{figure*}

\begin{figure*}
\centering
\vspace{0cm}
\begin{minipage}[t]{0.42\textwidth}
    \centering
    \centerline{\includegraphics[width=3in]{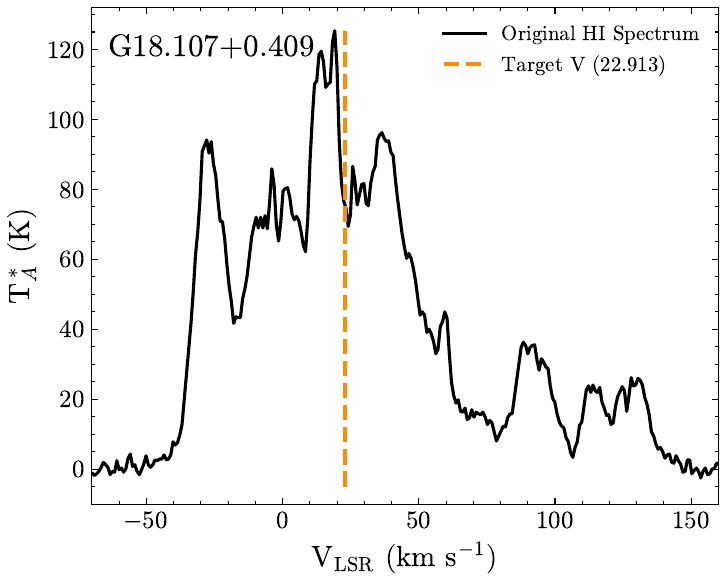}}
\end{minipage}
\begin{minipage}[t]{0.42\textwidth}
    \centering
    \centerline{\includegraphics[width=3in]{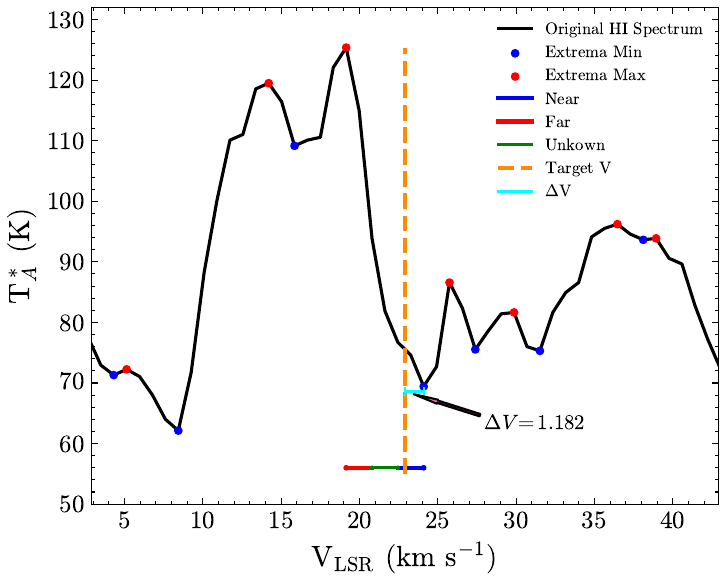}}
\end{minipage}
\caption{Example of automatic HISA identification for the source G18.107+0.409+22.913 \citep[filament Fil 7 from][]{DPConCFil}. Left panel: the black curve shows the signal-point HI spectrum of the target source from the VGPS \citep{VGPS} survey, with the orange vertical line indicating the velocity of the source. Right panel: this is a zoomed-in view of the left panel. Red points mark the local maxima of the HI profile, while blue points mark minima. The velocity range between adjacent extrema is divided into three segments: the red segment (near maximum) corresponds to the far kinematic distance solution, the blue segment (near minimum) corresponds to the near solution, and the green intermediate segment denotes ambiguous cases. The target's systemic velocity (orange line) falls within the blue segment. The velocity difference between the target and the nearest minimum (1.182 km s$^{-1}$) is smaller than the tolerance threshold (3 km s$^{-1}$), leading to its classification as near kinematic distance at this spectrum.}
\label{KDA_HI_Spectral}
\end{figure*}

The automated approach first identifies all local maxima and minima in the HI spectral line and divides the velocity range of local extrema around the target velocity into three equal segments. The distance is classified as near when the target velocity falls within a local minimum region with a velocity difference smaller than the tolerance \citep[typically 3 km s$^{-1}$;][]{HiGal_Dist}. When the target velocity occurs within a local maximum region with the same velocity tolerance, the classification is far. Target velocities falling within the intermediate range yield an unknown classification. 

Given that SRs typically exceed the spatial resolution of HI spectral lines, a single SR may correspond to multiple HI spectra. To enhance robustness against noise and fully leverage all available spectral information, we extract three distinct types of HI spectra:

\begin{itemize}
    \item Multiple single-point HI spectra (HILine1s) for each spatial point within the SR,
    \item A central single-point HI spectrum (HILine2) at the SR's center,
    \item An integrated HI spectrum (HILine3) averaged across the entire SR.
\end{itemize}

Each spectrum undergoes classification according to the method depicted in Figure \ref{KDA_HI_Spectral}. For HILine1s, majority voting determines the overall classification, with ties defaulting to unknown. When the predominant classification from HILine1s matches those derived from HILine2 or HILine3, we adopt that classification; otherwise, the distance remains undetermined.

\item Matching Infrared Dark Clouds (IRDCs) (6): The kinematic distance assignment for IRDCs relies on their mid-infrared extinction features against Galactic background emission \citep{IRDC_1}. Since detecting extinction requires mid-infrared radiation from the inner Galactic disk, this method tends to favor near-distance solutions \citep{IRDC_3,ATLASGAL_2018,SEDIGISM_2,HiGal_Dist}, with an associated uncertainty of approximately 10\% \citep{IRDC_3}.

If an IRDC lies within an SR, and the SR at the center of the IRDC exhibits the most velocity channels along the line of sight, the SR is classified as near. Due to line-of-sight blending, an IRDC may correspond to multiple SRs at different velocities; however, IRDCs generally correspond to the strongest velocity component \citep[e.g.,][]{IRDC_4}. This approach helps minimize the interference of weaker components by the stronger ones, thereby enhancing the accuracy of the classification.

\item Parameter relations method (7):
Distance determination based on the empirical $\sigma_v$-$R$ relation \citep{KDA_Method_PR_1} resolves near and far kinematic ambiguities by requiring consistency with established scaling laws, a methodology widely implemented in inner Galactic molecular cloud studies \citep{KDA_Method_PR_4,KDA_Method_PR_5,SEDIGISM_2}. However, as demonstrated in Section~\ref{PRs_DS} and corroborated by previous investigations \citep[e.g.,][]{KDA_Method_PR_2,Review_13,CFA_2017,MWISP_Analysis_Feng_2024}, the composite parameter relation \(\sigma_v\)–\(\Sigma R\) exhibits reduced scatter and demonstrates enhanced robustness across diverse dynamical environments within the Milky Way. Accordingly, we adopt this scaling relation as the premise for developing near- or far-distance discriminants \citep{CFA_2017}.

Building on obtained SR classifications, we propagate distance solutions to individual clumps through the process detailed in Section~\ref{KDA_Clump}, successfully determining distances for over 96\% of clumps in the FGQ. We construct a $\sigma_v$-$R\Sigma$ distribution plane using samples with known distances as described in Section~\ref{PRs_DS}. For each clump with undetermined classification, we calculate the Mahalanobis distance \citep{SciPy} between its derived ($\sigma_v$, $R\Sigma$) parameters and the multivariate distributions corresponding to both near and far kinematic solutions. This statistical metric, superior to Euclidean alternatives due to its covariance-matrix normalization, effectively accounts for intrinsic parameter correlations in the $\sigma_v$-$R\Sigma$ plane and accommodates varying measurement uncertainties. We then adopt the kinematic solution that minimizes this metric. 

We expect all clumps within a given SR to share identical distance classifications, but the results of this procedural step have not been unified. In the final stage of this step, a majority vote consolidates clump distances into their parent SR, with far-distance solutions taking precedence in tied cases, after which the clump classifications are updated according to their parent SR designation. 
\end{itemize}

\begin{figure}
\centering
\centerline{\includegraphics[width=3.2in]{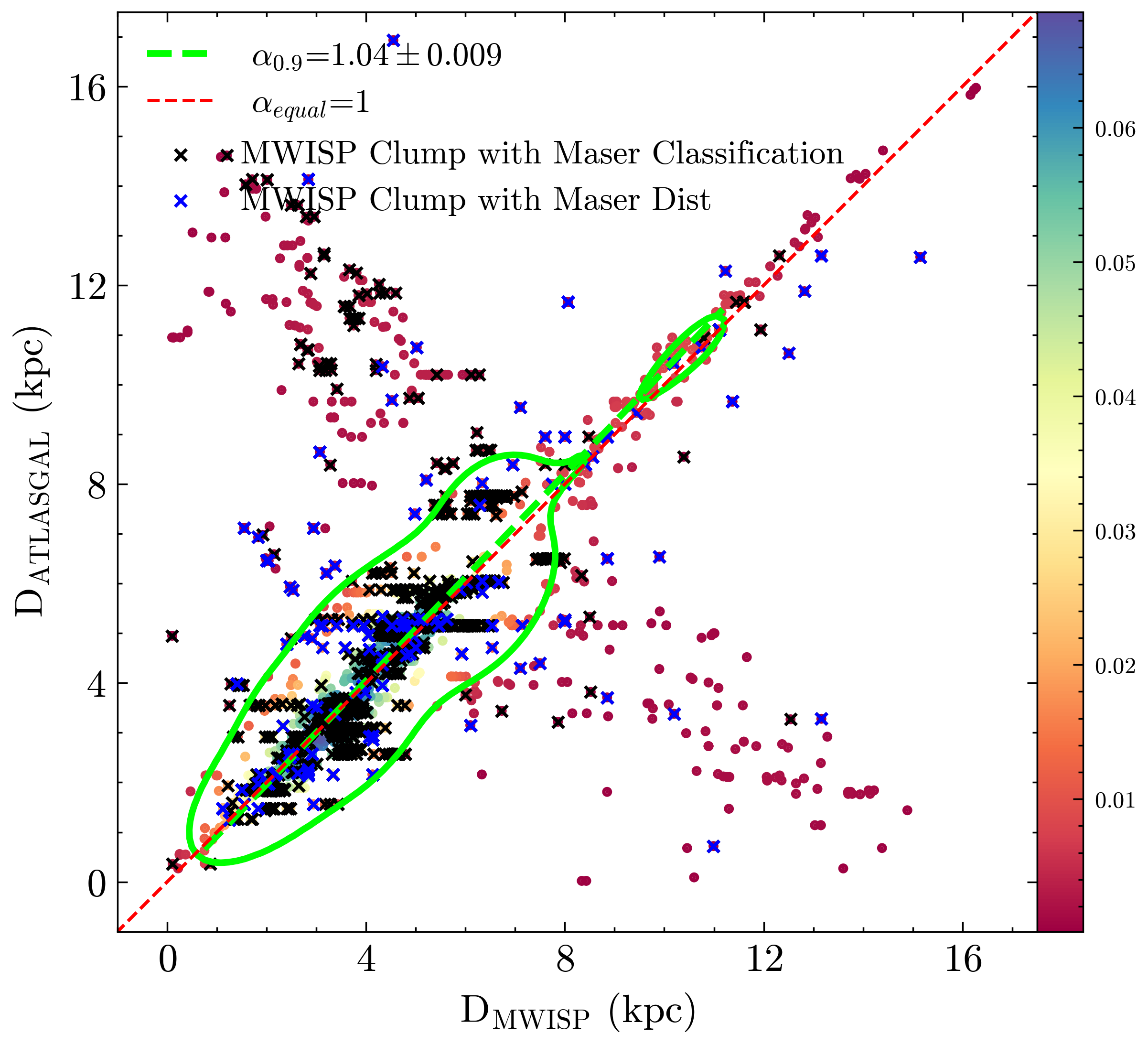}}
\caption{Distance comparison of paired ATLASGAL and MWISP clumps. Black crosses mark MWISP clumps whose distance classification is based on maser associations, while blue crosses mark MWISP clumps that utilize maser distance measurements. The density distribution of matched clumps is estimated via KDE, where the color scale represents the relative probability density of clump occurrences. The lime contour outlines the region where KDE values exceed 90\% of the distribution. Within this contour, the lime dotted line represents the fitted linear relationship to the clump distance pairs, with a slope of $1.04\pm0.009$. The red dashed line represents the $y = x$ reference line. }
\label{SMatch_Dist_Agal}
\end{figure}

\begin{figure*}
\centering
\centerline{\includegraphics[width=5in]{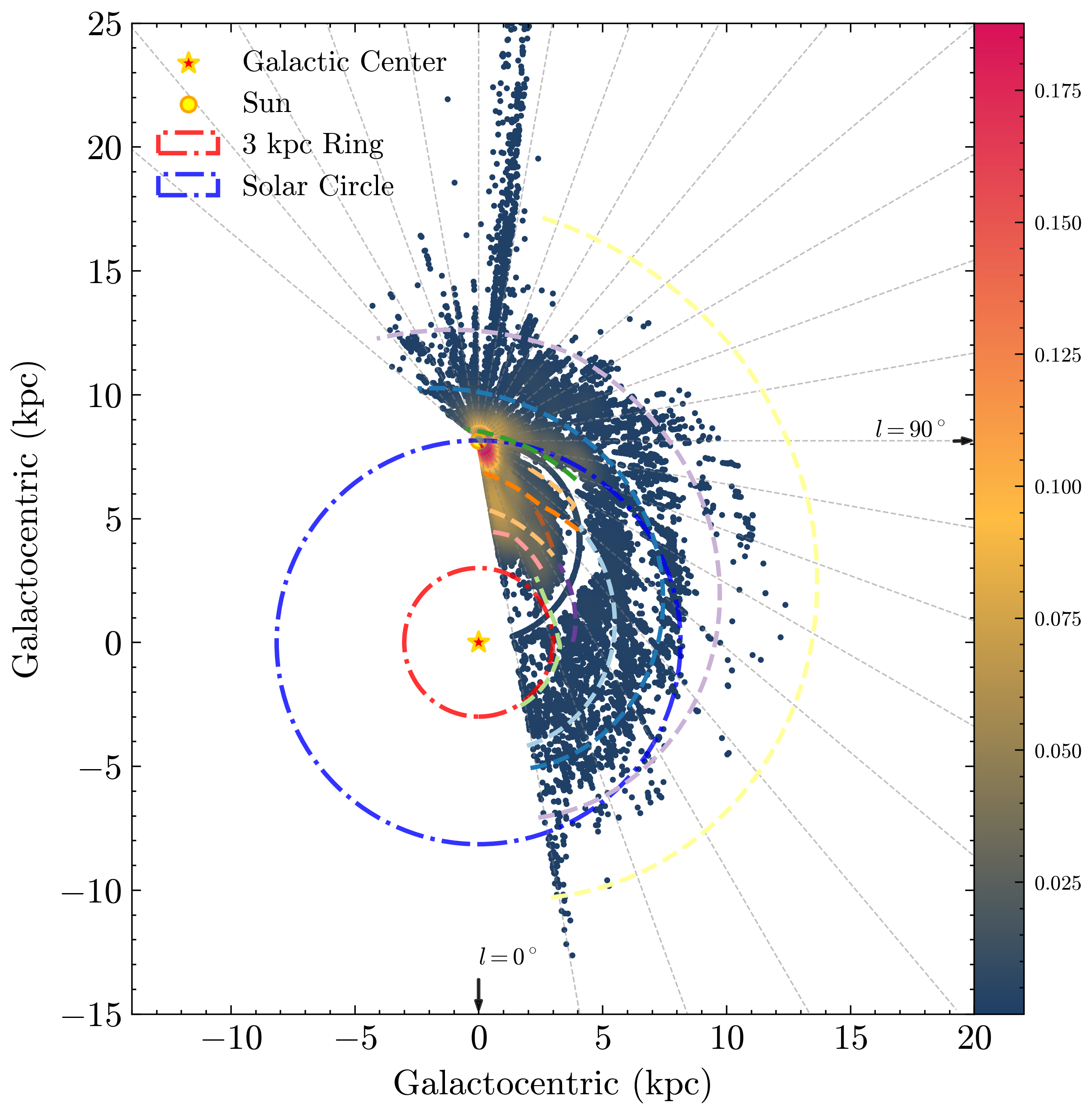}}
\caption{Face-on number density distribution map of clumps in the Galactic plane. The coordinate system centers on the Galactic center (red star) with the Sun positioned at $(0, 8.15)$~kpc (yellow circle). The red dashed-dot circle represents the 3 kpc ring around the Galactic Center, while the blue dashed-dot circle represents the solar circle. The number density distribution is calculated using KDE, where the color scale represents the density of clumps, with higher density regions depicted in greater estimated values. Gray dashed lines show Galactic longitude lines (10° intervals). Axes are centered at the Galactic center, with the negative $y$-axis toward $l = 0^\circ$ and the positive $x$-axis toward $l = 90^\circ$. The colored dashed lines depict the spiral arms, with colors matching those in the bottom panel of Figure \ref{MWISP_Clumps_LBV}, based on the model by \cite{Distance_1}.}
\label{Number_Density}
\end{figure*}

\subsection{Distance Assignment of Clumps}\label{KDA_Clump}
The hierarchical distance determination process assigns classifications to all SRs in the FGQ. Due to some SRs having large spatial scales, it is necessary to calculate individual distances for each clump within these SRs. We propagate these SR classifications to individual clumps and assign corresponding distances to each clump. For improved precision, clumps associated with masers and their neighbors receive the corresponding maser distance instead. The distance catalog is detailed in Table~\ref{Catalogue_Table_Distance} of Appendix~\ref{Tables}. 

Table~\ref{KDA_Summary} summarizes the KDA solution statistics. A total of 1718 clumps, approximately 2.4\%, have their distances determined using the maser distance, with the KDA method being \textit{Matching Maser} or located in the SGQ and TGQ. KDA resolution is applied to 72.44\% (51,908) of the total, with 73.7\% at near-distance and 22.7\% at far-distance, in agreement with previous studies \citep[e.g.][]{GRS_Dist_2,ATLASGAL_2018}. Furthermore, 3.6\% of the clumps have an unknown KDA classification, with these clumps assigned a tangent distance as indicated in Figure~\ref{KDA_Flow}. The final catalog of physical parameters encompasses 70,184 clumps, achieving a completeness of 97.94\%. Each clump has a validated distance greater than 0.1 kpc, with all unphysical solutions derived from rotation curve extrapolations excluded. 

To assess the validation of the distance, we match MWISP clumps with ATLASGAL clumps (\citealt{ATLASGAL_2018}; defined as ATLASGAL clumps located within MWISP clump masks). This cross-matching yields 2865 ATLASGAL clumps associated with 1938 MWISP clumps, of which 1161 MWISP clumps utilize maser associations for distance classification and 692 employ direct maser distance measurements. As shown in Figure~\ref{SMatch_Dist_Agal}, kernel density estimation (KDE) is applied to smoothly estimate the probability density function. Linear regression performed on clumps within regions exceeding 90\% KDE values produces a fitted slope of 1.04. Despite some clumps showing anti-correlated distances (a common phenomenon between catalogs due to different distance assignments; \citealt{HiGal_Dist}), the overall distribution and quantitative analysis reveal strong agreement between both catalogs, validating the SR-centric automated KDA methodology. This study achieves automated KDA solutions, incorporates more maser distances, and enables a maser to determine multiple clumps through the SR, substantially enhancing the reliability of distance classification. 

Figure~\ref{Number_Density} shows the face-on spatial distribution of molecular clumps in the Galactic plane using a Galactocentric coordinate system. Clump number density peaks in the solar neighborhood and declines with increasing heliocentric distance, mainly due to observational bias as discussed in Section \ref{PRs_Dist}: lower-mass clumps are harder to detect at greater distances. The increased discreteness in longitude ranges $l < 15^{\circ}$ and $165^{\circ} < l < 195^{\circ}$ stems from larger kinematic distance uncertainties in these regions \citep{Distance_2}. Within the visualization, a scatter-point arc demarcates the solar circle, while nearby clumps lacking definitive distance classifications have been assigned tangent distances. Our samples exhibit a cavity structure within 3 kpc of the Galactic center, supporting previous observational findings of molecular gas deficiency in this region \citep[e.g.,][]{Review_17}. The molecular gas deficiency in this region may be related to the combined effects of intense nuclear winds impacting the Galactic gas disk \citep{MWISP_Analysis_Su_2022} and the dynamical influence of the Galactic bar on gas motion \citep{MWISP_Analysis_Su_2024,MWISP_Analysis_Su_2025}.

\begin{table*}
\centering
\caption{Summary of the KDA methods and solutions of clumps. }
\begin{tabular}{c|c|cccc}\hline\hline
    Step&Method&Number (Percentage)&Near&Far&Unknown\\\hline
    (0)&Maser Distances &1718 (2.4\%)&...&...&...\\\hline
    (1)&Matching Masers &13,166 (18.4\%)&88.9\%&11.1\%&0\\\hline 
    (2)&Outside Solar Circle &5136 (7.2\%)&0&100\%&0\\\hline 
    (3)&Tangent Point Zone &5558 (7.8\%)&37.3\%&33.3\%&29.4\%\\\hline 
    (4)&Height Constraint&14,765 (20.6\%)&100\%&0&0\\\hline 
    (5)&Automatic HISA Method &9853 (13.7\%)&70.6\%&29.4\%&0\\\hline 
    (6)&Matching IRDCs &1348 (1.9\%)&100\%&0&0\\\hline 
    (7)&Parameter Relations Method&2078 (2.9\%)&68.2\%&31.8\%&0\\\hline 
    (8)&SGQ and TGQ& 19,757 (27.56\%)&...&...&...\\\hline
    Summary&Clumps with KDA& 51,908 (72.44\%)&73.7\%&22.7\%&3.6\%\\\hline
\end{tabular}
\begin{tablenotes}
\item \textbf{Note.} [1] \textit{Maser Distances} refers to the direct application of parallax distances measured from associated masers, while \textit{Matching Masers} refers to adopting KDA solutions from parent SRs that contain masers. \textit{Maser Distances} is a subset of \textit{Matching Masers}.

[2] SGQ and TGQ refer to clumps that do not require KDA classification. 
\end{tablenotes}
\label{KDA_Summary}
\end{table*}

\begin{figure*}
\centering
\vspace{0cm}
\begin{minipage}[t]{0.4\textwidth}
    \centering
    \centerline{\includegraphics[width=2.5in]{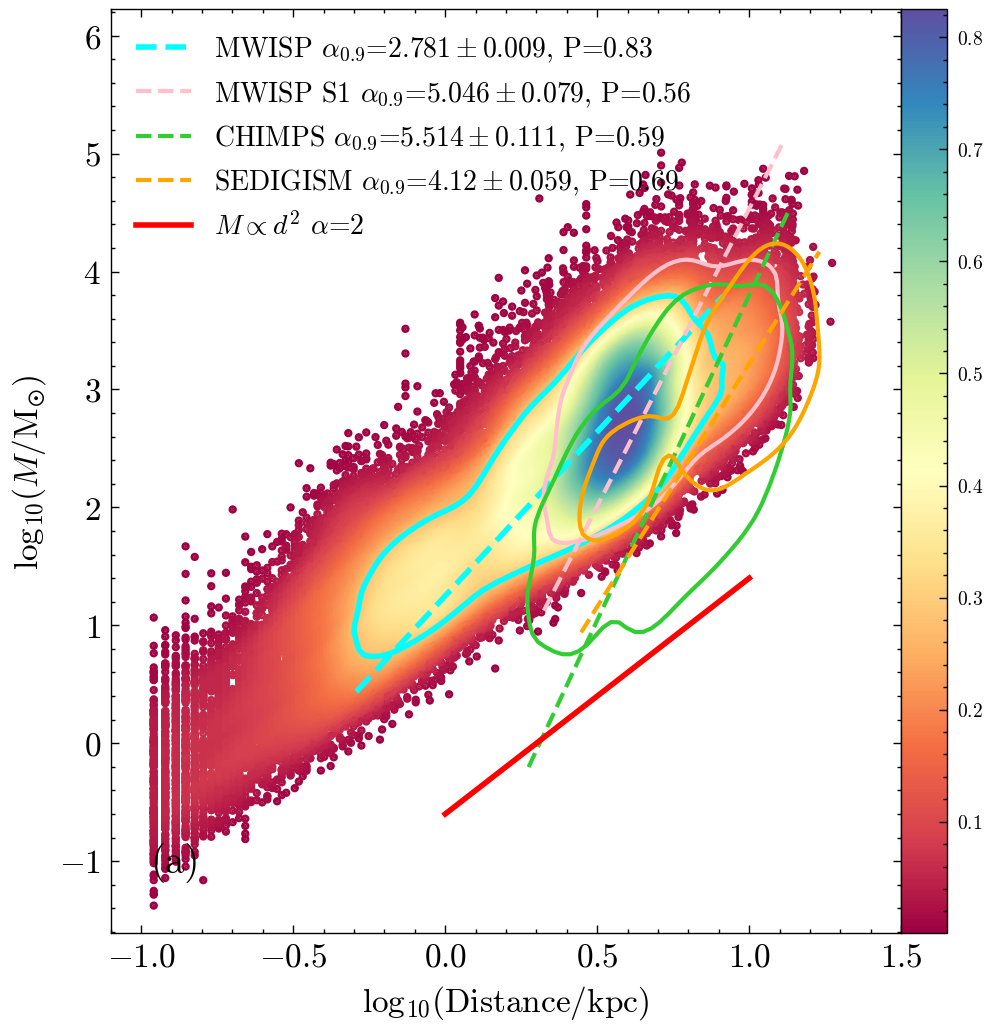}}
\end{minipage}
\begin{minipage}[t]{0.4\textwidth}
    \centering
    \centerline{\includegraphics[width=2.5in]{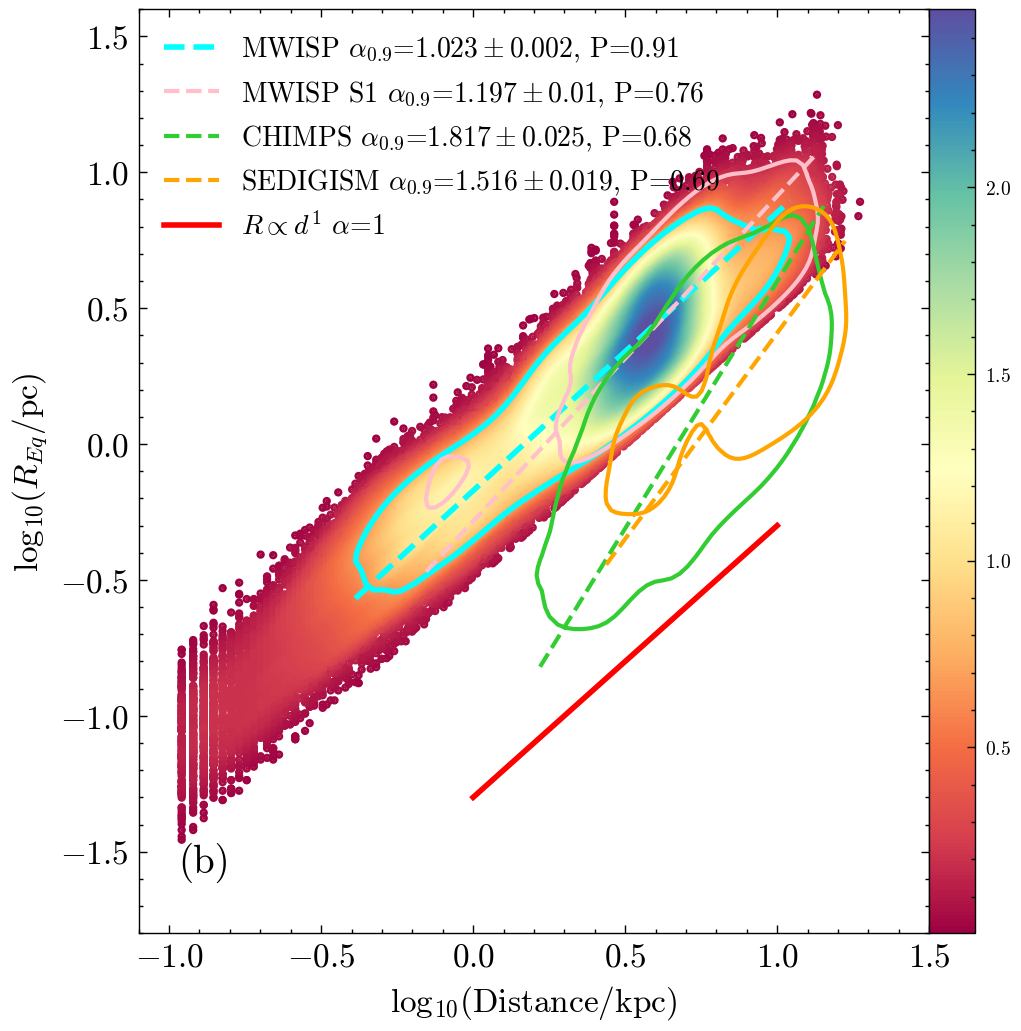}}
\end{minipage}

\begin{minipage}[t]{0.4\textwidth}
    \centering
    \centerline{\includegraphics[width=2.5in]{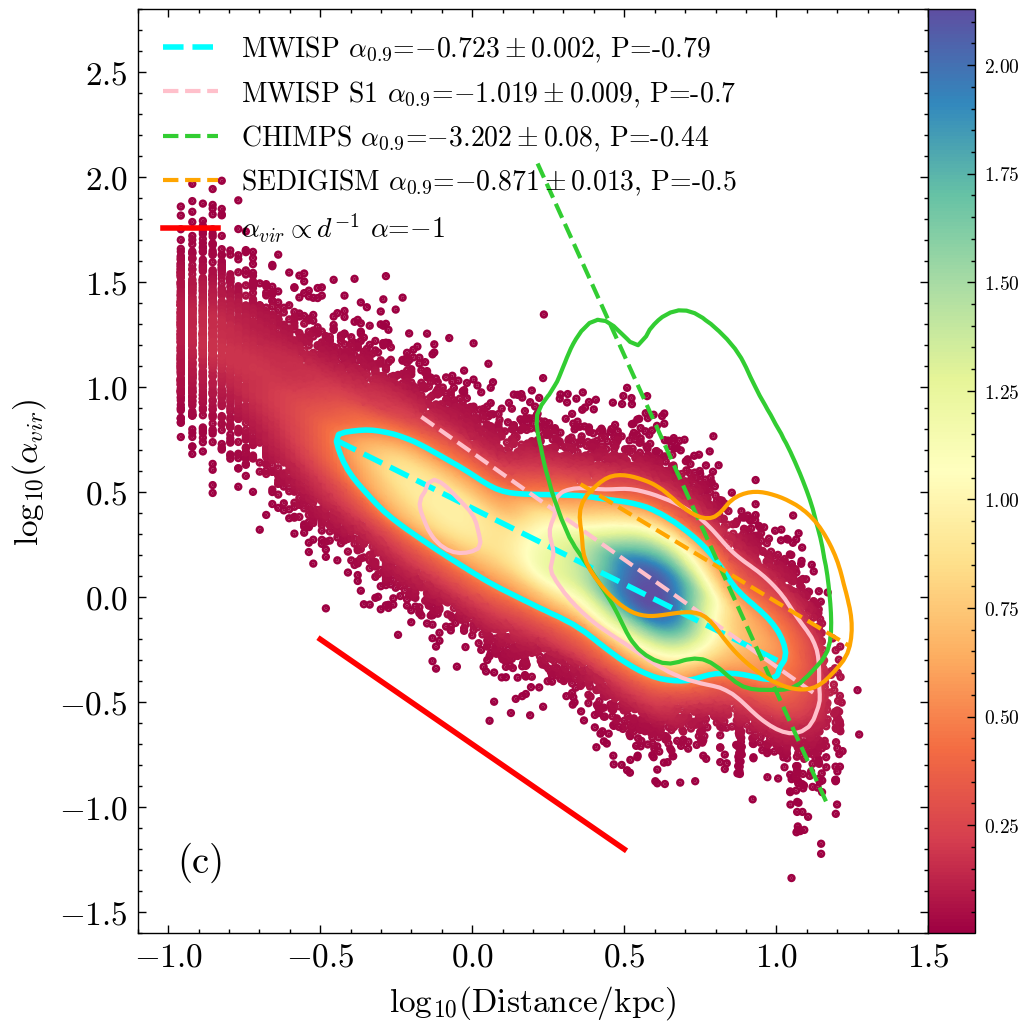}}
\end{minipage}
\begin{minipage}[t]{0.4\textwidth}
    \centering
    \centerline{\includegraphics[width=2.5in]{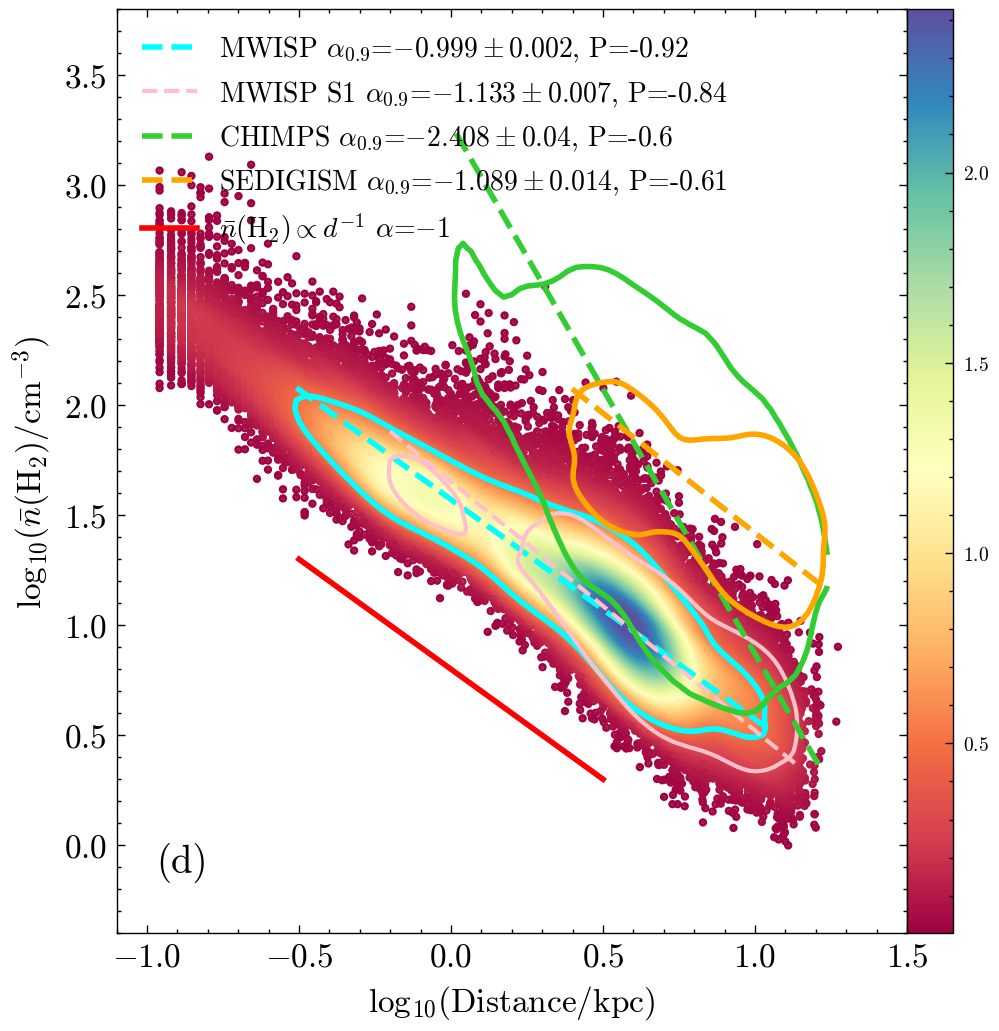}}
\end{minipage}
\caption{Relationships between distance-dependent parameters and heliocentric distance. Panels show (a) $M-d$; (b) $R_{eq}-d$; (c) $\alpha_{\text{vir}}-d$; and (d) $n(\mathrm{H}_2)-d$. Scatter points from the MWISP project are color-coded according to their KDE values, as indicated by the color bar, reflecting the local data density. MWISP S1 represents a subset of the MWISP survey that covers a similar Galactic sector to CHIMPS. Contour lines, labeled for MWISP (cyan), MWISP S1 (pink), CHIMPS (lime green), SEDIGISM (orange), highlight regions with KDE values exceeding 90\%. The red lines represent the reference lines showing the theoretical parameter-distance relationships. For each survey's high-density region, $\alpha_{0.9}$ denotes the fitted power-law slope of the relationship, presented alongside its uncertainty and $P$.}
\label{Imgs_PR_1}
\end{figure*}

\begin{figure*}
\centering
\vspace{0cm}
\begin{minipage}[t]{0.3\textwidth}
    \centering
    \centerline{\includegraphics[width=2.2in]{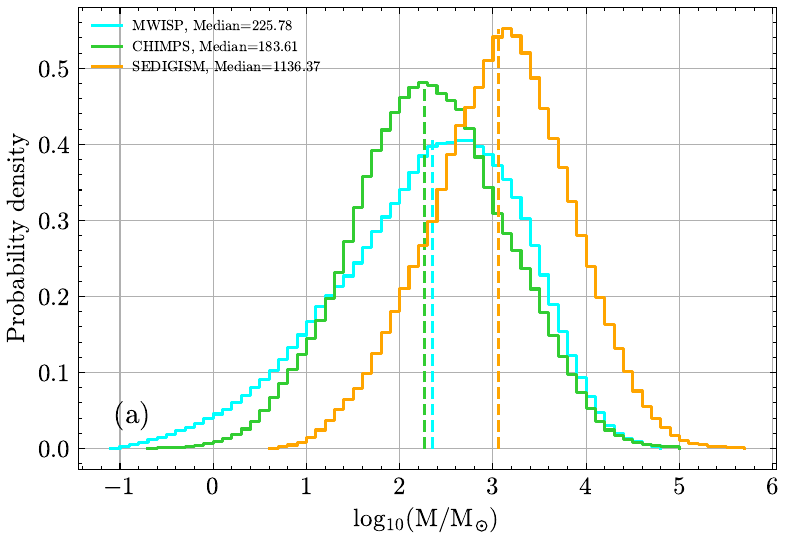}}
\end{minipage}
\begin{minipage}[t]{0.3\textwidth}
    \centering
    \centerline{\includegraphics[width=2.2in]{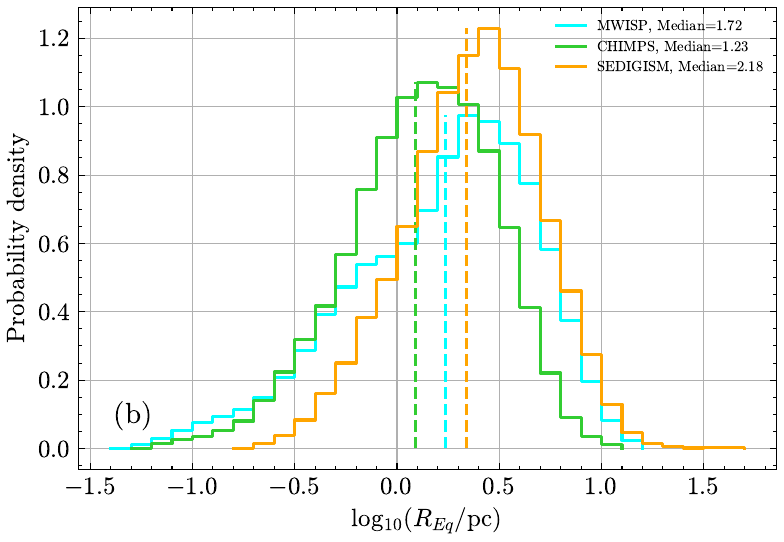}}
\end{minipage}
\begin{minipage}[t]{0.3\textwidth}
    \centering
    \centerline{\includegraphics[width=2.2in]{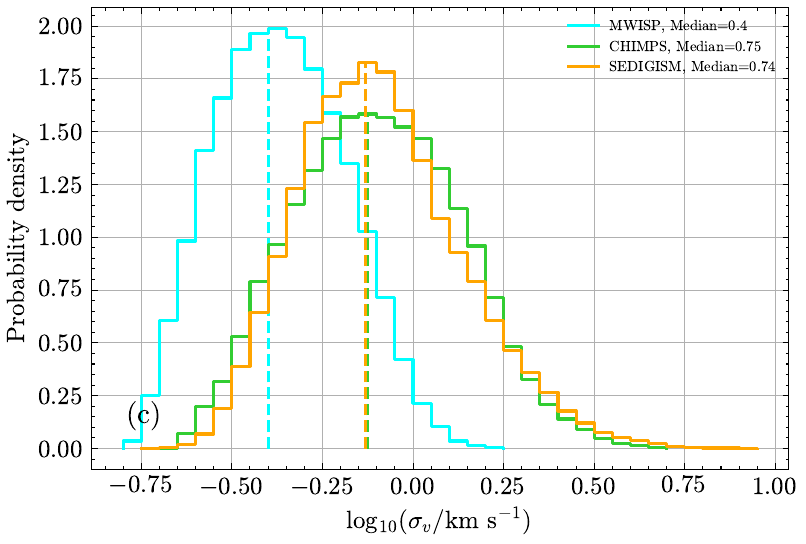}}
\end{minipage}

\begin{minipage}[t]{0.3\textwidth}
    \centering
    \centerline{\includegraphics[width=2.2in]{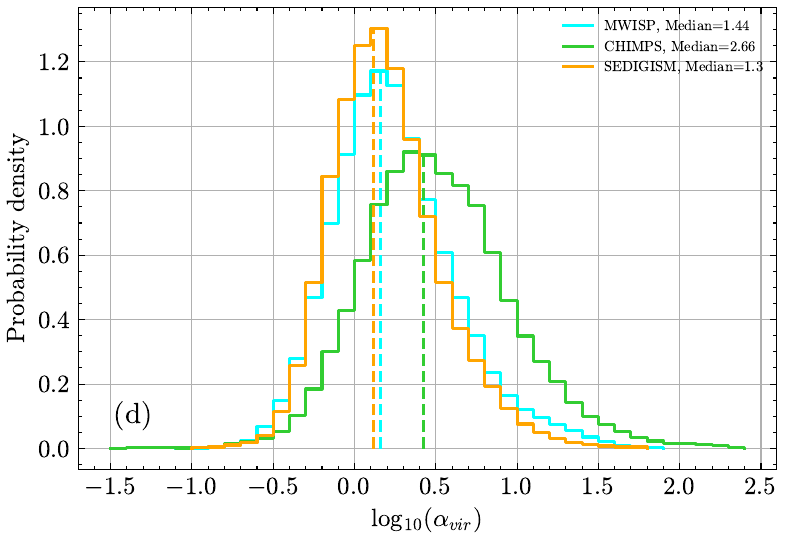}}
\end{minipage}
\begin{minipage}[t]{0.3\textwidth}
    \centering
    \centerline{\includegraphics[width=2.2in]{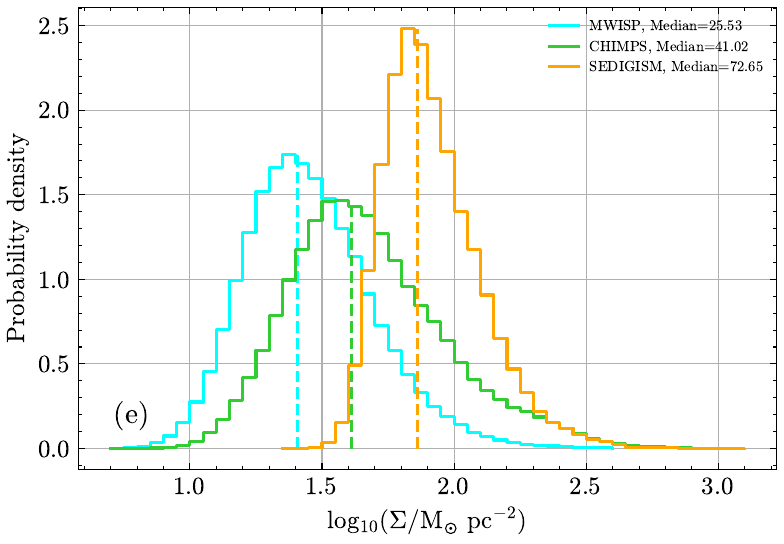}}
\end{minipage}
\begin{minipage}[t]{0.3\textwidth}
    \centering
    \centerline{\includegraphics[width=2.2in]{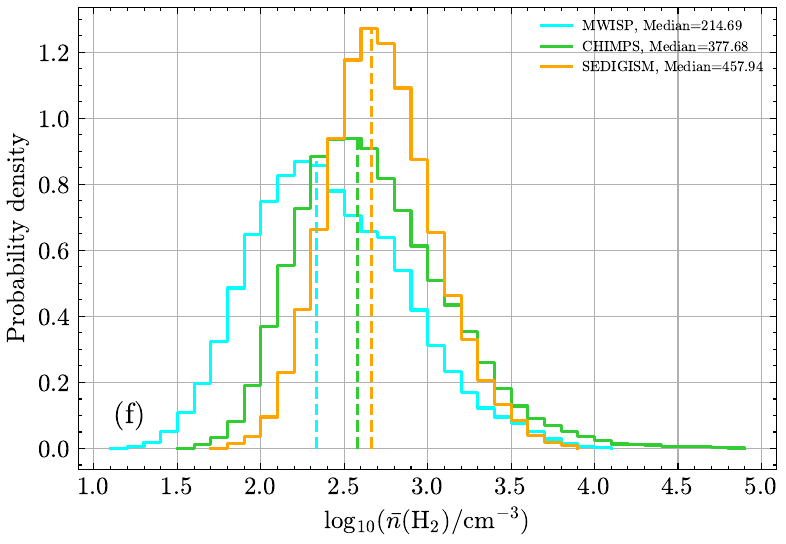}}
\end{minipage}
\caption{Probability distribution function histograms for six physical parameters. Panels show (a) $M$; (b) $R_{eq}$; (c) $\sigma_v$; (d) $\alpha_{\text{vir}}$; (e) $\Sigma$; and (f) $n(\mathrm{H}_2)$. Each subplot compares data from MWISP, CHIMPS, and SEDIGISM, with their respective median values marked. }
\label{Imgs_Hist_Pars}
\end{figure*}

\section{Physical Properties and Scaling Relations}\label{PProperties}
Utilizing multiple transitions of $^{12}$CO and $^{13}$CO spectral lines, we conduct radiative transfer analysis under the LTE assumption, with the detailed calculation methods presented in Appendix \ref{Cal_Equations}. The analytical process begins with calculating the excitation temperature ($T_{\text{ex}}$) from the brightness temperature ratio of isotopic lines, followed by constraining the optical depth ($\tau$) through radiative transfer balance equations, and subsequently deriving the column density ($N(^{13}\mathrm{CO})$) via integration of the population distribution over the mask velocity space. From the $N(^{13}\mathrm{CO})$ measurements, we derive distance-dependent physical parameters: mass ($M$), equivalent radius ($R_{\rm eq}$), surface density ($\Sigma$), and volume density ($n(\mathrm{H}_2)$). Additionally, we calculate the virial parameter ($\alpha_{\rm vir}$) with velocity dispersion ($\sigma_v$) obtained from spectral line measurements to assess the dynamical state of clumps. 

The catalog of physical parameters is detailed in Table \ref{Catalogue_Table_Physical} of Appendix \ref{Tables}, with representative parameter maps of exemplary clumps displayed in Appendix \ref{Clump_Examples}. A quantitative assessment of parameter biases stemming from sensitivity thresholds is presented in Appendix \ref{ExtMethod}. To minimize measurement uncertainties, we exclude samples with kinematic distances and their corresponding SRs located within the longitude ranges $l < 15^{\circ}$ and $165^{\circ} < l < 195^{\circ}$. This refined dataset comprises 61,202 clumps, representing 85.4\% of the total observed population. 

To establish broader context for MWISP survey findings, we compare our results with those of two other Galactic molecular cloud surveys. The CHIMPS survey \citep{CHIMPS_1} mapped $^{13}$CO (3--2) and C$^{18}$O (3--2) emission within $27.8^\circ < |l| < 46.2^\circ$ and $|b| < 0.5^\circ$, with a spatial resolution of 15$^{\prime\prime}$ and a channel width of 0.5 km~s$^{-1}$, and cataloged 4999 $^{13}$CO (3--2) clumps using FellWalker \citep{CHIMPS_2,FellWalker}; and the SEDIGISM survey \citep{SEDIGISM_1} observing $^{13}$CO (2--1) emission across $-60^\circ < |l| < +18^\circ$ and $|b| < 0.5^\circ$, with a spatial resolution of 28$^{\prime\prime}$ and a channel width of 0.25 km~s$^{-1}$, and cataloged 10,663 molecular clouds using SCIMES \citep{SEDIGISM_2,SCIMES}. 

In the following sections, we first analyze distance-dependent observational biases that influence clump parameter measurements, then examine physical parameter distributions across different surveys, and finally investigate scaling relations among clump populations, with particular attention to maser-associated clumps as tracers of active star formation phases. The detailed scaling relations between extrapolated physical parameters are presented in Appendix \ref{PRs_Ext}. 

\subsection{Distance-Dependent Observational Biases}\label{PRs_Dist}

Figure \ref{Imgs_PR_1} evaluates the power-law relationships between distance-dependent physical parameters ($M$, $R_{\text{eq}}$, $\alpha_{\text{vir}}$, and $n_{\text{H}_2}$) and heliocentric distance. MWISP S1 represents a subset of the MWISP survey that encompasses a similar Galactic sector to CHIMPS, spanning the range $27^\circ < l < 47^\circ$ and $|b| < 0.5^\circ$. The scatter points are color-coded according to their KDE values, reflecting local data density, while contour lines highlight high-density regions with KDE values exceeding 90\%. The fitted power-law slopes ($\alpha_{0.9}$) are calculated through orthogonal distance regression \citep{CHIMPS_2,SciPy,MWISP_Analysis_Feng_2024}, and Pearson correlation coefficients ($P$) quantify the correlation significance. The red lines represent theoretical distance-parameter relationships as described in Appendix \ref{Cal_Equations_2} for comparison. 

Figures \ref{Imgs_PR_1}(a) and (b) reveal parallel behavior in mass and size parameters across different surveys. Figure \ref{Imgs_PR_1}(a) shows positive correlations between mass and distance with slopes steeper than the theoretical geometric effect $M \propto d^2$ expectation (MWISP: $\alpha_{0.9} = 2.781\pm0.009$, $P = 0.83$; MWISP S1: $\alpha_{0.9} = 5.046\pm0.079$; CHIMPS: $\alpha_{0.9} = 5.514$; SEDIGISM: $\alpha_{0.9} = 4.12$). Figure \ref{Imgs_PR_1}(b) demonstrates that MWISP exhibits excellent agreement with the theoretical $R_{\text{eq}} \propto d$ relationship ($\alpha_{0.9} = 1.023\pm0.002$, $P = 0.91$), showing minimal deviation from expected scaling, while MWISP S1, CHIMPS, and SEDIGISM display departures ($\alpha_{0.9} = 1.197\pm0.01$, $1.817$, and $1.516$, respectively). These deviations arise from observational selection effects that systematically bias measurements toward larger values at greater distances: (1) sensitivity-limited Malmquist bias that excludes faint, low-mass clumps \citep{GRS_2}; (2) beam dilution and surface brightness thresholds that preferentially detect bright, extended sources; and (3) angular resolution limitations causing beam confusion where multiple small clumps appear as single massive sources. 

Figure \ref{Imgs_PR_1}(c) shows that the virial parameter follows the expected negative distance dependence, with varying degrees of deviation from the $\alpha_{\text{vir}} \propto d^{-1}$ scaling. MWISP displays a moderate slope ($\alpha_{0.9} = -0.723\pm0.002$, $P = -0.79$), while SEDIGISM ($\alpha_{0.9} = -0.871$) shows closer agreement with the expectation. CHIMPS exhibits the steepest deviation ($\alpha_{0.9} = -3.202$), and MWISP S1 shows intermediate behavior ($\alpha_{0.9} = -1.019\pm0.009$). The deviation from $d^{-1}$ scaling reflects the combined effects of mass and size selection biases discussed above. 

Figure \ref{Imgs_PR_1}(d) presents volume density relationships, where MWISP ($\alpha_{0.9} = -0.999\pm0.002$, $P = 0.92$) closely follows the $n(\text{H}_2) \propto d^{-1}$ expectation with remarkable precision. SEDIGISM ($\alpha_{0.9} = -1.089$) also shows reasonable agreement with the prediction, while MWISP S1 ($\alpha_{0.9} = -1.133\pm0.007$) and CHIMPS ($\alpha_{0.9} = -2.408$) exhibit steeper slopes. These slopes suggest that observational biases preferentially detect denser structures at greater distances, consistent with surface brightness selection effects. 

These correlations between clump parameters and distance \citep[e.g.,][]{GRS_2,SEDIGISM_2,SN_Map_2} are primarily driven by observational constraints rather than intrinsic physical variations. The universal nature of these biases reflects limitations in sensitivity, angular resolution, and surface brightness detection thresholds that affect millimeter and submillimeter surveys. MWISP S1 shows steeper slopes than the complete MWISP sample across all parameters, indicating that observational selection effects vary across different Galactic regions. The comparison between MWISP S1 and CHIMPS, which probe the similar Galactic regions and eliminate potential confusion from different Galactic environments, reveals that MWISP S1 exhibits more moderate slopes than CHIMPS. 

MWISP clumps exhibit slopes closer to theoretical geometric effect reference lines across most parameters, potentially indicating less severe additional distance-related effects than other catalogs. Beyond differences in instrumentation and observational coverage, these reduced biases likely originate from both distance calculation methods and the FacetClumps detection algorithm, which adopts pixel distance-independent parameter constraints and extracts homogeneous clump characteristics, yielding a more uniform population of molecular structures across varying distances. The resulting dataset provides a valuable resource for investigating the intrinsic physical properties of clump structures. 

\subsection{Clump Physical Parameter Distributions}\label{PRs_Hist}
Figure~\ref{Imgs_Hist_Pars} presents probability distribution function histograms for six physical parameters across different molecular cloud/clump catalogs, revealing distinct populations of structures that trace different physical regimes. 

Figure \ref{Imgs_Hist_Pars}(a) shows that SEDIGISM detects the most massive structures (median $1136.37~M_{\odot}$), while CHIMPS ($183.61~M_{\odot}$) and MWISP ($225.78~M_{\odot}$) trace less massive objects. Figure \ref{Imgs_Hist_Pars}(b) differentiates structural information, with SEDIGISM identifying the most spatially extended objects (median $2.18$ pc), followed by MWISP ($1.72$ pc) and CHIMPS ($1.23$ pc). SEDIGISM's catalog with the SCIMES algorithm specifically targets cloud-scale structures, accounting for its higher median $M$ and $R_{\text{eq}}$, while both MWISP and CHIMPS detect smaller clumps within larger complexes. 

Figure \ref{Imgs_Hist_Pars}(c) shows the $\sigma_v$ distribution, with MWISP exhibiting a narrow distribution centered at $0.4$ km s$^{-1}$, while CHIMPS ($0.75$ km s$^{-1}$) and SEDIGISM ($0.74$ km s$^{-1}$) display broader distributions with higher median values and extended tails. The smaller velocity dispersions in MWISP structures reflect superior velocity resolution capabilities.

Figure \ref{Imgs_Hist_Pars}(d) presents the $\alpha_{\text{vir}}$ distribution, revealing the dynamical states of these populations. CHIMPS shows the highest median value ($2.66$), indicating less gravitationally bound objects, while MWISP ($1.44$) and SEDIGISM ($1.3$) exhibit lower values. The MWISP median approaches the critical threshold of $\alpha_{\text{vir}} = 2$ \citep{Virial_Analysis_4,Virial_Analysis_1}, with 65.3\% of clumps having $\alpha_{\text{vir}} < 2$, indicating that the majority are gravitationally bound and account for approximately 96.3\% of the total mass in the statistical sample. However, it should be noted that $\alpha_{\text{vir}}$ is susceptible to distance-related observational effects, potentially limiting the accuracy of this threshold in determining the true gravitational binding state. 

Figures \ref{Imgs_Hist_Pars}(e) and (f) reveal consistent density trends. SEDIGISM exhibits the highest values in both $\Sigma$ (median  $72.65~M_{\odot}\text{pc}^{-2}$) and $n(\text{H}_2)$ (median $457.94$ cm$^{-3}$), followed by CHIMPS ($41.02~M_{\odot}\text{pc}^{-2}$ and $377.68$ cm$^{-3}$) and MWISP ($25.53~M_{\odot}\text{pc}^{-2}$ and $214.69$ cm$^{-3}$). The lower densities in MWISP reflect its sensitivity to extended, diffuse structures traced by $^{13}$CO ($J=1\text{--}0$) emission, which samples gas at lower densities than the higher-$J$ transitions used by other surveys. 

These distinct parameter distributions illustrate how the catalogs systematically target different components of the molecular clump population: MWISP characterizes extended, moderately massive structures with intermediate virial parameters, low velocity dispersions, and low densities; CHIMPS captures similar-sized regions with comparable masses but higher virial parameters and intermediate densities; and SEDIGISM detects the most massive cloud-scale structures with larger extents and higher densities. The magnitude of these differences exceeds what would be expected from distance-dependent biases alone, highlighting the genuine physical distinctions between structures targeted by each catalog. 

\begin{figure*}
\centering
\vspace{0cm}
\begin{minipage}[t]{0.3\textwidth}
    \centering
    \centerline{\includegraphics[width=2.2in]{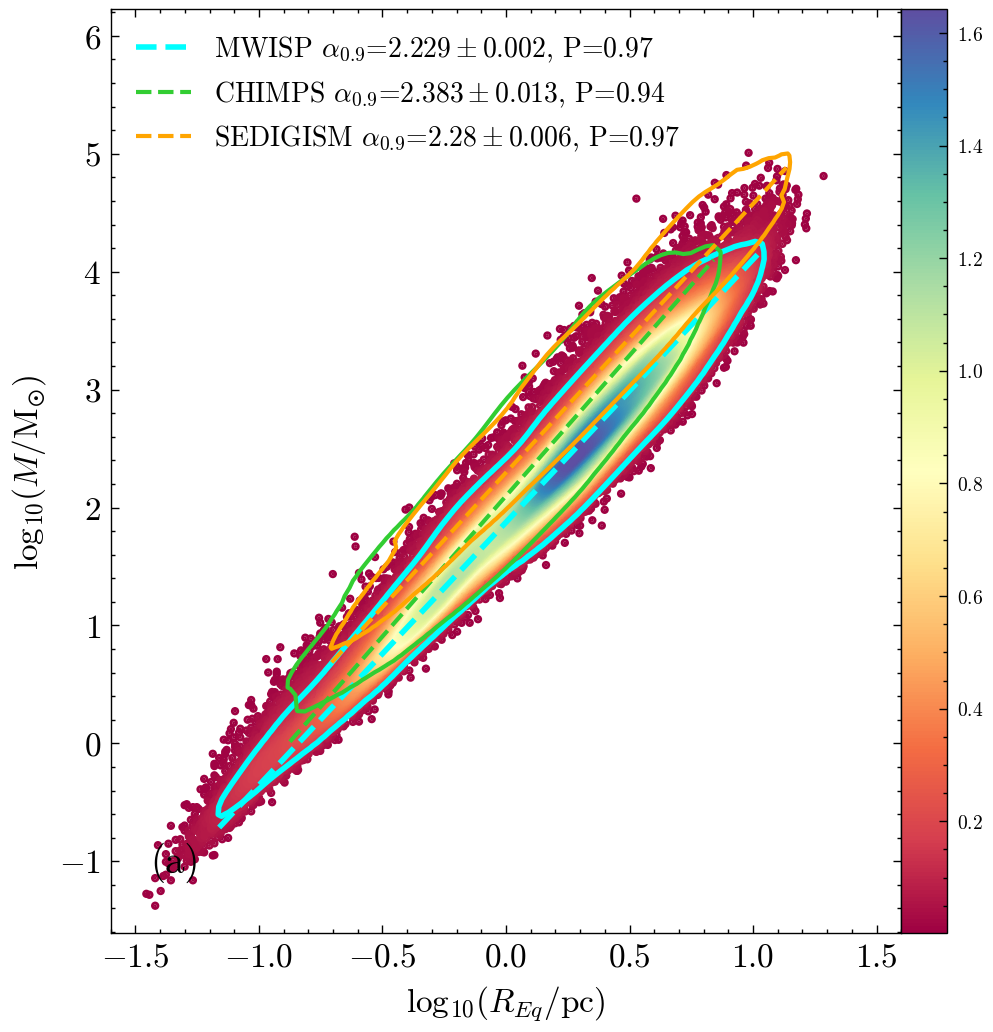}}
\end{minipage}
\begin{minipage}[t]{0.3\textwidth}
    \centering
    \centerline{\includegraphics[width=2.2in]{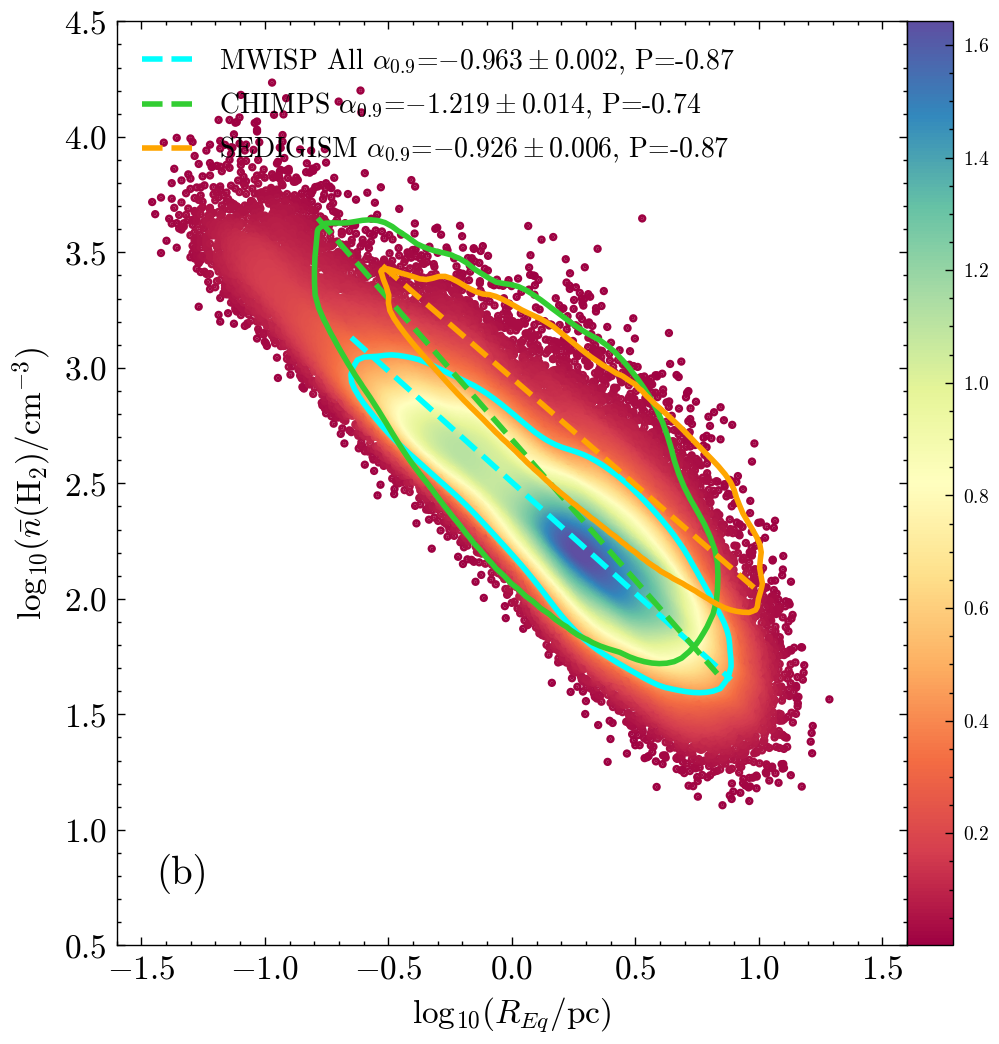}}
\end{minipage}
\begin{minipage}[t]{0.3\textwidth}
    \centering
    \centerline{\includegraphics[width=2.2in]{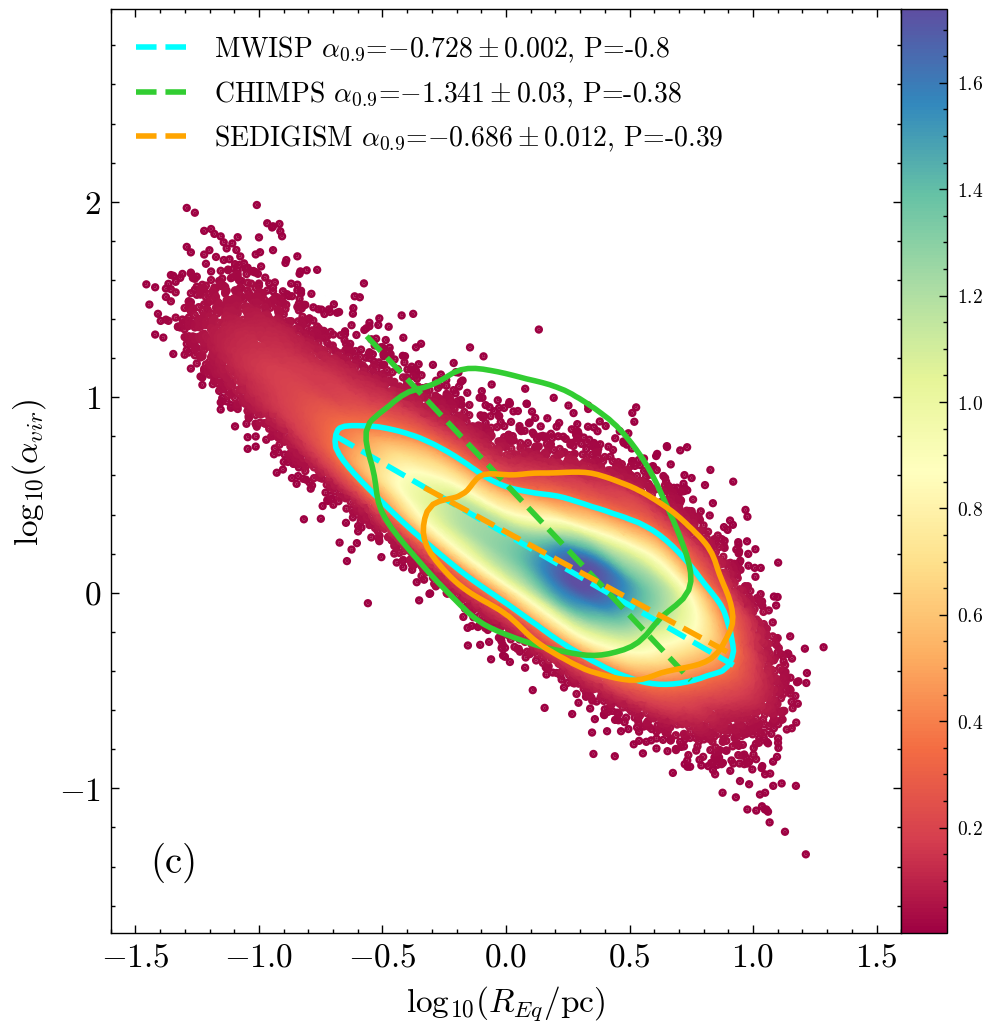}}
\end{minipage}

\begin{minipage}[t]{0.3\textwidth}
    \centering
    \centerline{\includegraphics[width=2.2in]{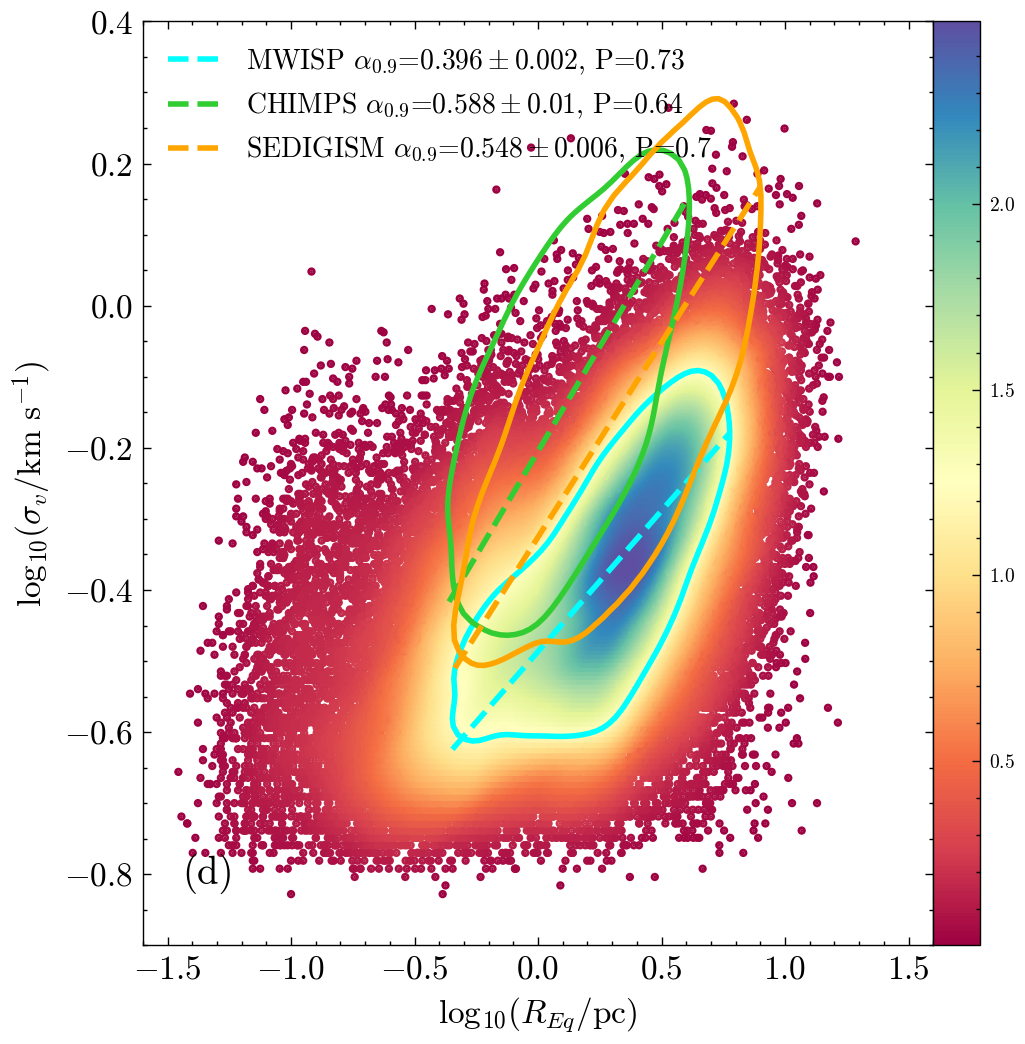}}
\end{minipage}
\begin{minipage}[t]{0.3\textwidth}
    \centering
    \centerline{\includegraphics[width=2.2in]{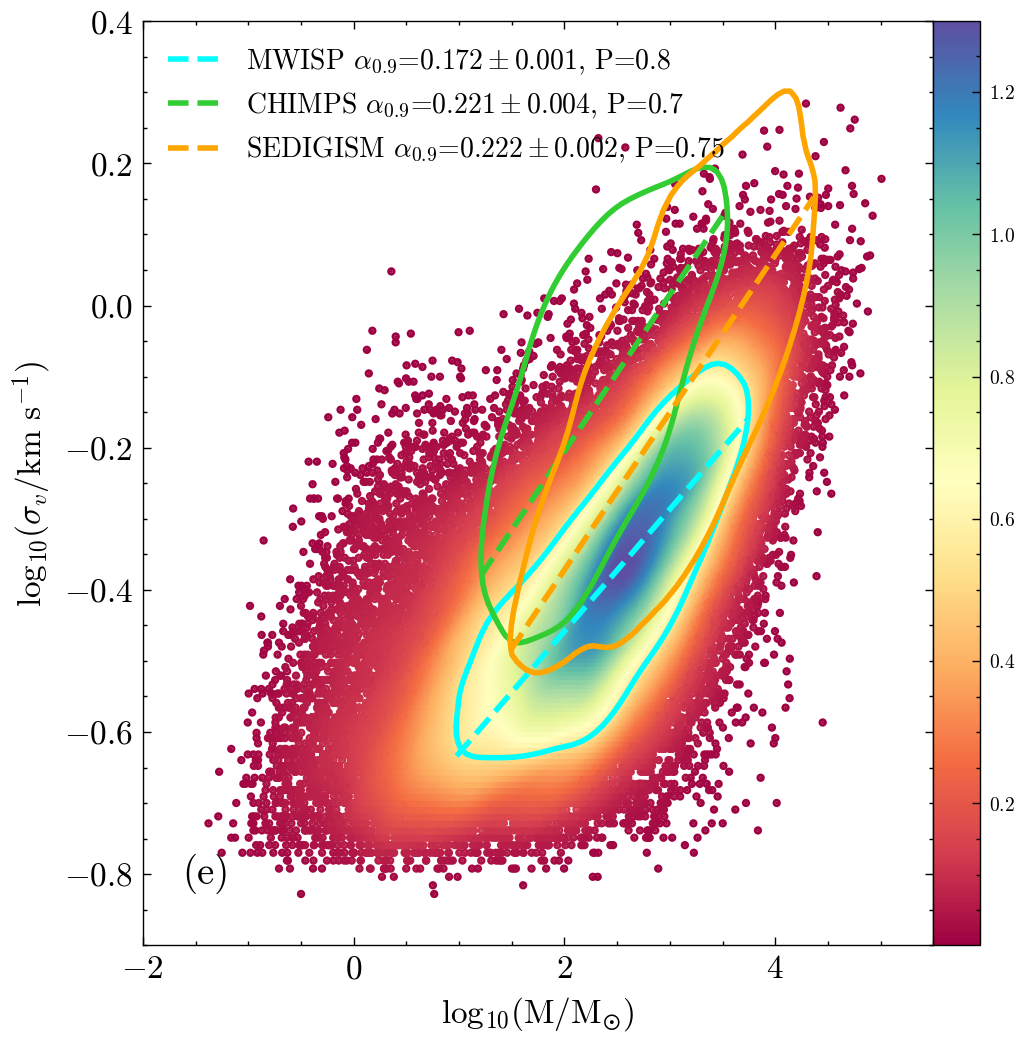}}
\end{minipage}
\begin{minipage}[t]{0.3\textwidth}
    \centering
    \centerline{\includegraphics[width=2.2in]{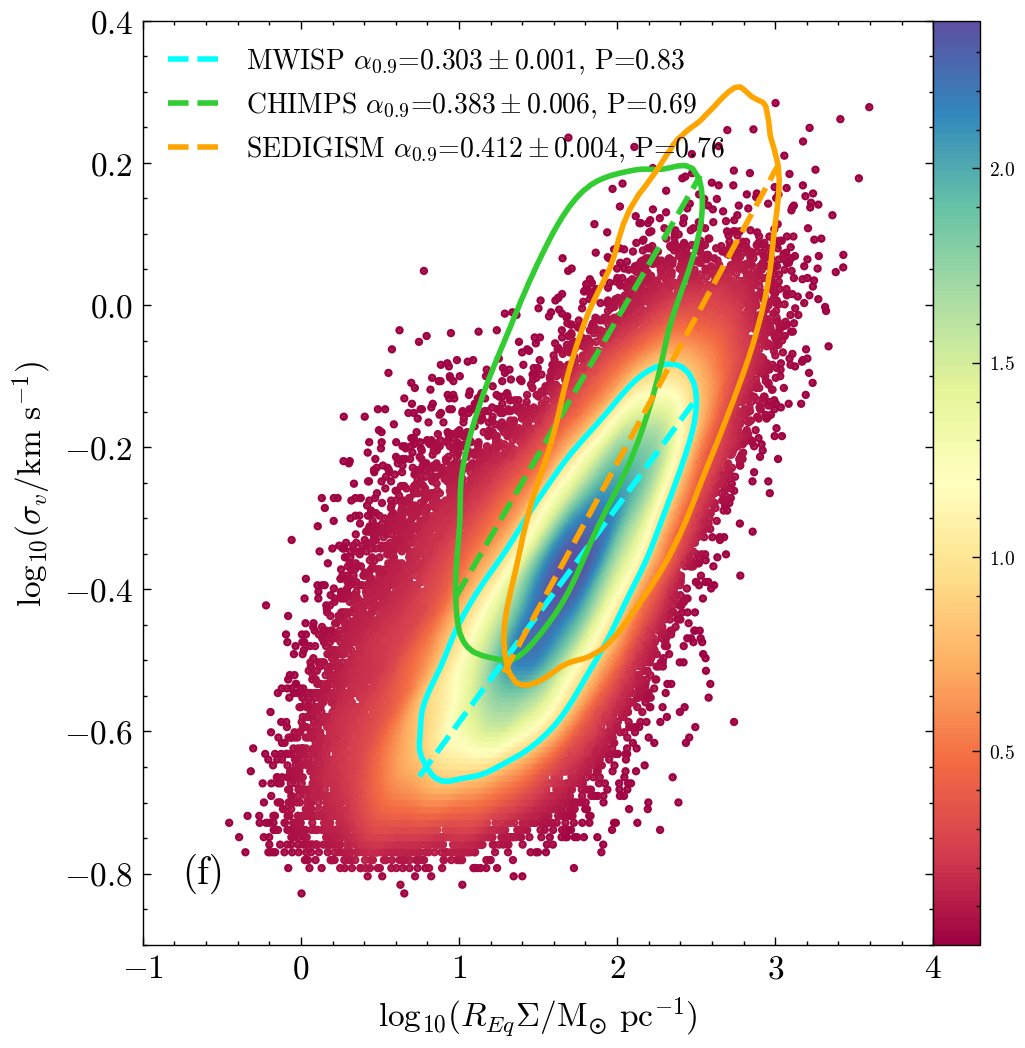}}
\end{minipage}
\caption{Scaling relations between physical parameters of different surveys. Panels show (a) $M$-$R_{\rm eq}$; (b) $n({\rm H_2})$-$R_{\rm eq}$; (c) $\alpha_{\rm vir}$-$R_{\rm eq}$; (d) $\sigma_v$-$R_{\rm eq}$; (e) $\sigma_v$-$M$; and (f) $\sigma_v$-$R_{\rm eq}\Sigma$. The color scale represents the KDE of MWISP clumps. Colored dashed lines indicate power-law fits to the main distributions outlined by corresponding contours, with derived power-law indices ($\alpha_{0.9} \pm$ uncertainty) and $P$ labeled for each survey.}
\label{Imgs_PR_2}
\end{figure*}

\subsection{Scaling Relations Across Clump Populations}\label{PRs_XX}
The scaling relations reveal key structural and dynamical properties of molecular clumps, focusing on the most reliable data with KDE confidence levels exceeding 90\%. We first conduct a cross-survey comparison of parameter relations among MWISP, CHIMPS, and SEDIGISM datasets, then analyze maser-associated subsamples to understand how active star formation environments differ from the general population. Through quantitative analysis of power-law indices and correlation strengths, this approach reveals both universal aspects of clump scaling behavior and the specific modifications associated with enhanced star formation activity.

\subsubsection{Cross-survey Comparison of Clump Scaling Relations}\label{PRs_DS}
Figure \ref{Imgs_PR_2} presents a detailed comparison of molecular clump scaling relations across catalogs, which can be directly assessed against Larson's classical scaling laws \citep{KDA_Method_PR_1}, which have served as key references in molecular cloud dynamics. 

The $M$ - $R_{\text{eq}}$ relation for molecular clumps (Figure \ref{Imgs_PR_2}a) exhibits a power-law index of $\alpha_{0.9} = 2.229\pm0.002$ ($P = 0.97$), with CHIMPS ($\alpha_{0.9} = 2.383$) and SEDIGISM ($\alpha_{0.9} = 2.28$) displaying nearly identical slopes. These slopes, significantly steeper than the classical $M \propto R^{2.0}$ relation, indicate that surface density increases with size rather than remaining constant \citep{KDA_Method_PR_5,PRelation_4}. This departure from constant surface density spans multiple scales may reflect hierarchical structure formation, where material becomes increasingly concentrated at larger scales \citep{Gravitation_2,Turbulence_4,Hierarchical_Collapse_2}, potentially arising from the intrinsic structure of supersonic, intermittent turbulent flows \citep{GRS_2}. 

The $n({\text H}_2)$ - $R_{\text{eq}}$ relation (Figure~\ref{Imgs_PR_2}b) for MWISP clumps shows an inverse relationship ($\alpha_{0.9} = -0.963\pm0.002$, $P = -0.87$) and remains shallower than Larson's density-size relation of $n(\text{H}_2) \propto R^{-1.1}$. SEDIGISM ($\alpha_{0.9} = -0.926$, $P = -0.87$) demonstrates a comparable trend to MWISP, while CHIMPS ($\alpha_{0.9} = -1.219$, $P = -0.74$) displays a steeper slope. The observed power-law indices across all surveys deviate from the constant column density scenario expected from simple scaling models \citep{KDA_Method_PR_1}, suggesting complex density distribution in molecular clumps. 

The $\alpha_{\mathrm{vir}}$ - $R_{\mathrm{eq}}$ relationship (Figure \ref{Imgs_PR_2}c) exhibits a significant negative correlation for MWISP clumps ($\alpha_{0.9} = -0.728\pm0.002$, $P = -0.8$). Among the surveys, CHIMPS demonstrates a more pronounced negative slope ($\alpha_{0.9} = -1.341$, $P = -0.38$), while SEDIGISM ($\alpha_{0.9} = -0.686$, $P = -0.39$) shows a trend similar to MWISP. This decrease in virial parameter with increasing clump radius provides evidence that gravitational binding becomes more dominant in larger molecular structures. 

The $\sigma_v$ - $R_{\text{eq}}$ relation (Figure~\ref{Imgs_PR_2}d) yields $\alpha_{0.9} = 0.396\pm0.002$ ($P = 0.73$) for MWISP clumps, slightly greater than Larson's original relation of $\sigma_v \propto R^{0.38}$ and replicating findings from previous MWISP investigations \citep{MWISP_Analysis_Feng_2024}. Larson's original relation falls between predictions for Kolmogorov turbulence in incompressible fluids ($\sigma_v \propto R^{1/3}$) and Burgers turbulence for strongly shock-dominated supersonic turbulence ($\sigma_v \propto R^{1/2}$) \citep{Review_10,Review_5}. In contrast, CHIMPS ($\alpha_{0.9} = 0.588$) and SEDIGISM ($\alpha_{0.9} = 0.548$) show scaling relations closer to the Burgers turbulence regime, while MWISP occupies an intermediate position between these two turbulent scenarios. 

The $\sigma_v$ - $M$ relation (Figure~\ref{Imgs_PR_2}e) yields $\alpha_{0.9} = 0.172\pm0.001$ ($P = 0.8$) for MWISP clumps, with CHIMPS and SEDIGISM exhibiting steeper scaling relations ($\alpha_{0.9} = 0.221$ and $\alpha_{0.9} = 0.222$, with $P = 0.7$ and $P = 0.75$ respectively). All three populations follow Larson's original mass-velocity relation of $\sigma_v \propto M^{0.2}$. The weak scaling relations across all surveys indicate that velocity dispersion increases only modestly with clump mass, suggesting that molecular clumps maintain similar dynamical states across a wide range of masses. 

Figure \ref{Imgs_PR_2}(f) presents the $\sigma_v$ - $R_{\mathrm{eq}}\Sigma$ relationship, yielding a power-law index of $\alpha_{0.9} = 0.303\pm0.001$ with strong correlation ($P = 0.83$) for MWISP clumps, paralleling earlier MWISP datasets \citep{MWISP_Analysis_Feng_2024}. Both CHIMPS ($\alpha_{0.9} = 0.383$, $P = 0.69$) and SEDIGISM ($\alpha_{0.9} = 0.412$, $P = 0.76$)  display slightly higher indices compared to MWISP. This relation demonstrates superior statistical robustness compared to the $\sigma_v$ - $R_{\mathrm{eq}}$ relation ($P = 0.83$ versus $P = 0.73$; \citep{Review_13,CFA_2017}. For clumps in virial equilibrium, the virial theorem predicts $\sigma_v \propto (R \times \Sigma)^{0.5}$. The observed exponents remain below the virial prediction, indicating subvirial states where turbulent support scales more weakly with gravitational potential than expected in idealized equilibrium models. 

While Larson's original relations were based on relatively limited samples and the technologies available over 40 years ago \citep{KDA_Method_PR_1}, the comprehensive coverage and high precision of modern surveys like MWISP reveal nuances in molecular clump structure and dynamics that could not be captured previously \citep[e.g. ][]{KDA_Method_PR_5,KDA_Method_PR_2,PRelation_4,Turbulence_6}. The MWISP survey's unique characteristics, including large-scale coverage, an extensive sample, and uniform detection and distance analysis methods, make it well suited for analyzing correlations between parameters. These minimized systematic effects across diverse Galactic environments allow observed scaling relations to reflect more genuine physical processes. Rather than challenging the importance of Larson's pioneering work, these results represent a natural advancement in our understanding of interstellar matter distribution enabled by modern observational capabilities.

\begin{figure*}
\centering
\vspace{0cm}
\begin{minipage}[t]{0.3\textwidth}
    \centering
    \centerline{\includegraphics[width=2.2in]{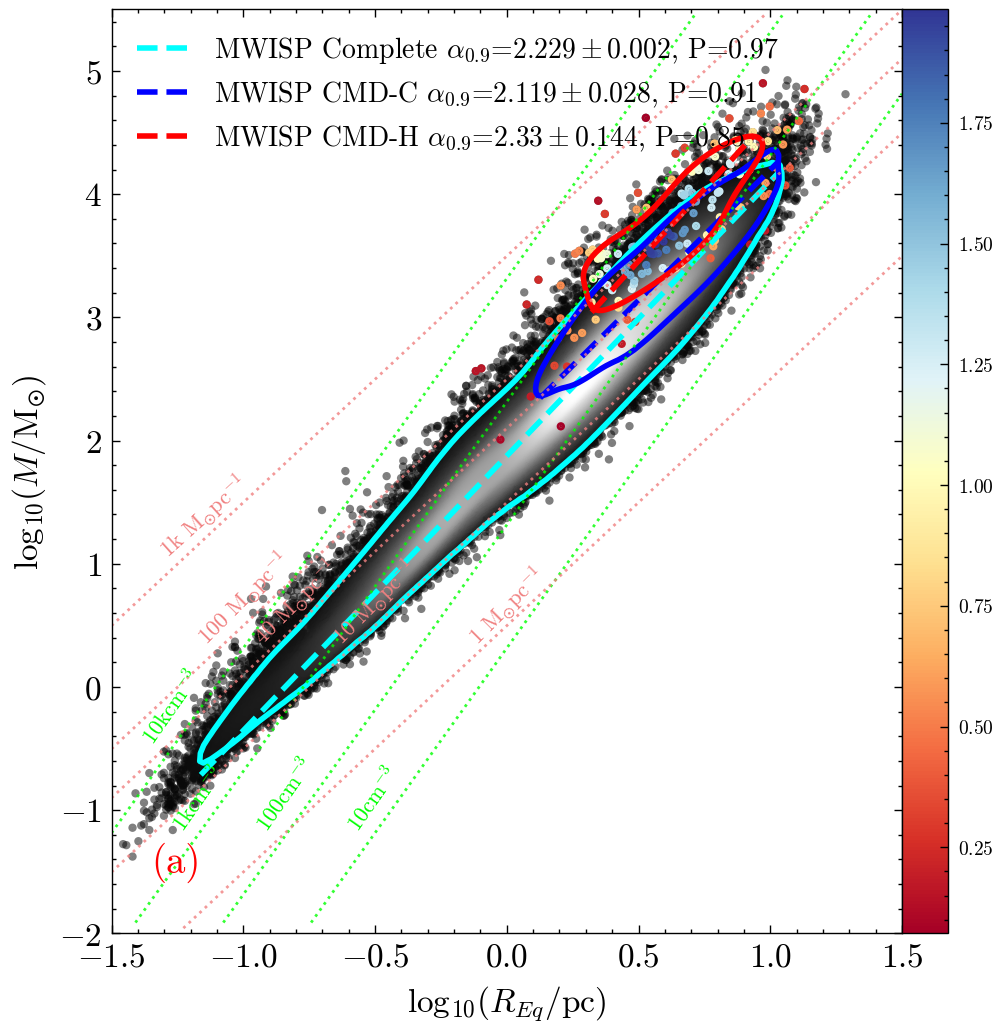}}
\end{minipage}
\begin{minipage}[t]{0.3\textwidth}
    \centering
    \centerline{\includegraphics[width=2.2in]{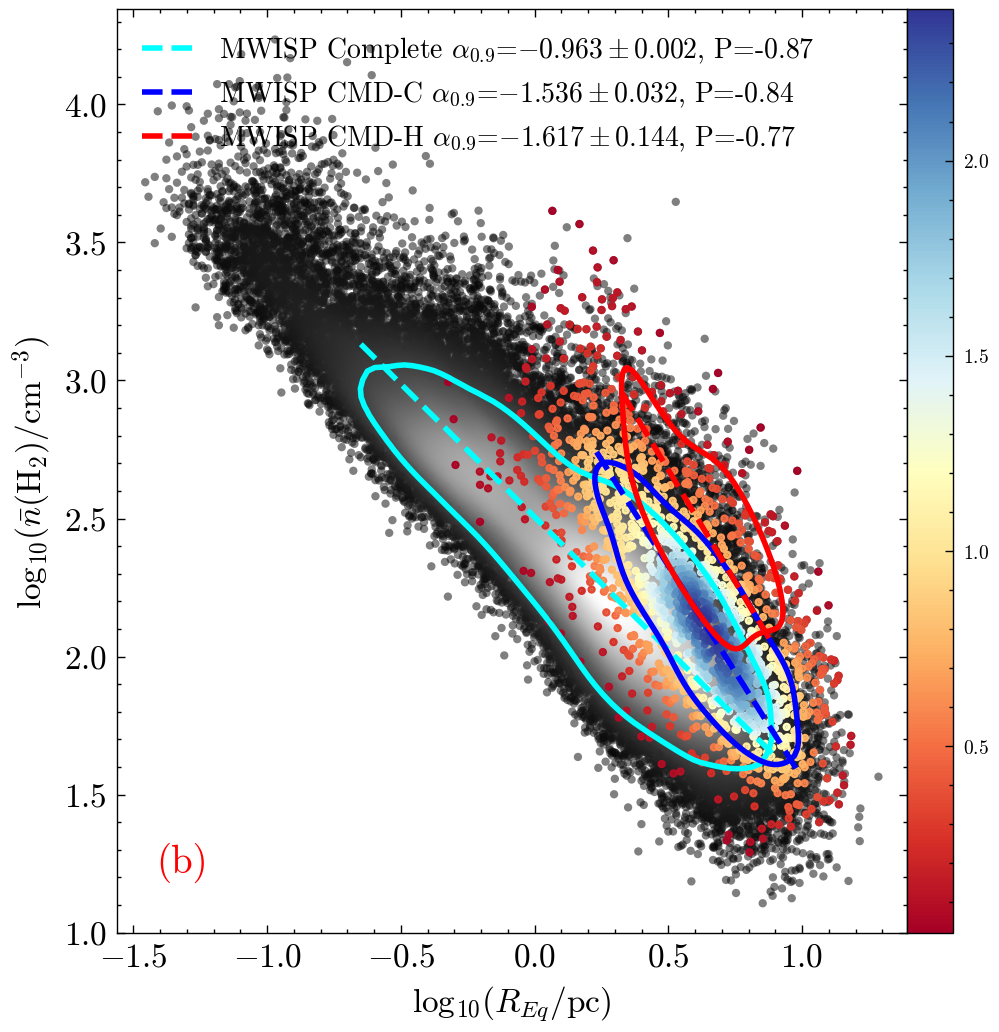}}
\end{minipage}
\begin{minipage}[t]{0.3\textwidth}
    \centering
    \centerline{\includegraphics[width=2.2in]{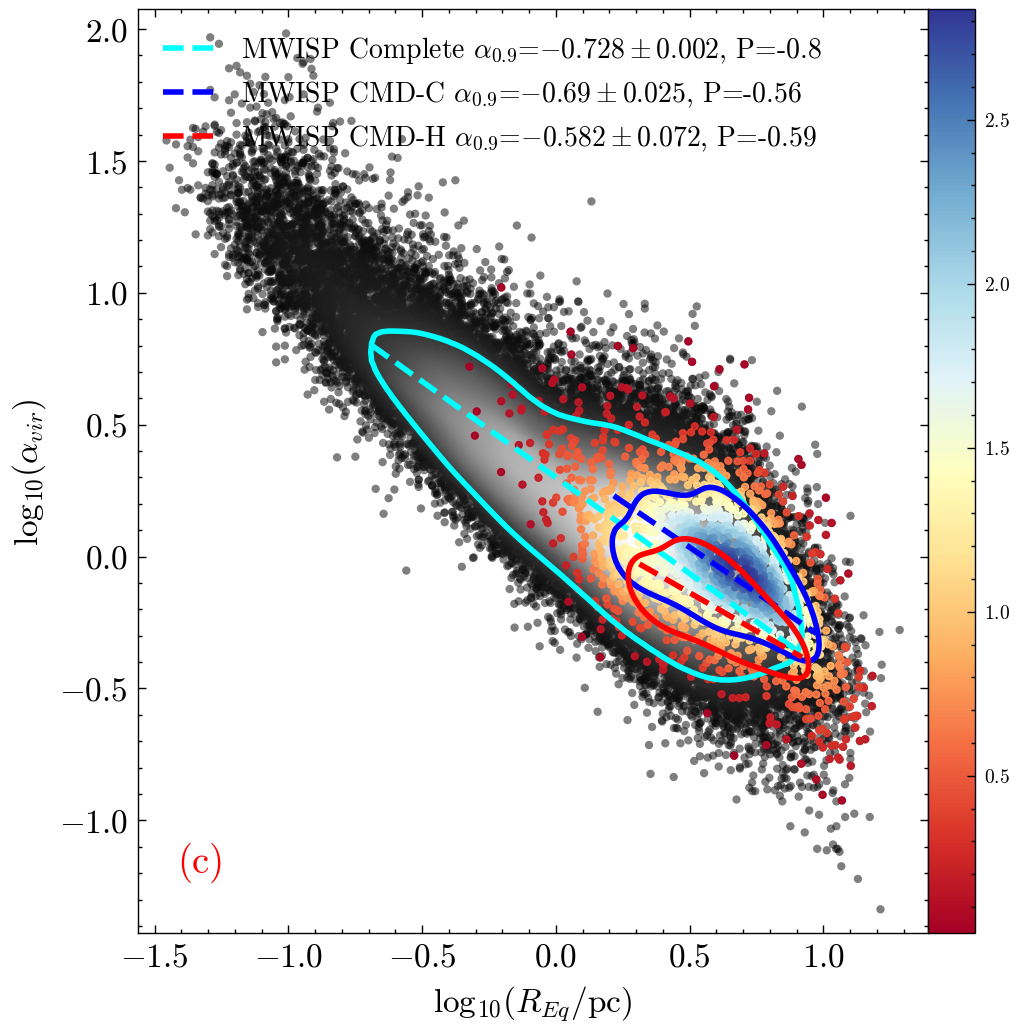}}
\end{minipage}

\begin{minipage}[t]{0.3\textwidth}
    \centering
    \centerline{\includegraphics[width=2.2in]{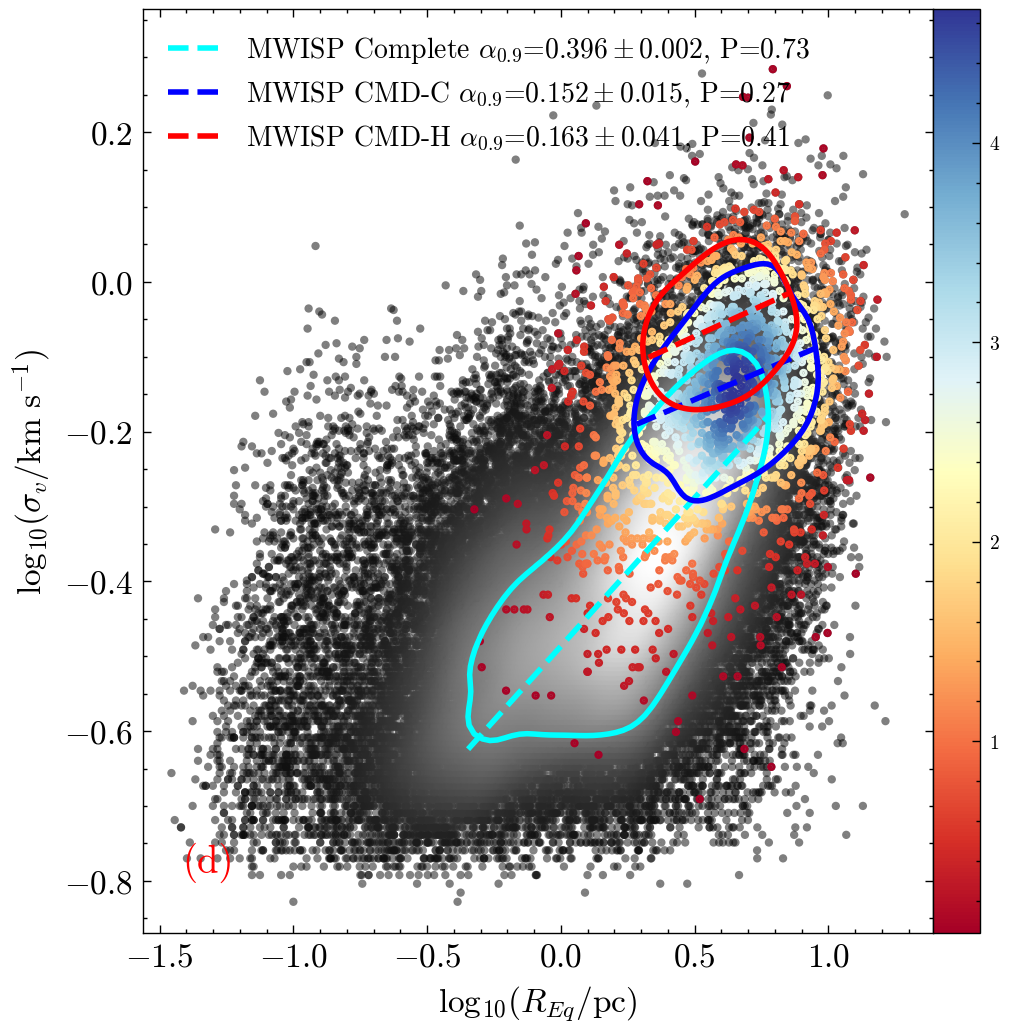}}
\end{minipage}
\begin{minipage}[t]{0.3\textwidth}
    \centering
    \centerline{\includegraphics[width=2.2in]{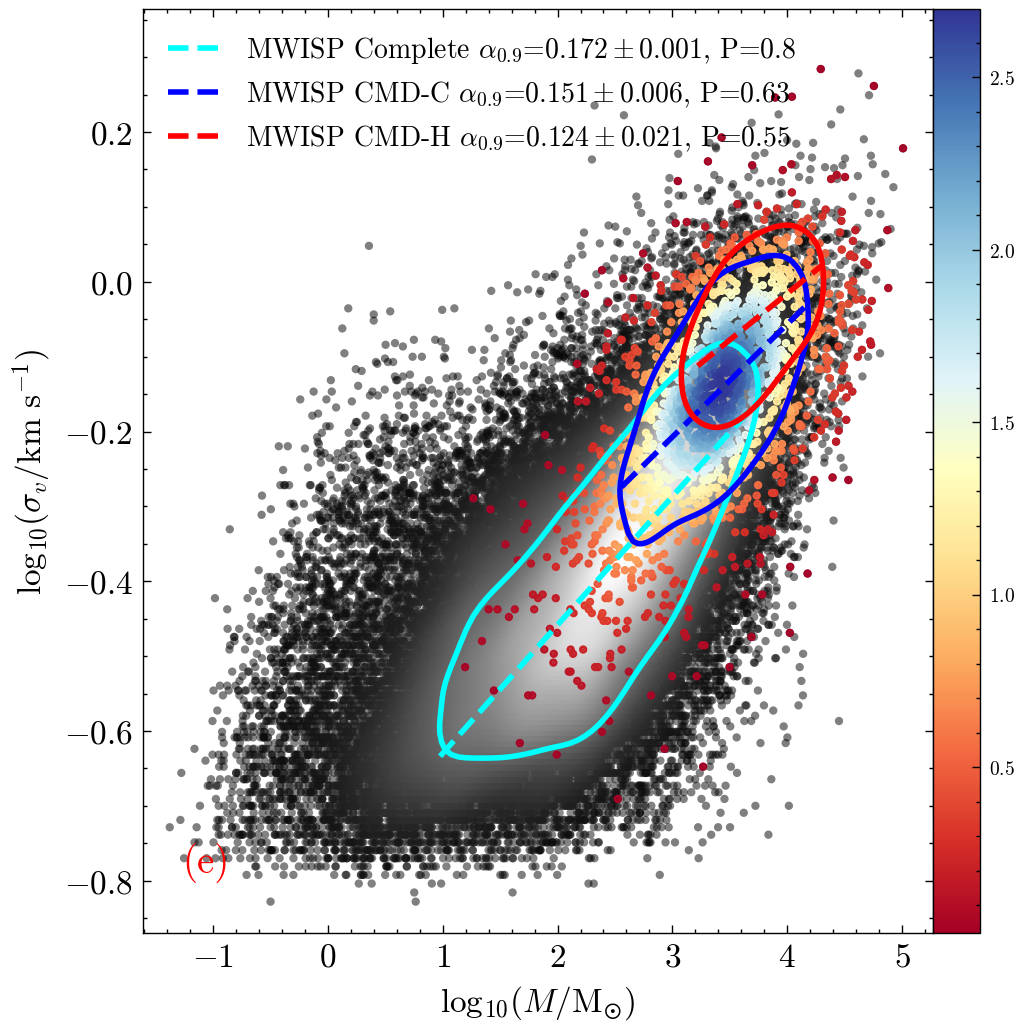}}
\end{minipage}
\begin{minipage}[t]{0.3\textwidth}
    \centering
    \centerline{\includegraphics[width=2.2in]{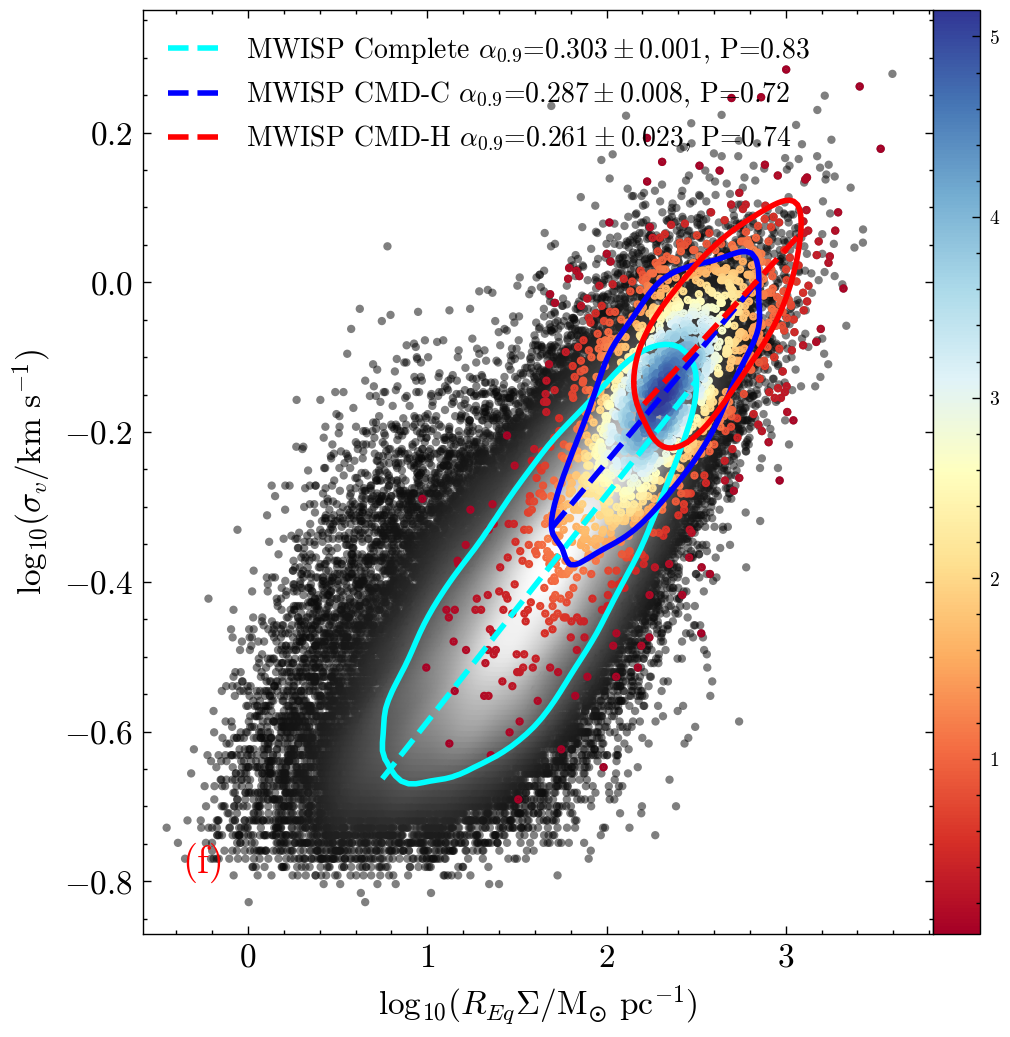}}
\end{minipage}

\begin{minipage}[t]{0.3\textwidth}
    \centering
    \centerline{\includegraphics[width=2.2in]{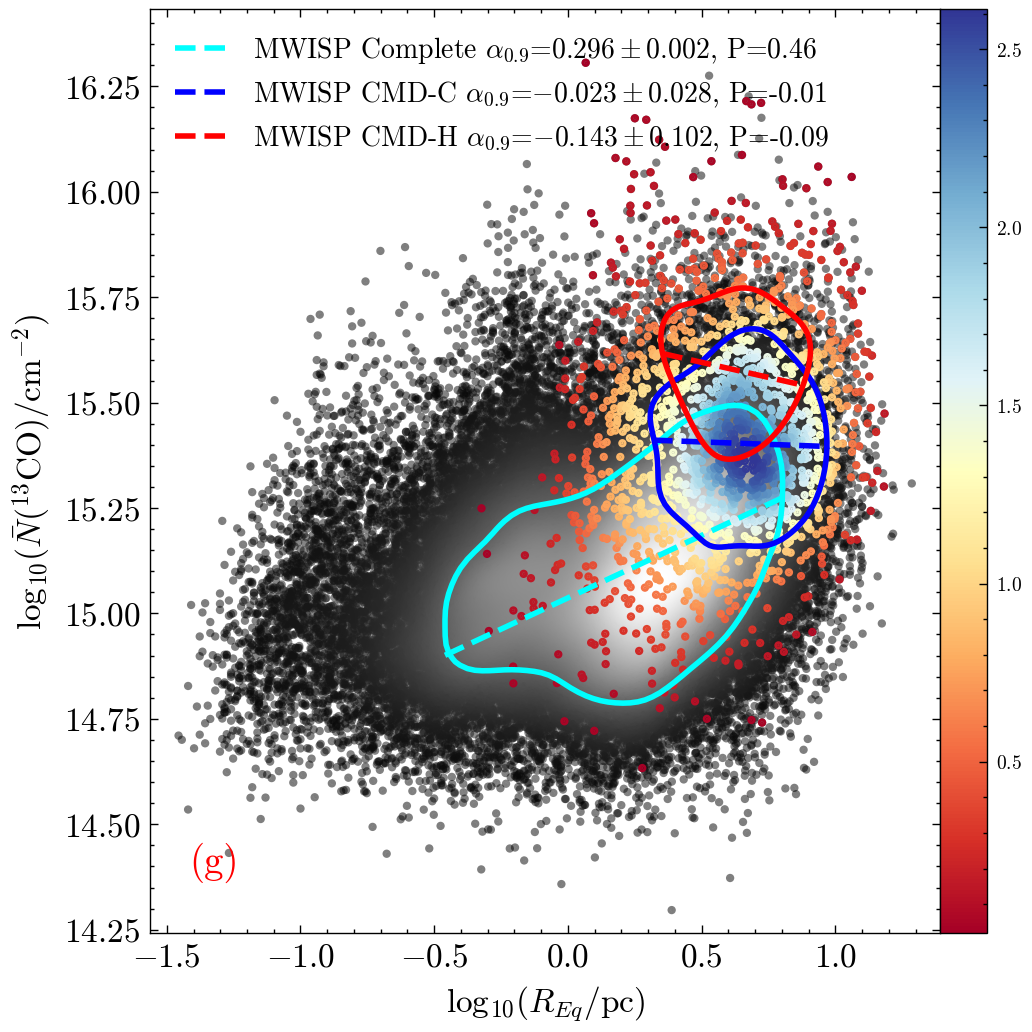}}
\end{minipage}
\begin{minipage}[t]{0.3\textwidth}
    \centering
    \centerline{\includegraphics[width=2.2in]{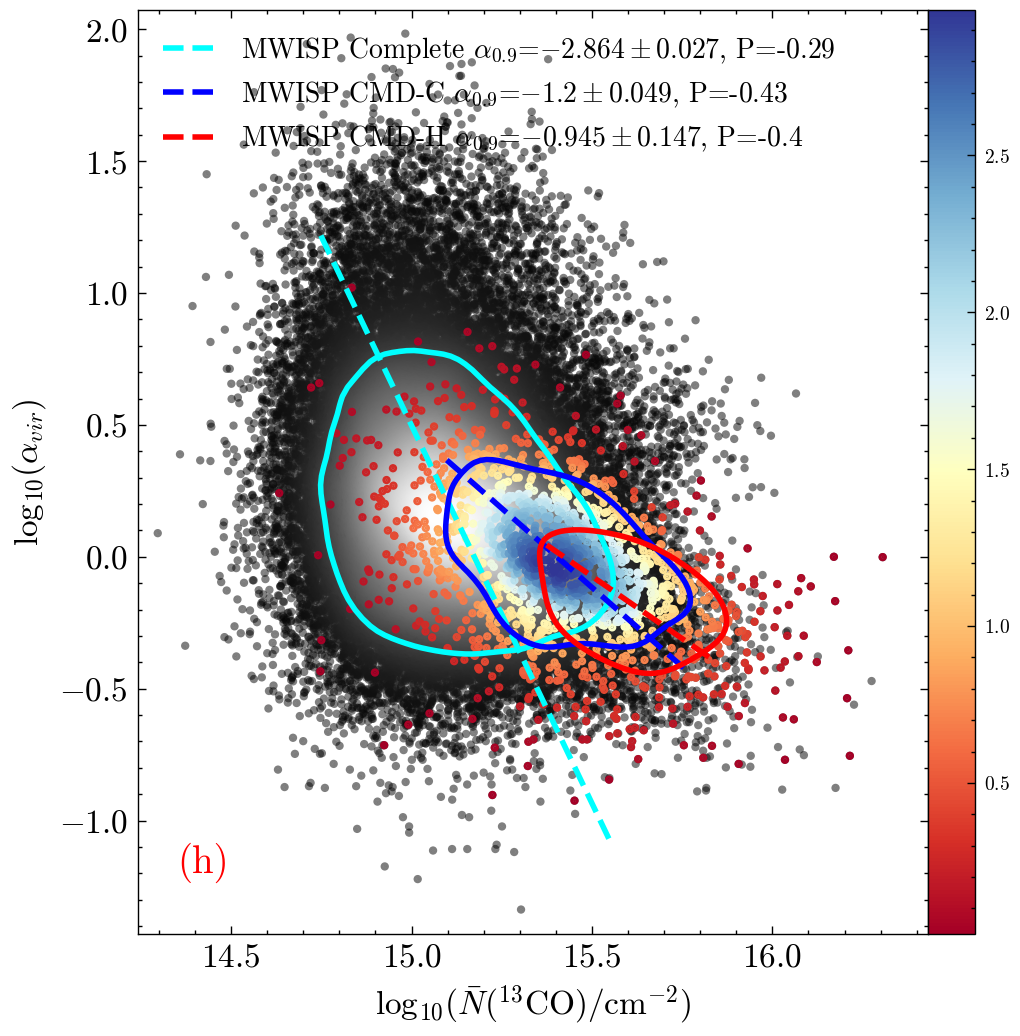}}
\end{minipage}
\begin{minipage}[t]{0.3\textwidth}
    \centering
    \centerline{\includegraphics[width=2.2in]{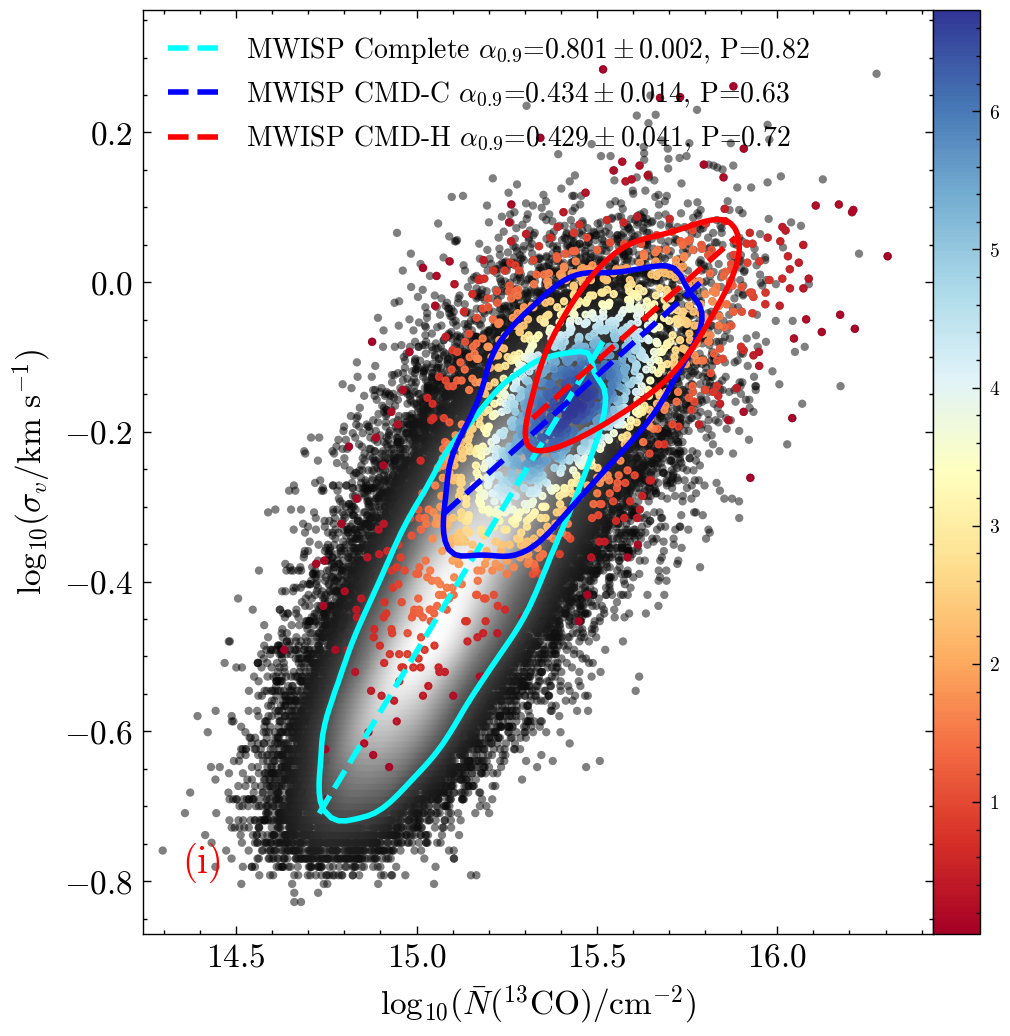}}
\end{minipage}
\caption{Scaling relations between physical parameters of MWISP. Panel shows: (a) $M$-$R_{\rm eq}$, where light coral dotted lines indicate constant $\Sigma$ = 1, 10, 40, 100, 1000 M$_{\odot}$pc$^{-2}$ and lime dotted lines indicate constant $n({\rm H_2})$ = 10, 100, 1000, 10,000 cm$^{-3}$; (b) $n({\rm H_2})$-$R_{\rm eq}$; (c) $\alpha_{\rm vir}$-$R_{\rm eq}$; (d) $\sigma_v$-$R_{\rm eq}$; (e) $\sigma_v$-$M$; (f) $\sigma_v$-$R_{\rm eq}\Sigma$; (g) $N(^{13}\mathrm{CO})$-$R_{\rm eq}$; (h) $\alpha_{\rm vir}$-$N(^{13}\mathrm{CO})$; and (i) $\sigma_v$-$N(^{13}\mathrm{CO})$. The distributions are shown for two subsets: MWISP complete clumps (gray points with cyan contours) and MWISP CMDs. Among the CMD subset, CMD-C, indicated by colored points with blue contours, represents clumps in the close vicinity of masers, while CMD-H, marked by red contours, represents clumps that host masers. For each relation, the best-fit power-law index and corresponding $P$ are indicated in the legend.}
\label{Imgs_PR_Maser}
\end{figure*}

\subsubsection{Impact of Star Formation Activity on Clump Scaling Relations}\label{PRs_Maser}
Figure \ref{Imgs_PR_Maser} contrasts scaling relations between the complete MWISP sample and a subset of clumps with maser-determined distances (CMDs), further subdivided into clumps close to maser emissions (CMD-C, 1566 sources) and clumps directly hosting masers (CMD-H, 152 sources). These subsets represent different levels of star formation activity: CMD-H captures clumps with ongoing maser excitation processes indicative of active star formation \citep[e.g.,][]{Maser_Star_Formation_1,Maser_Star_Formation_3,Maser_Star_Formation_4,Maser_Star_Formation_2}, CMD-C represents regions influenced by nearby star formation activity, while the complete sample encompasses diverse star formation environments ranging from quiescent to active phases. 

The $M$--$R_{\rm eq}$ relation (Figure \ref{Imgs_PR_Maser}a) shows that CMD-C exhibits a slightly shallower slope ($\alpha_{0.9}=2.119$, $P=0.91$) compared to the complete sample ($\alpha_{0.9}=2.229$), while CMD-H shows a marginally steeper relation ($\alpha_{0.9}=2.33$, $P=0.85$). The median mass of CMD-H clumps (4576 $M_{\odot}$) is approximately twice that of CMD-C clumps (2327 M$_{\odot}$), while their median radii are comparable: 4.2 pc for CMD-H versus 4 pc for CMD-C. The overlaid density lines show that CMD-H clumps preferentially occupy higher density regions (both surface density $>$ 40 $M_{\odot}$pc$^{-2}$ and volume density $>$ 100 cm$^{-3}$), while CMD-C spans a broader density parameter space. Although these variations exist, the relatively small differences among power-law indices suggest that the mass-radius scaling remains preserved across different star formation environments. This modest variation indicates that while local physical conditions may introduce structural modifications, they do not largely alter the mass distribution behavior across spatial scales. 

Figure \ref{Imgs_PR_Maser}(b) reveals a progressive steepening of the $n(H_2)$--$R_{\rm eq}$ relation from the complete sample ($\alpha_{0.9}=-0.963$) to CMD-C ($\alpha_{0.9}=-1.536$, $P=-0.83$) to CMD-H ($\alpha_{0.9}=-1.617$, $P=-0.77$). This trend indicates increasingly pronounced density contrasts in maser-associated regions, with dense cores embedded within more extended, lower-density envelopes. Such structural characteristics are consistent with the physical requirements for maser excitation, which typically requires specific density and temperature conditions \citep{Review_9,Maser_Star_Formation_3}. The steep density gradients may reflect enhanced central concentration processes \citep{Cloud_Evolution_4} driven by gravitational collapse and stellar feedback in active star formation regions. 

The $\alpha_{\rm vir}$-$R_{\rm eq}$ relation (Figure \ref{Imgs_PR_Maser}c) displays negative correlations, with changes in slope from the complete sample ($\alpha_{0.9}=-0.728$) to CMD-C ($\alpha_{0.9}=-0.69$, $P=-0.56$) to CMD-H ($\alpha_{0.9}=-0.582$, $P=-0.59$). This progressive flattening of the power-law slope, combined with the rightward and downward displacement of CMD-H distribution, indicates that maser-hosting clumps are systematically larger and more gravitationally bound. The flattening trend suggests that virial parameters become less sensitive to radius variations in active star formation environments, possibly reflecting stellar feedback processes that provide scale-independent support mechanisms across different clump sizes, partially balancing gravitational binding with nonthermal support to maintain relatively uniform dynamical states in these regions. 

Panels (d)-(f) in Figure \ref{Imgs_PR_Maser} demonstrate changes in velocity dispersion scaling relations across different star formation environments. In the $\sigma_v$–$R_{\rm eq}$ relation (panel (d)), both maser-associated populations display flatter slopes ($\alpha_{0.9} = 0.152, 0.161$, $P = 0.27, 0.41$) compared to the complete sample ($\alpha_{0.9} = 0.396$). This pattern continues in the $\sigma_v$-$M$ relation (panel (e)), where power-law indices decrease from the complete sample ($\alpha_{0.9}=0.172$) to CMD-C ($\alpha_{0.9}=0.151$, $P=0.63$) and CMD-H ($\alpha_{0.9}=0.124$, $P=0.55$). Similarly, the $\sigma_v$-$R_{\rm eq}\Sigma$ relation (panel (f)) shows a decline from the complete sample ($\alpha_{0.9}=0.303$) to CMD-C ($\alpha_{0.9}=0.287$, $P=0.72$) and CMD-H ($\alpha_{0.9}=0.261$, $P=0.74$). The overall flattening of these scaling relations indicates that velocity dispersion becomes less sensitive to physical parameters in active star formation environments. The relatively small differences between CMD-C and CMD-H coefficients suggest that these modifications to turbulent properties occur in association with star formation activity, irrespective of whether clumps directly host masers.

The $N(^{13}\mathrm{CO})$-$R_{\rm eq}$ relation (Figure \ref{Imgs_PR_Maser}g) shows notable differences across populations, with the complete sample exhibiting a positive correlation ($\alpha_{0.9}=0.296$, $P=0.46$), CMD-C showing a nearly flat relationship ($\alpha_{0.9}=-0.023$, $P=-0.01$), and CMD-H displaying a slightly negative correlation ($\alpha_{0.9}=-0.143$, $P=-0.09$). The low correlation coefficients in CMD-C and CMD-H populations indicate weaker size-column density relationships in maser-associated regions compared to the general population. This progression from weakly positive to negligible correlations suggests that enhanced star formation activity may diminish the dependence between clump size and column density. 

The $\alpha_{\rm vir}$-$N(^{13}{\rm CO})$ relation (Figure \ref{Imgs_PR_Maser}h) shows systematically shallower slopes from the complete sample ($\alpha_{0.9}=-2.864$) through CMD-C ($\alpha_{0.9}=-1.2$, $P=-0.43$) to CMD-H ($\alpha_{0.9}=-0.945$, $P=-0.4$). While all populations become more gravitationally bound with increasing column density, the slope flattening indicates that virial parameters may become less sensitive to column density variations in maser-associated environments. The moderate correlation coefficients further support this trend, suggesting that column density alone is a weaker predictor of virial state in regions with enhanced star formation activity. 

The $\sigma_v$-$N(^{13}\mathrm{CO})$ relation (Figure \ref{Imgs_PR_Maser}i) shows substantial reduction in slopes from the complete sample ($\alpha_{0.9}=0.801$, $P=0.82$) to identical values for both CMD-C ($\alpha_{0.9}=0.434$, $P=0.63$) and CMD-H ($\alpha_{0.9}=0.429$, $P=0.72$). This relation provides particularly valuable insights as both parameters are distance-measurement-independent quantities, eliminating potential biases from distance uncertainties. The power-law indices demonstrate higher-density regions in general clumps maintain proportionally higher turbulent motions. The identical slopes between CMD-C and CMD-H populations indicate that these relationships converge to similar characteristics in maser-associated environments, suggesting that turbulent properties may be regulated by similar physical processes \citep{Turbulence_1} in maser-associated regions. 

All panels exhibit a consistent pattern where CMD population distributions are displaced rightward and vertically (either upward or downward) compared to the complete sample, with CMD-H positioned at the distribution extremes. This indicates that maser-associated clumps occupy distinct regions characterized by larger physical scales, higher masses at given radii, and modified internal structure. The power-law indices of CMD-C and CMD-H show two behavioral patterns: nearly identical coefficients in velocity dispersion relations ($\sigma_v$–$R_{\rm eq}$, $\sigma_v$-$M$, $\sigma_v$-$R_{\rm eq}\Sigma$, and $\sigma_v$-$N(^{13}\mathrm{CO})$), and systematic progressions in structural relations ($\alpha_{\rm vir}$-$R_{\rm eq}$, $n(H_2)$-$R_{\rm eq}$, $\alpha_{\rm vir}$-$N(^{13}\mathrm{CO})$, and $N(^{13}\mathrm{CO})$-$R_{\rm eq}$). The similarity in velocity dispersion relations suggests that turbulent properties achieve similar characteristics in maser-associated regions, while the progressions in structural relations indicate that these physical properties respond more sensitively to different levels of star formation activity. 

These differences offer observational constraints for theoretical models \citep{Review_5,Review_4}, revealing how physical mechanisms governing clump properties might be altered in active star formation environments \citep[e.g.][]{Cloud_Evolution_1,Review_8,Cloud_Evolution_3,PRelation_1,Hierarchical_Collapse_1,ATLASGAL_2022}. The notable scaling behaviors across populations suggest that different physical processes may dominate under varying star formation conditions, potentially serving as diagnostic indicators of star formation activity levels or evolutionary sequence \citep[e.g.][]{Cloud_Evolution_4} within clump populations. 

\section{Summary}\label{Summary}
We present an equivalent global detection scheme that significantly reduces computational resource requirements while preserving detection integrity for processing large astronomical datasets. By integrating this scheme with FacetClumps, we successfully extract 71,661 molecular clumps across the MWISP Phase I survey's high-resolution $^{13}$CO data spanning 2310 deg$^2$. The designed automatic hierarchical distance decision method, using SRs as fundamental objects, effectively resolves the KDA problem, offering reliable distance measurements for 97.94\% of clumps. 

The MWISP clumps have median values of 225.78 $M_{\odot}$ in mass, 1.72 pc in radius, 0.4 km s$^{-1}$ in velocity dispersion, 25.53 $M_{\odot}$ pc$^{-2}$ in surface density, and 214.69 cm$^{-3}$ in volume density. With a median virial parameter of $\alpha_{\text{vir}} = 1.44$, these structures are predominantly gravitationally bound, as 65.3\% satisfy the criterion $\alpha_{\text{vir}} < 2$ and account for 96.3\% of the total statistical mass. The MWISP clump catalog uniquely traces intermediate-mass and intermediate-size structures with narrow velocity dispersions and low densities. These clumps, characterized by extended, diffuse molecular distributions, represent an important component of the Galactic molecular clump population. 

The analysis of scaling relations reveals nine power-law relationships that characterize key aspects of clump physics:
\begin{enumerate}
\item The mass - radius relation ($M \propto R^{2.229}$) indicates surface density increases with clump size, deviating from constant surface density.
\item The H$_2$ column density - radius relation ($N(H_2) \propto R^{-0.963}$) shows density decline approximately linearly with increasing size. 
\item The virial parameter - radius relation ($\alpha_{\text{vir}} \propto R^{-0.728}$) demonstrates that larger clumps are more gravitationally bound.
\item The velocity dispersion - radius relation ($\sigma_v \propto R^{0.396}$) suggests a turbulent cascade a turbulent cascade between Kolmogorov and Burgers regimes.
\item The velocity dispersion - mass relation ($\sigma_v \propto M^{0.172}$) indicates that velocity dispersion scales weakly with mass.
\item The velocity dispersion - radius $\times$ surface density relation ($\sigma_v \propto (R\Sigma)^{0.303}$) indicates sub-virial states where turbulent support is insufficient for equilibrium.
\item The $^{13}$CO column density - radius relation ($\bar{N}(^{13}\mathrm{CO}) \propto R^{0.296}$) shows a gradual column density increase with clump size.
\item The virial parameter - $^{13}$CO column density relation ($\alpha_{\text{vir}} \propto \bar{N}(^{13}\mathrm{CO})^{-2.864}$) demonstrates stronger gravitational binding in denser regions.
\item The velocity dispersion - $^{13}$CO column density relation ($\sigma_v \propto \bar{N}(^{13}\mathrm{CO})^{0.801}$) represents a distance-independent measure of the strong correlation between turbulent motions and molecular gas density.
\end{enumerate}

Through comparative analysis of maser-associated clumps, we reveal differences in clump parameter distributions and scaling relations across different star formation environments. The contrast between general population and maser-hosting clumps shows distinct structural characteristics, with steeper density gradients and modified dynamical states in maser-associated regions. Velocity dispersion relations exhibit similar power-law indices between CMD-C and CMD-H populations, while structural relations display systematic progressions, indicating different responses to enhanced star formation activity. These systematic differences provide observational constraints for understanding clump properties in active star formation environments. 

The quantification of detection sensitivity biases through extrapolation demonstrates underestimations in parameters (12.4\% in flux, 5.7\% in velocity dispersion, and 6.1\% in radius), while confirming the robustness of established scaling relations. This validation strengthens confidence in the physical interpretations and the catalog's utility for future research. 

The MWISP clump catalogs feature broad coverage, effectively capturing the critical transition from diffuse molecular gas to gravitationally collapsing structures. Despite intrinsic survey limitations, completeness limits, and distance effects, these catalogs achieve exceptional statistical robustness and represent the most extensive study of molecular clumps to date. This unique characteristic establishes these catalogs the basis for a series of investigations, including: (1) analyzing the distributions of clump parameters across the Galactic disk and their trends with Galactocentric distance \citep{MWISP_Analysis_Sun_2024_1}; (2) identifying filaments using clump masks, directional properties, and positional properties \citep{DPConCFil}; (3) integrating multiwavelength observations to accurately classify the evolutionary stages of star-forming regions \citep{ATLASGAL_2022} and derive their associated scaling relations; and numerous additional applications. These investigations will significantly advance our understanding of the complex interplay between molecular clumps, star formation, and Galactic evolution. 

\section*{Acknowledgements}
We are grateful to the anonymous referee for the invaluable insights and comments, which enabled us to refine and enhance this work. This work is supported by the National Key R\&D Program of China (grant No. 2023YFA1608001) and the National Natural Science Foundation of China (grant No. U2031202). This research makes use of the data from the Milky Way Imaging Scroll Painting (MWISP) project, which is a multiline survey in $^{12}$CO/$^{13}$CO/C$^{18}$O along the northern Galactic plane with the PMO-13.7m telescope. We are grateful to all the members of the MWISP working group, particularly the staff members at PMO 13.7m telescope, for their long-term support. MWISP was sponsored by the National Key R\&D Program of China with grants 2023YFA1608000, 2017YFA0402701, and the CAS Key Research Program of Frontier Sciences with grant QYZDJSSW-SLH047. 

$Software$: NumPy \citep{Numpy}, Astropy \citep{Astropy_1,Astropy_2,Astropy_3}, Matplotlib \citep{Matplotlib}, SciPy \citep{SciPy}, Scikit-learn \citep{scikit-learn}, Scikit-image \citep{scikit-image}, FacetClumps \citep{FacetClumps_Sfot}, DPConCFil \citep{DPConCFil_Sfot}. 

\bibliography{MWISP_Clumps.bib}{}

\begin{thebibliography}{}
\expandafter\ifx\csname natexlab\endcsname\relax\def\natexlab#1{#1}\fi
\providecommand{\url}[1]{\href{#1}{#1}}
\providecommand{\dodoi}[1]{doi:~\href{http://doi.org/#1}{\nolinkurl{#1}}}
\providecommand{\doeprint}[1]{\href{http://ascl.net/#1}{\nolinkurl{http://ascl.net/#1}}}
\providecommand{\doarXiv}[1]{\href{https://arxiv.org/abs/#1}{\nolinkurl{https://arxiv.org/abs/#1}}}

\bibitem[{{Andr{\'e}} {et~al.}(2014){Andr{\'e}}, {Di Francesco}, {Ward-Thompson}, {Inutsuka}, {Pudritz}, \& {Pineda}}]{Review_16}
{Andr{\'e}}, P., {Di Francesco}, J., {Ward-Thompson}, D., {et~al.} 2014, in Protostars and Planets VI, ed. H.~{Beuther}, R.~S. {Klessen}, C.~P. {Dullemond}, \& T.~{Henning}, 27--51, \dodoi{10.2458/azu_uapress_9780816531240-ch002}

\bibitem[{{Astropy Collaboration} {et~al.}(2013){Astropy Collaboration}, {Robitaille}, {Tollerud}, {Greenfield}, {Droettboom}, {Bray}, {Aldcroft}, {Davis}, {Ginsburg}, {Price-Whelan}, {Kerzendorf}, {Conley}, {Crighton}, {Barbary}, {Muna}, {Ferguson}, {Grollier}, {Parikh}, {Nair}, {Unther}, {Deil}, {Woillez}, {Conseil}, {Kramer}, {Turner}, {Singer}, {Fox}, {Weaver}, {Zabalza}, {Edwards}, {Azalee Bostroem}, {Burke}, {Casey}, {Crawford}, {Dencheva}, {Ely}, {Jenness}, {Labrie}, {Lim}, {Pierfederici}, {Pontzen}, {Ptak}, {Refsdal}, {Servillat}, \& {Streicher}}]{Astropy_1}
{Astropy Collaboration}, {Robitaille}, T.~P., {Tollerud}, E.~J., {et~al.} 2013, \aap, 558, A33, \dodoi{10.1051/0004-6361/201322068}

\bibitem[{{Astropy Collaboration} {et~al.}(2018){Astropy Collaboration}, {Price-Whelan}, {Sip{\H{o}}cz}, {G{\"u}nther}, {Lim}, {Crawford}, {Conseil}, {Shupe}, {Craig}, {Dencheva}, {Ginsburg}, {VanderPlas}, {Bradley}, {P{\'e}rez-Su{\'a}rez}, {de Val-Borro}, {Aldcroft}, {Cruz}, {Robitaille}, {Tollerud}, {Ardelean}, {Babej}, {Bach}, {Bachetti}, {Bakanov}, {Bamford}, {Barentsen}, {Barmby}, {Baumbach}, {Berry}, {Biscani}, {Boquien}, {Bostroem}, {Bouma}, {Brammer}, {Bray}, {Breytenbach}, {Buddelmeijer}, {Burke}, {Calderone}, {Cano Rodr{\'\i}guez}, {Cara}, {Cardoso}, {Cheedella}, {Copin}, {Corrales}, {Crichton}, {D'Avella}, {Deil}, {Depagne}, {Dietrich}, {Donath}, {Droettboom}, {Earl}, {Erben}, {Fabbro}, {Ferreira}, {Finethy}, {Fox}, {Garrison}, {Gibbons}, {Goldstein}, {Gommers}, {Greco}, {Greenfield}, {Groener}, {Grollier}, {Hagen}, {Hirst}, {Homeier}, {Horton}, {Hosseinzadeh}, {Hu}, {Hunkeler}, {Ivezi{\'c}}, {Jain}, {Jenness}, {Kanarek}, {Kendrew}, {Kern}, {Kerzendorf}, {Khvalko}, {King}, {Kirkby}, {Kulkarni},
  {Kumar}, {Lee}, {Lenz}, {Littlefair}, {Ma}, {Macleod}, {Mastropietro}, {McCully}, {Montagnac}, {Morris}, {Mueller}, {Mumford}, {Muna}, {Murphy}, {Nelson}, {Nguyen}, {Ninan}, {N{\"o}the}, {Ogaz}, {Oh}, {Parejko}, {Parley}, {Pascual}, {Patil}, {Patil}, {Plunkett}, {Prochaska}, {Rastogi}, {Reddy Janga}, {Sabater}, {Sakurikar}, {Seifert}, {Sherbert}, {Sherwood-Taylor}, {Shih}, {Sick}, {Silbiger}, {Singanamalla}, {Singer}, {Sladen}, {Sooley}, {Sornarajah}, {Streicher}, {Teuben}, {Thomas}, {Tremblay}, {Turner}, {Terr{\'o}n}, {van Kerkwijk}, {de la Vega}, {Watkins}, {Weaver}, {Whitmore}, {Woillez}, {Zabalza}, \& {Astropy Contributors}}]{Astropy_2}
{Astropy Collaboration}, {Price-Whelan}, A.~M., {Sip{\H{o}}cz}, B.~M., {et~al.} 2018, \aj, 156, 123, \dodoi{10.3847/1538-3881/aabc4f}

\bibitem[{{Astropy Collaboration} {et~al.}(2022){Astropy Collaboration}, {Price-Whelan}, {Lim}, {Earl}, {Starkman}, {Bradley}, {Shupe}, {Patil}, {Corrales}, {Brasseur}, {N{\"o}the}, {Donath}, {Tollerud}, {Morris}, {Ginsburg}, {Vaher}, {Weaver}, {Tocknell}, {Jamieson}, {van Kerkwijk}, {Robitaille}, {Merry}, {Bachetti}, {G{\"u}nther}, {Aldcroft}, {Alvarado-Montes}, {Archibald}, {B{\'o}di}, {Bapat}, {Barentsen}, {Baz{\'a}n}, {Biswas}, {Boquien}, {Burke}, {Cara}, {Cara}, {Conroy}, {Conseil}, {Craig}, {Cross}, {Cruz}, {D'Eugenio}, {Dencheva}, {Devillepoix}, {Dietrich}, {Eigenbrot}, {Erben}, {Ferreira}, {Foreman-Mackey}, {Fox}, {Freij}, {Garg}, {Geda}, {Glattly}, {Gondhalekar}, {Gordon}, {Grant}, {Greenfield}, {Groener}, {Guest}, {Gurovich}, {Handberg}, {Hart}, {Hatfield-Dodds}, {Homeier}, {Hosseinzadeh}, {Jenness}, {Jones}, {Joseph}, {Kalmbach}, {Karamehmetoglu}, {Ka{\l}uszy{\'n}ski}, {Kelley}, {Kern}, {Kerzendorf}, {Koch}, {Kulumani}, {Lee}, {Ly}, {Ma}, {MacBride}, {Maljaars}, {Muna}, {Murphy}, {Norman},
  {O'Steen}, {Oman}, {Pacifici}, {Pascual}, {Pascual-Granado}, {Patil}, {Perren}, {Pickering}, {Rastogi}, {Roulston}, {Ryan}, {Rykoff}, {Sabater}, {Sakurikar}, {Salgado}, {Sanghi}, {Saunders}, {Savchenko}, {Schwardt}, {Seifert-Eckert}, {Shih}, {Jain}, {Shukla}, {Sick}, {Simpson}, {Singanamalla}, {Singer}, {Singhal}, {Sinha}, {Sip{\H{o}}cz}, {Spitler}, {Stansby}, {Streicher}, {{\v{S}}umak}, {Swinbank}, {Taranu}, {Tewary}, {Tremblay}, {de Val-Borro}, {Van Kooten}, {Vasovi{\'c}}, {Verma}, {de Miranda Cardoso}, {Williams}, {Wilson}, {Winkel}, {Wood-Vasey}, {Xue}, {Yoachim}, {Zhang}, {Zonca}, \& {Astropy Project Contributors}}]{Astropy_3}
{Astropy Collaboration}, {Price-Whelan}, A.~M., {Lim}, P.~L., {et~al.} 2022, \apj, 935, 167, \dodoi{10.3847/1538-4357/ac7c74}

\bibitem[{{Ballesteros-Paredes} {et~al.}(2011){Ballesteros-Paredes}, {V{\'a}zquez-Semadeni}, {Gazol}, {Hartmann}, {Heitsch}, \& {Col{\'\i}n}}]{Turbulence_4}
{Ballesteros-Paredes}, J., {V{\'a}zquez-Semadeni}, E., {Gazol}, A., {et~al.} 2011, \mnras, 416, 1436, \dodoi{10.1111/j.1365-2966.2011.19141.x}

\bibitem[{{Ballesteros-Paredes} {et~al.}(2018){Ballesteros-Paredes}, {V{\'a}zquez-Semadeni}, {Palau}, \& {Klessen}}]{Turbulence_6}
{Ballesteros-Paredes}, J., {V{\'a}zquez-Semadeni}, E., {Palau}, A., \& {Klessen}, R.~S. 2018, \mnras, 479, 2112, \dodoi{10.1093/mnras/sty1515}

\bibitem[{{Ballesteros-Paredes} {et~al.}(2020){Ballesteros-Paredes}, {Andr{\'e}}, {Hennebelle}, {Klessen}, {Kruijssen}, {Chevance}, {Nakamura}, {Adamo}, \& {V{\'a}zquez-Semadeni}}]{Review_12}
{Ballesteros-Paredes}, J., {Andr{\'e}}, P., {Hennebelle}, P., {et~al.} 2020, \ssr, 216, 76, \dodoi{10.1007/s11214-020-00698-3}

\bibitem[{{Bally}(2016)}]{Review_11}
{Bally}, J. 2016, \araa, 54, 491, \dodoi{10.1146/annurev-astro-081915-023341}

\bibitem[{{Barnes} {et~al.}(2015){Barnes}, {Muller}, {Indermuehle}, {O'Dougherty}, {Lowe}, {Cunningham}, {Hernandez}, \& {Fuller}}]{ThrUMMS}
{Barnes}, P.~J., {Muller}, E., {Indermuehle}, B., {et~al.} 2015, \apj, 812, 6, \dodoi{10.1088/0004-637X/812/1/6}

\bibitem[{{Benedettini} {et~al.}(2020){Benedettini}, {Molinari}, {Baldeschi}, {Beltr{\'a}n}, {Brand}, {Cesaroni}, {Elia}, {Fontani}, {Merello}, {Olmi}, {Pezzuto}, {Rygl}, {Schisano}, {Testi}, \& {Traficante}}]{FQS}
{Benedettini}, M., {Molinari}, S., {Baldeschi}, A., {et~al.} 2020, \aap, 633, A147, \dodoi{10.1051/0004-6361/201936096}

\bibitem[{{Bergin} \& {Tafalla}(2007)}]{Review_15}
{Bergin}, E.~A., \& {Tafalla}, M. 2007, \araa, 45, 339, \dodoi{10.1146/annurev.astro.45.071206.100404}

\bibitem[{{Berry}(2015)}]{FellWalker}
{Berry}, D.~S. 2015, Astronomy and Computing, 10, 22, \dodoi{10.1016/j.ascom.2014.11.004}

\bibitem[{{Bertoldi} \& {McKee}(1992)}]{Virial_Analysis_1}
{Bertoldi}, F., \& {McKee}, C.~F. 1992, \apj, 395, 140, \dodoi{10.1086/171638}

\bibitem[{{Bian} {et~al.}(2022){Bian}, {Xu}, {Li}, {Wu}, {Zhang}, {Chen}, {Li}, {Lin}, {Hao}, \& {Liu}}]{Maser_4}
{Bian}, S.~B., {Xu}, Y., {Li}, J.~J., {et~al.} 2022, \aj, 163, 54, \dodoi{10.3847/1538-3881/ac3d90}

\bibitem[{{Bian} {et~al.}(2024){Bian}, {Wu}, {Xu}, {Reid}, {Li}, {Zhang}, {Menten}, {Moscadelli}, \& {Brunthaler}}]{Maser_7}
{Bian}, S.~B., {Wu}, Y.~W., {Xu}, Y., {et~al.} 2024, \aj, 167, 267, \dodoi{10.3847/1538-3881/ad4030}

\bibitem[{{Bourke} {et~al.}(1997){Bourke}, {Garay}, {Lehtinen}, {K{\"o}hnenkamp}, {Launhardt}, {Nyman}, {May}, {Robinson}, \& {Hyland}}]{Tex}
{Bourke}, T.~L., {Garay}, G., {Lehtinen}, K.~K., {et~al.} 1997, \apj, 476, 781, \dodoi{10.1086/303642}

\bibitem[{{Breen} {et~al.}(2010){Breen}, {Caswell}, {Ellingsen}, \& {Phillips}}]{Maser_Star_Formation_4}
{Breen}, S.~L., {Caswell}, J.~L., {Ellingsen}, S.~P., \& {Phillips}, C.~J. 2010, \mnras, 406, 1487, \dodoi{10.1111/j.1365-2966.2010.16791.x}

\bibitem[{{Burton}(1976)}]{Review_17}
{Burton}, W.~B. 1976, \araa, 14, 275, \dodoi{10.1146/annurev.aa.14.090176.001423}

\bibitem[{{Colombo} {et~al.}(2015){Colombo}, {Rosolowsky}, {Ginsburg}, {Duarte-Cabral}, \& {Hughes}}]{SCIMES}
{Colombo}, D., {Rosolowsky}, E., {Ginsburg}, A., {Duarte-Cabral}, A., \& {Hughes}, A. 2015, \mnras, 454, 2067, \dodoi{10.1093/mnras/stv2063}

\bibitem[{{Cragg} {et~al.}(2005){Cragg}, {Sobolev}, \& {Godfrey}}]{Maser_Star_Formation_3}
{Cragg}, D.~M., {Sobolev}, A.~M., \& {Godfrey}, P.~D. 2005, \mnras, 360, 533, \dodoi{10.1111/j.1365-2966.2005.09077.x}

\bibitem[{{Dame} {et~al.}(2001){Dame}, {Hartmann}, \& {Thaddeus}}]{CO_Survey_3}
{Dame}, T.~M., {Hartmann}, D., \& {Thaddeus}, P. 2001, \apj, 547, 792, \dodoi{10.1086/318388}

\bibitem[{{Dame} \& {Thaddeus}(1985)}]{KDA_Method_PR_4}
{Dame}, T.~M., \& {Thaddeus}, P. 1985, \apj, 297, 751, \dodoi{10.1086/163573}

\bibitem[{{Dame} {et~al.}(1987){Dame}, {Ungerechts}, {Cohen}, {de Geus}, {Grenier}, {May}, {Murphy}, {Nyman}, \& {Thaddeus}}]{CO_Survey_2}
{Dame}, T.~M., {Ungerechts}, H., {Cohen}, R.~S., {et~al.} 1987, \apj, 322, 706, \dodoi{10.1086/165766}

\bibitem[{{Dong} {et~al.}(2023){Dong}, {Sun}, {Xu}, {Lin}, {Bian}, {Hao}, {Liu}, {Li}, {Yang}, {Su}, {Zhou}, {Zhang}, {Yan}, \& {Chen}}]{MWISP_Analysis_Dong_2023}
{Dong}, Y., {Sun}, Y., {Xu}, Y., {et~al.} 2023, \apjs, 268, 1, \dodoi{10.3847/1538-4365/acde81}

\bibitem[{{Duarte-Cabral} {et~al.}(2021){Duarte-Cabral}, {Colombo}, {Urquhart}, {Ginsburg}, {Russeil}, {Schuller}, {Anderson}, {Barnes}, {Beltr{\'a}n}, {Beuther}, {Bontemps}, {Bronfman}, {Csengeri}, {Dobbs}, {Eden}, {Giannetti}, {Kauffmann}, {Mattern}, {Medina}, {Menten}, {Lee}, {Pettitt}, {Riener}, {Rigby}, {Traficante}, {Veena}, {Wienen}, {Wyrowski}, {Agurto}, {Azagra}, {Cesaroni}, {Finger}, {Gonzalez}, {Henning}, {Hernandez}, {Kainulainen}, {Leurini}, {Lopez}, {Mac-Auliffe}, {Mazumdar}, {Molinari}, {Motte}, {Muller}, {Nguyen-Luong}, {Parra}, {Perez-Beaupuits}, {Montenegro-Montes}, {Moore}, {Ragan}, {S{\'a}nchez-Monge}, {Sanna}, {Schilke}, {Schisano}, {Schneider}, {Suri}, {Testi}, {Torstensson}, {Venegas}, {Wang}, \& {Zavagno}}]{SEDIGISM_2}
{Duarte-Cabral}, A., {Colombo}, D., {Urquhart}, J.~S., {et~al.} 2021, {The SEDIGISM survey: molecular clouds in the inner Galaxy},  OUP, \dodoi{10.1093/mnras/staa2480}

\bibitem[{{Eden} {et~al.}(2012){Eden}, {Moore}, {Plume}, \& {Morgan}}]{GRS_Dist_2}
{Eden}, D.~J., {Moore}, T.~J.~T., {Plume}, R., \& {Morgan}, L.~K. 2012, \mnras, 422, 3178, \dodoi{10.1111/j.1365-2966.2012.20840.x}

\bibitem[{{Eden} {et~al.}(2017){Eden}, {Moore}, {Plume}, {Urquhart}, {Thompson}, {Parsons}, {Dempsey}, {Rigby}, {Morgan}, {Thomas}, {Berry}, {Buckle}, {Brunt}, {Butner}, {Carretero}, {Chrysostomou}, {Currie}, {deVilliers}, {Fich}, {Gibb}, {Hoare}, {Jenness}, {Manser}, {Mottram}, {Natario}, {Olguin}, {Peretto}, {Pestalozzi}, {Polychroni}, {Redman}, {Salji}, {Summers}, {Tahani}, {Traficante}, {diFrancesco}, {Evans}, {Fuller}, {Johnstone}, {Joncas}, {Longmore}, {Martin}, {Richer}, {Weferling}, {White}, \& {Zhu}}]{SN_Map_1}
{Eden}, D.~J., {Moore}, T.~J.~T., {Plume}, R., {et~al.} 2017, \mnras, 469, 2163, \dodoi{10.1093/mnras/stx874}

\bibitem[{{Elitzur}(1992)}]{Review_9}
{Elitzur}, M. 1992, \araa, 30, 75, \dodoi{10.1146/annurev.aa.30.090192.000451}

\bibitem[{{Elmegreen} \& {Scalo}(2004)}]{Review_10}
{Elmegreen}, B.~G., \& {Scalo}, J. 2004, \araa, 42, 211, \dodoi{10.1146/annurev.astro.41.011802.094859}

\bibitem[{{Feng} {et~al.}(2024){Feng}, {Chen}, {Jiang}, {Ma}, {Yang}, {Yu}, {Ge}, {Zhou}, {Du}, {Wang}, {Zhang}, {Su}, \& {Yang}}]{MWISP_Analysis_Feng_2024}
{Feng}, H., {Chen}, Z., {Jiang}, Z., {et~al.} 2024, Research in Astronomy and Astrophysics, 24, 115018, \dodoi{10.1088/1674-4527/ad89a9}

\bibitem[{{Frerking} {et~al.}(1982){Frerking}, {Langer}, \& {Wilson}}]{Ratio_CO12_H2}
{Frerking}, M.~A., {Langer}, W.~D., \& {Wilson}, R.~W. 1982, \apj, 262, 590, \dodoi{10.1086/160451}

\bibitem[{{Ge} \& {Wang}(2022)}]{MST_Apply_1}
{Ge}, Y., \& {Wang}, K. 2022, \apjs, 259, 36, \dodoi{10.3847/1538-4365/ac4a76}

\bibitem[{{Giannetti} {et~al.}(2015){Giannetti}, {Wyrowski}, {Leurini}, {Urquhart}, {Csengeri}, {Menten}, {Bronfman}, \& {van der Tak}}]{IRDC_3}
{Giannetti}, A., {Wyrowski}, F., {Leurini}, S., {et~al.} 2015, \aap, 580, L7, \dodoi{10.1051/0004-6361/201526474}

\bibitem[{{Giannetti} {et~al.}(2013){Giannetti}, {Brand}, {S{\'a}nchez-Monge}, {Fontani}, {Cesaroni}, {Beltr{\'a}n}, {Molinari}, {Dodson}, \& {Rioja}}]{Cloud_Evolution_4}
{Giannetti}, A., {Brand}, J., {S{\'a}nchez-Monge}, {\'A}., {et~al.} 2013, \aap, 556, A16, \dodoi{10.1051/0004-6361/201321456}

\bibitem[{{Hacar} {et~al.}(2013){Hacar}, {Tafalla}, {Kauffmann}, \& {Kov{\'a}cs}}]{FIVe}
{Hacar}, A., {Tafalla}, M., {Kauffmann}, J., \& {Kov{\'a}cs}, A. 2013, \aap, 554, A55, \dodoi{10.1051/0004-6361/201220090}

\bibitem[{{Heitsch} {et~al.}(2009){Heitsch}, {Ballesteros-Paredes}, \& {Hartmann}}]{Gravitation_2}
{Heitsch}, F., {Ballesteros-Paredes}, J., \& {Hartmann}, L. 2009, \apj, 704, 1735, \dodoi{10.1088/0004-637X/704/2/1735}

\bibitem[{{Hennebelle} \& {Falgarone}(2012)}]{Review_8}
{Hennebelle}, P., \& {Falgarone}, E. 2012, \aapr, 20, 55, \dodoi{10.1007/s00159-012-0055-y}

\bibitem[{{Hennebelle} \& {Inutsuka}(2019)}]{Review_4}
{Hennebelle}, P., \& {Inutsuka}, S.-i. 2019, Frontiers in Astronomy and Space Sciences, 6, 5, \dodoi{10.3389/fspas.2019.00005}

\bibitem[{{Heyer} \& {Dame}(2015)}]{Review_13}
{Heyer}, M., \& {Dame}, T.~M. 2015, \araa, 53, 583, \dodoi{10.1146/annurev-astro-082214-122324}

\bibitem[{{Heyer} {et~al.}(2009){Heyer}, {Krawczyk}, {Duval}, \& {Jackson}}]{KDA_Method_PR_2}
{Heyer}, M., {Krawczyk}, C., {Duval}, J., \& {Jackson}, J.~M. 2009, \apj, 699, 1092, \dodoi{10.1088/0004-637X/699/2/1092}

\bibitem[{{Heyer} \& {Brunt}(2004)}]{Turbulence_1}
{Heyer}, M.~H., \& {Brunt}, C.~M. 2004, \apjl, 615, L45, \dodoi{10.1086/425978}

\bibitem[{{HI4PI Collaboration} {et~al.}(2016){HI4PI Collaboration}, {Ben Bekhti}, {Fl{\"o}er}, {Keller}, {Kerp}, {Lenz}, {Winkel}, {Bailin}, {Calabretta}, {Dedes}, {Ford}, {Gibson}, {Haud}, {Janowiecki}, {Kalberla}, {Lockman}, {McClure-Griffiths}, {Murphy}, {Nakanishi}, {Pisano}, \& {Staveley-Smith}}]{HI4PI}
{HI4PI Collaboration}, {Ben Bekhti}, N., {Fl{\"o}er}, L., {et~al.} 2016, \aap, 594, A116, \dodoi{10.1051/0004-6361/201629178}

\bibitem[{{Hunter}(2007)}]{Matplotlib}
{Hunter}, J.~D. 2007, Computing in Science and Engineering, 9, 90, \dodoi{10.1109/MCSE.2007.55}

\bibitem[{{Jackson} {et~al.}(2006){Jackson}, {Rathborne}, {Shah}, {Simon}, {Bania}, {Clemens}, {Chambers}, {Johnson}, {Dormody}, {Lavoie}, \& {Heyer}}]{GRS_1}
{Jackson}, J.~M., {Rathborne}, J.~M., {Shah}, R.~Y., {et~al.} 2006, \apjs, 163, 145, \dodoi{10.1086/500091}

\bibitem[{{Jiang}(2023)}]{FacetClumps_Sfot}
{Jiang}, Y. 2023, {FacetClumps: A Facet-based Molecular Clump Detection Algorithm}, v0.0.4,  Zenodo, \dodoi{10.5281/zenodo.7991006}

\bibitem[{{Jiang}(2024)}]{DPConCFil_Sfot}
---. 2024, {DPConCFil: A Collection of Filament Identification and Analysis Algorithms}, Software, \dodoi{10.5281/zenodo.13853675}

\bibitem[{{Jiang} {et~al.}(2023){Jiang}, {Chen}, {Zheng}, {Jiang}, {Huang}, {Zeng}, {Zeng}, \& {Luo}}]{FacetClumps}
{Jiang}, Y., {Chen}, Z., {Zheng}, S., {et~al.} 2023, \apjs, 267, 32, \dodoi{10.3847/1538-4365/acda89}

\bibitem[{{Jiang} {et~al.}(2022){Jiang}, {Zheng}, {Jiang}, {Zeng}, {Chen}, {Zeng}, {Luo}, \& {Huang}}]{ConBased}
{Jiang}, Y., {Zheng}, S., {Jiang}, Z., {et~al.} 2022, Astronomy and Computing, 40, 100613, \dodoi{10.1016/j.ascom.2022.100613}

\bibitem[{{Jiang} {et~al.}(2025){Jiang}, {Chen}, {Zheng}, {Jiang}, {Chen}, {Huang}, {Su}, {Sun}, {Feng}, {Feng}, \& {Yang}}]{DPConCFil}
{Jiang}, Y., {Chen}, X., {Zheng}, S., {et~al.} 2025, \apjs, 276, 27, \dodoi{10.3847/1538-4365/ad91a8}

\bibitem[{{Kainulainen} \& {Tan}(2013)}]{IRDC_4}
{Kainulainen}, J., \& {Tan}, J.~C. 2013, \aap, 549, A53, \dodoi{10.1051/0004-6361/201219526}

\bibitem[{{Kauffmann} {et~al.}(2010){Kauffmann}, {Pillai}, {Shetty}, {Myers}, \& {Goodman}}]{PRelation_1}
{Kauffmann}, J., {Pillai}, T., {Shetty}, R., {Myers}, P.~C., \& {Goodman}, A.~A. 2010, \apj, 712, 1137, \dodoi{10.1088/0004-637X/712/2/1137}

\bibitem[{{Krumholz}(2014)}]{Review_6}
{Krumholz}, M.~R. 2014, \physrep, 539, 49, \dodoi{10.1016/j.physrep.2014.02.001}

\bibitem[{{Krumholz} {et~al.}(2005){Krumholz}, {McKee}, \& {Klein}}]{Gravitation_1}
{Krumholz}, M.~R., {McKee}, C.~F., \& {Klein}, R.~I. 2005, \nat, 438, 332, \dodoi{10.1038/nature04280}

\bibitem[{{Larson}(1981)}]{KDA_Method_PR_1}
{Larson}, R.~B. 1981, \mnras, 194, 809, \dodoi{10.1093/mnras/194.4.809}

\bibitem[{{Li} {et~al.}(2023){Li}, {Wang}, {Ma}, {Lin}, {Su}, {Li}, {Sun}, {Zhou}, \& {Yang}}]{MWISP_Analysis_Li_2023}
{Li}, C., {Wang}, H., {Ma}, Y., {et~al.} 2023, \apjs, 267, 30, \dodoi{10.3847/1538-4365/acd9a7}

\bibitem[{{Li} {et~al.}(2022){Li}, {Immer}, {Reid}, {Sanna}, {Rygl}, {Xu}, {Zhang}, {Brunthaler}, \& {Menten}}]{Maser_6}
{Li}, J.~J., {Immer}, K., {Reid}, M.~J., {et~al.} 2022, \apjs, 262, 42, \dodoi{10.3847/1538-4365/ac8791}

\bibitem[{{Lombardi} {et~al.}(2010){Lombardi}, {Alves}, \& {Lada}}]{PRelation_4}
{Lombardi}, M., {Alves}, J., \& {Lada}, C.~J. 2010, \aap, 519, L7, \dodoi{10.1051/0004-6361/201015282}

\bibitem[{{Long} {et~al.}(2024){Long}, {Zheng}, {Huang}, {Zeng}, {Jiang}, {Chen}, {Luo}, {Jiang}, \& {Zeng}}]{MWISP_Analysis_Long_2024}
{Long}, C., {Zheng}, S., {Huang}, Y., {et~al.} 2024, \na, 110, 102215, \dodoi{10.1016/j.newast.2024.102215}

\bibitem[{{Luo} {et~al.}(2022){Luo}, {Zheng}, {Huang}, {Zeng}, {Zeng}, {Jiang}, \& {Chen}}]{LDC}
{Luo}, X., {Zheng}, S., {Huang}, Y., {et~al.} 2022, Research in Astronomy and Astrophysics, 22, 015003, \dodoi{10.1088/1674-4527/ac321d}

\bibitem[{{Luo} {et~al.}(2024{\natexlab{a}}){Luo}, {Zheng}, {Jiang}, {Chen}, {Huang}, {Zeng}, \& {Zeng}}]{MWISP_Analysis_Luo_2024_2}
{Luo}, X., {Zheng}, S., {Jiang}, Z., {et~al.} 2024{\natexlab{a}}, \aap, 683, A104, \dodoi{10.1051/0004-6361/202347341}

\bibitem[{{Luo} {et~al.}(2024{\natexlab{b}}){Luo}, {Zheng}, {Jiang}, {Chen}, {Huang}, {Zeng}, {Zeng}, {Zhang}, {Long}, {Zhou}, \& {Hu}}]{MWISP_Analysis_Luo_2024}
---. 2024{\natexlab{b}}, Research in Astronomy and Astrophysics, 24, 055018, \dodoi{10.1088/1674-4527/ad3d12}

\bibitem[{{MacLaren} {et~al.}(1988){MacLaren}, {Richardson}, \& {Wolfendale}}]{Virial_Analysis_4}
{MacLaren}, I., {Richardson}, K.~M., \& {Wolfendale}, A.~W. 1988, \apj, 333, 821, \dodoi{10.1086/166791}

\bibitem[{{McClure-Griffiths} {et~al.}(2005){McClure-Griffiths}, {Dickey}, {Gaensler}, {Green}, {Haverkorn}, \& {Strasser}}]{SGPS}
{McClure-Griffiths}, N.~M., {Dickey}, J.~M., {Gaensler}, B.~M., {et~al.} 2005, \apjs, 158, 178, \dodoi{10.1086/430114}

\bibitem[{{McKee} \& {Ostriker}(2007)}]{Review_5}
{McKee}, C.~F., \& {Ostriker}, E.~C. 2007, \araa, 45, 565, \dodoi{10.1146/annurev.astro.45.051806.110602}

\bibitem[{{M{\`e}ge} {et~al.}(2021){M{\`e}ge}, {Russeil}, {Zavagno}, {Elia}, {Molinari}, {Brunt}, {Butora}, {Cambresy}, {Di Giorgio}, {Fenouillet}, {Fukui}, {Lambert}, {Makai}, {Merello}, {Meunier}, {Molinaro}, {Moreau}, {Pezzuto}, {Poulin}, {Schisano}, \& {Schuller}}]{HiGal_Dist}
{M{\`e}ge}, P., {Russeil}, D., {Zavagno}, A., {et~al.} 2021, \aap, 646, A74, \dodoi{10.1051/0004-6361/202038956}

\bibitem[{{Mei} {et~al.}(2024){Mei}, {Chen}, {Jiang}, {Zheng}, \& {Feng}}]{Emission_D2}
{Mei}, J., {Chen}, Z., {Jiang}, Z., {Zheng}, S., \& {Feng}, H. 2024, \aap, 685, A39, \dodoi{10.1051/0004-6361/202347952}

\bibitem[{{Menten}(1991)}]{Maser_Star_Formation_1}
{Menten}, K.~M. 1991, \apjl, 380, L75, \dodoi{10.1086/186177}

\bibitem[{{Milam} {et~al.}(2005){Milam}, {Savage}, {Brewster}, {Ziurys}, \& {Wyckoff}}]{Trans_Factor_With_RGC}
{Milam}, S.~N., {Savage}, C., {Brewster}, M.~A., {Ziurys}, L.~M., \& {Wyckoff}, S. 2005, \apj, 634, 1126, \dodoi{10.1086/497123}

\bibitem[{{Miville-Desch{\^e}nes} {et~al.}(2017){Miville-Desch{\^e}nes}, {Murray}, \& {Lee}}]{CFA_2017}
{Miville-Desch{\^e}nes}, M.-A., {Murray}, N., \& {Lee}, E.~J. 2017, \apj, 834, 57, \dodoi{10.3847/1538-4357/834/1/57}

\bibitem[{{Molinari} {et~al.}(2016){Molinari}, {Merello}, {Elia}, {Cesaroni}, {Testi}, \& {Robitaille}}]{Cloud_Evolution_3}
{Molinari}, S., {Merello}, M., {Elia}, D., {et~al.} 2016, \apjl, 826, L8, \dodoi{10.3847/2041-8205/826/1/L8}

\bibitem[{{Molinari} {et~al.}(2010){Molinari}, {Swinyard}, {Bally}, {Barlow}, {Bernard}, {Martin}, {Moore}, {Noriega-Crespo}, {Plume}, {Testi}, {Zavagno}, {Abergel}, {Ali}, {Andr{\'e}}, {Baluteau}, {Benedettini}, {Bern{\'e}}, {Billot}, {Blommaert}, {Bontemps}, {Boulanger}, {Brand}, {Brunt}, {Burton}, {Campeggio}, {Carey}, {Caselli}, {Cesaroni}, {Cernicharo}, {Chakrabarti}, {Chrysostomou}, {Codella}, {Cohen}, {Compiegne}, {Davis}, {de Bernardis}, {de Gasperis}, {Di Francesco}, {di Giorgio}, {Elia}, {Faustini}, {Fischera}, {Fukui}, {Fuller}, {Ganga}, {Garcia-Lario}, {Giard}, {Giardino}, {Glenn}, {Goldsmith}, {Griffin}, {Hoare}, {Huang}, {Jiang}, {Joblin}, {Joncas}, {Juvela}, {Kirk}, {Lagache}, {Li}, {Lim}, {Lord}, {Lucas}, {Maiolo}, {Marengo}, {Marshall}, {Masi}, {Massi}, {Matsuura}, {Meny}, {Minier}, {Miville-Desch{\^e}nes}, {Montier}, {Motte}, {M{\"u}ller}, {Natoli}, {Neves}, {Olmi}, {Paladini}, {Paradis}, {Pestalozzi}, {Pezzuto}, {Piacentini}, {Pomar{\`e}s}, {Popescu}, {Reach}, {Richer}, {Ristorcelli},
  {Roy}, {Royer}, {Russeil}, {Saraceno}, {Sauvage}, {Schilke}, {Schneider-Bontemps}, {Schuller}, {Schultz}, {Shepherd}, {Sibthorpe}, {Smith}, {Smith}, {Spinoglio}, {Stamatellos}, {Strafella}, {Stringfellow}, {Sturm}, {Taylor}, {Thompson}, {Tuffs}, {Umana}, {Valenziano}, {Vavrek}, {Viti}, {Waelkens}, {Ward-Thompson}, {White}, {Wyrowski}, {Yorke}, \& {Zhang}}]{HiGal_0}
{Molinari}, S., {Swinyard}, B., {Bally}, J., {et~al.} 2010, \pasp, 122, 314, \dodoi{10.1086/651314}

\bibitem[{{Motte} {et~al.}(2018){Motte}, {Bontemps}, \& {Louvet}}]{Review_3}
{Motte}, F., {Bontemps}, S., \& {Louvet}, F. 2018, \araa, 56, 41, \dodoi{10.1146/annurev-astro-091916-055235}

\bibitem[{{Oliphant}(2007)}]{Numpy}
{Oliphant}, T.~E. 2007, Computing in Science and Engineering, 9, 10, \dodoi{10.1109/MCSE.2007.58}

\bibitem[{Pedregosa {et~al.}(2011)Pedregosa, Varoquaux, Gramfort, Michel, Thirion, Grisel, Blondel, Prettenhofer, Weiss, Dubourg, Vanderplas, Passos, Cournapeau, Brucher, Perrot, \& Duchesnay}]{scikit-learn}
Pedregosa, F., Varoquaux, G., Gramfort, A., {et~al.} 2011, Journal of Machine Learning Research, 12, 2825

\bibitem[{{Peek} {et~al.}(2011){Peek}, {Heiles}, {Douglas}, {Lee}, {Grcevich}, {Stanimirovi{\'c}}, {Putman}, {Korpela}, {Gibson}, {Begum}, {Saul}, {Robishaw}, \& {Kr{\v{c}}o}}]{GalFa}
{Peek}, J.~E.~G., {Heiles}, C., {Douglas}, K.~A., {et~al.} 2011, \apjs, 194, 20, \dodoi{10.1088/0067-0049/194/2/20}

\bibitem[{{Peretto} \& {Fuller}(2009)}]{IRDC_2}
{Peretto}, N., \& {Fuller}, G.~A. 2009, \aap, 505, 405, \dodoi{10.1051/0004-6361/200912127}

\bibitem[{{Pineda} {et~al.}(2023){Pineda}, {Arzoumanian}, {Andre}, {Friesen}, {Zavagno}, {Clarke}, {Inoue}, {Chen}, {Lee}, {Soler}, \& {Kuffmeier}}]{Review_2}
{Pineda}, J.~E., {Arzoumanian}, D., {Andre}, P., {et~al.} 2023, in Astronomical Society of the Pacific Conference Series, Vol. 534, Protostars and Planets VII, ed. S.~{Inutsuka}, Y.~{Aikawa}, T.~{Muto}, K.~{Tomida}, \& M.~{Tamura}, 233, \dodoi{10.48550/arXiv.2205.03935}

\bibitem[{{Pineda} {et~al.}(2010){Pineda}, {Goldsmith}, {Chapman}, {Snell}, {Li}, {Cambr{\'e}sy}, \& {Brunt}}]{Col_Density}
{Pineda}, J.~L., {Goldsmith}, P.~F., {Chapman}, N., {et~al.} 2010, \apj, 721, 686, \dodoi{10.1088/0004-637X/721/1/686}

\bibitem[{{Rani} {et~al.}(2023){Rani}, {Moore}, {Eden}, {Rigby}, {Duarte-Cabral}, \& {Lee}}]{SN_Map_2}
{Rani}, R., {Moore}, T. J.~T., {Eden}, D.~J., {et~al.} 2023, \mnras, 523, 1832, \dodoi{10.1093/mnras/stad1507}

\bibitem[{{Reid} {et~al.}(2019){Reid}, {Menten}, {Brunthaler}, {Zheng}, {Dame}, {Xu}, {Li}, {Sakai}, {Wu}, {Immer}, {Zhang}, {Sanna}, {Moscadelli}, {Rygl}, {Bartkiewicz}, {Hu}, {Quiroga-Nu{\~n}ez}, \& {van Langevelde}}]{Distance_1}
{Reid}, M.~J., {Menten}, K.~M., {Brunthaler}, A., {et~al.} 2019, \apj, 885, 131, \dodoi{10.3847/1538-4357/ab4a11}

\bibitem[{{Rigby} {et~al.}(2016){Rigby}, {Moore}, {Plume}, {Eden}, {Urquhart}, {Thompson}, {Mottram}, {Brunt}, {Butner}, {Dempsey}, {Gibson}, {Hatchell}, {Jenness}, {Kuno}, {Longmore}, {Morgan}, {Polychroni}, {Thomas}, {White}, \& {Zhu}}]{CHIMPS_1}
{Rigby}, A.~J., {Moore}, T.~J.~T., {Plume}, R., {et~al.} 2016, \mnras, 456, 2885, \dodoi{10.1093/mnras/stv2808}

\bibitem[{{Rigby} {et~al.}(2019){Rigby}, {Moore}, {Eden}, {Urquhart}, {Ragan}, {Peretto}, {Plume}, {Thompson}, {Currie}, \& {Park}}]{CHIMPS_2}
{Rigby}, A.~J., {Moore}, T.~J.~T., {Eden}, D.~J., {et~al.} 2019, \aap, 632, A58, \dodoi{10.1051/0004-6361/201935236}

\bibitem[{{Rigby} {et~al.}(2025){Rigby}, {Thompson}, {Eden}, {Moore}, {Mutale}, {Peretto}, {Plume}, {Urquhart}, {Williams}, \& {Currie}}]{PAMS}
{Rigby}, A.~J., {Thompson}, M.~A., {Eden}, D.~J., {et~al.} 2025, \mnras, 538, 198, \dodoi{10.1093/mnras/staf278}

\bibitem[{{Roman-Duval} {et~al.}(2009){Roman-Duval}, {Jackson}, {Heyer}, {Johnson}, {Rathborne}, {Shah}, \& {Simon}}]{GRS_Dist_1}
{Roman-Duval}, J., {Jackson}, J.~M., {Heyer}, M., {et~al.} 2009, \apj, 699, 1153, \dodoi{10.1088/0004-637X/699/2/1153}

\bibitem[{{Roman-Duval} {et~al.}(2010){Roman-Duval}, {Jackson}, {Heyer}, {Rathborne}, \& {Simon}}]{GRS_2}
{Roman-Duval}, J., {Jackson}, J.~M., {Heyer}, M., {Rathborne}, J., \& {Simon}, R. 2010, \apj, 723, 492, \dodoi{10.1088/0004-637X/723/1/492}

\bibitem[{{Rosolowsky} \& {Leroy}(2006)}]{Extrapolation}
{Rosolowsky}, E., \& {Leroy}, A. 2006, \pasp, 118, 590, \dodoi{10.1086/502982}

\bibitem[{{Sakai} {et~al.}(2022){Sakai}, {Nakanishi}, {Kurahara}, {Sakai}, {Hachisuka}, {Kim}, \& {Kameya}}]{Maser_5}
{Sakai}, N., {Nakanishi}, H., {Kurahara}, K., {et~al.} 2022, \pasj, 74, 209, \dodoi{10.1093/pasj/psab118}

\bibitem[{{Sanders} {et~al.}(1986){Sanders}, {Clemens}, {Scoville}, \& {Solomon}}]{CO_Survey_1}
{Sanders}, D.~B., {Clemens}, D.~P., {Scoville}, N.~Z., \& {Solomon}, P.~M. 1986, \apjs, 60, 1, \dodoi{10.1086/191086}

\bibitem[{{Schuller} {et~al.}(2009){Schuller}, {Menten}, {Contreras}, {Wyrowski}, {Schilke}, {Bronfman}, {Henning}, {Walmsley}, {Beuther}, {Bontemps}, {Cesaroni}, {Deharveng}, {Garay}, {Herpin}, {Lefloch}, {Linz}, {Mardones}, {Minier}, {Molinari}, {Motte}, {Nyman}, {Reveret}, {Risacher}, {Russeil}, {Schneider}, {Testi}, {Troost}, {Vasyunina}, {Wienen}, {Zavagno}, {Kovacs}, {Kreysa}, {Siringo}, \& {Wei{\ss}}}]{ATLASGAL_2009}
{Schuller}, F., {Menten}, K.~M., {Contreras}, Y., {et~al.} 2009, \aap, 504, 415, \dodoi{10.1051/0004-6361/200811568}

\bibitem[{{Schuller} {et~al.}(2017){Schuller}, {Csengeri}, {Urquhart}, {Duarte-Cabral}, {Barnes}, {Giannetti}, {Hernandez}, {Leurini}, {Mattern}, {Medina}, {Agurto}, {Azagra}, {Anderson}, {Beltr{\'a}n}, {Beuther}, {Bontemps}, {Bronfman}, {Dobbs}, {Dumke}, {Finger}, {Ginsburg}, {Gonzalez}, {Henning}, {Kauffmann}, {Mac-Auliffe}, {Menten}, {Montenegro-Montes}, {Moore}, {Muller}, {Parra}, {Perez-Beaupuits}, {Pettitt}, {Russeil}, {S{\'a}nchez-Monge}, {Schilke}, {Schisano}, {Suri}, {Testi}, {Torstensson}, {Venegas}, {Wang}, {Wienen}, {Wyrowski}, \& {Zavagno}}]{SEDIGISM_1}
{Schuller}, F., {Csengeri}, T., {Urquhart}, J.~S., {et~al.} 2017, \aap, 601, A124, \dodoi{10.1051/0004-6361/201628933}

\bibitem[{{Shan} {et~al.}(2012){Shan}, {Yang}, {Shi}, {Yao}, {Zuo}, {Lin}, {Chen}, {Zhang}, {Duan}, {Cao}, {Li}, {Li}, {Liu}, \& {Zhong}}]{CO_Survey_MWISP_2}
{Shan}, W., {Yang}, J., {Shi}, S., {et~al.} 2012, IEEE Transactions on Terahertz Science and Technology, 2, 593, \dodoi{10.1109/TTHZ.2012.2213818}

\bibitem[{{Simon} {et~al.}(2006){Simon}, {Jackson}, {Rathborne}, \& {Chambers}}]{IRDC_1}
{Simon}, R., {Jackson}, J.~M., {Rathborne}, J.~M., \& {Chambers}, E.~T. 2006, \apj, 639, 227, \dodoi{10.1086/499342}

\bibitem[{{Solomon} {et~al.}(1987){Solomon}, {Rivolo}, {Barrett}, \& {Yahil}}]{KDA_Method_PR_5}
{Solomon}, P.~M., {Rivolo}, A.~R., {Barrett}, J., \& {Yahil}, A. 1987, \apj, 319, 730, \dodoi{10.1086/165493}

\bibitem[{{Stil} {et~al.}(2006){Stil}, {Taylor}, {Dickey}, {Kavars}, {Martin}, {Rothwell}, {Boothroyd}, {Lockman}, \& {McClure-Griffiths}}]{VGPS}
{Stil}, J.~M., {Taylor}, A.~R., {Dickey}, J.~M., {et~al.} 2006, \aj, 132, 1158, \dodoi{10.1086/505940}

\bibitem[{{Su} {et~al.}(2019){Su}, {Yang}, {Zhang}, {Gong}, {Wang}, {Zhou}, {Wang}, {Chen}, {Sun}, {Chen}, {Xu}, \& {Jiang}}]{CO_Survey_MWISP_1}
{Su}, Y., {Yang}, J., {Zhang}, S., {et~al.} 2019, \apjs, 240, 9, \dodoi{10.3847/1538-4365/aaf1c8}

\bibitem[{{Su} {et~al.}(2020){Su}, {Yang}, {Yan}, {Gong}, {Chen}, {Zhang}, {Sun}, {Zhang}, {Chen}, {Zhou}, {Wang}, {Wang}, {Xu}, \& {Jiang}}]{MWISP_Analysis_Su_2020}
{Su}, Y., {Yang}, J., {Yan}, Q.-Z., {et~al.} 2020, \apj, 893, 91, \dodoi{10.3847/1538-4357/ab7fff}

\bibitem[{{Su} {et~al.}(2021){Su}, {Yang}, {Yan}, {Zhang}, {Wang}, {Sun}, {Chen}, {Wang}, {Zhou}, {Chen}, {Jiang}, \& {Wang}}]{KDA_Method_Height_2}
---. 2021, \apj, 910, 131, \dodoi{10.3847/1538-4357/abe5ab}

\bibitem[{{Su} {et~al.}(2022){Su}, {Zhang}, {Yang}, {Yan}, {Sun}, {Wang}, {Zhang}, {Chen}, {Chen}, {Zhou}, \& {Yuan}}]{MWISP_Analysis_Su_2022}
{Su}, Y., {Zhang}, S., {Yang}, J., {et~al.} 2022, \apj, 930, 112, \dodoi{10.3847/1538-4357/ac63b3}

\bibitem[{{Su} {et~al.}(2024){Su}, {Zhang}, {Sun}, {Yang}, {Yan}, {Zhang}, {Chen}, {Chen}, {Zhou}, \& {Yuan}}]{MWISP_Analysis_Su_2024}
{Su}, Y., {Zhang}, S., {Sun}, Y., {et~al.} 2024, \apjl, 971, L6, \dodoi{10.3847/2041-8213/ad656d}

\bibitem[{{Su} {et~al.}(2025){Su}, {Zhang}, {Sun}, {Yang}, {Du}, {Fang}, {Yan}, {Zhang}, {Chen}, {Chen}, {Zhou}, {Yuan}, \& {Ma}}]{MWISP_Analysis_Su_2025}
---. 2025, \apj, 984, 109, \dodoi{10.3847/1538-4357/adc38e}

\bibitem[{{Sun} {et~al.}(2024{\natexlab{a}}){Sun}, {Chen}, {Fang}, {Zhang}, {Gong}, {Feng}, {Li}, {Yan}, \& {Yang}}]{MWISP_Analysis_SunL_2024}
{Sun}, L., {Chen}, X., {Fang}, M., {et~al.} 2024{\natexlab{a}}, \aj, 167, 176, \dodoi{10.3847/1538-3881/ad2ea3}

\bibitem[{{Sun} {et~al.}(2021){Sun}, {Yang}, {Yan}, {Lin}, {Zhang}, {Su}, {Xu}, {Chen}, {Wang}, \& {Zhou}}]{MWISP_Analysis_Sun_2021}
{Sun}, Y., {Yang}, J., {Yan}, Q.-Z., {et~al.} 2021, \apjs, 256, 32, \dodoi{10.3847/1538-4365/ac11fe}

\bibitem[{{Sun} {et~al.}(2024{\natexlab{b}}){Sun}, {Yang}, {Yan}, {Zhang}, {Su}, {Chen}, {Zhou}, {Ma}, \& {Yuan}}]{MWISP_Analysis_Sun_2024_1}
---. 2024{\natexlab{b}}, \apjs, 275, 35, \dodoi{10.3847/1538-4365/ad8237}

\bibitem[{{Taylor} {et~al.}(2003){Taylor}, {Gibson}, {Peracaula}, {Martin}, {Landecker}, {Brunt}, {Dewdney}, {Dougherty}, {Gray}, {Higgs}, {Kerton}, {Knee}, {Kothes}, {Purton}, {Uyaniker}, {Wallace}, {Willis}, \& {Durand}}]{CGPS}
{Taylor}, A.~R., {Gibson}, S.~J., {Peracaula}, M., {et~al.} 2003, \aj, 125, 3145, \dodoi{10.1086/375301}

\bibitem[{{Umemoto} {et~al.}(2017){Umemoto}, {Minamidani}, {Kuno}, {Fujita}, {Matsuo}, {Nishimura}, {Torii}, {Tosaki}, {Kohno}, {Kuriki}, {Tsuda}, {Hirota}, {Ohashi}, {Yamagishi}, {Handa}, {Nakanishi}, {Omodaka}, {Koide}, {Matsumoto}, {Onishi}, {Tokuda}, {Seta}, {Kobayashi}, {Tachihara}, {Sano}, {Hattori}, {Onodera}, {Oasa}, {Kamegai}, {Tsuboi}, {Sofue}, {Higuchi}, {Chibueze}, {Mizuno}, {Honma}, {Muller}, {Inoue}, {Morokuma-Matsui}, {Shinnaga}, {Ozawa}, {Takahashi}, {Yoshiike}, {Costes}, \& {Kuwahara}}]{FUGIN_1}
{Umemoto}, T., {Minamidani}, T., {Kuno}, N., {et~al.} 2017, \pasj, 69, 78, \dodoi{10.1093/pasj/psx061}

\bibitem[{{Urquhart} {et~al.}(2011{\natexlab{a}}){Urquhart}, {Moore}, {Hoare}, {Lumsden}, {Oudmaijer}, {Rathborne}, {Mottram}, {Davies}, \& {Stead}}]{KDA_Method_Height_1}
{Urquhart}, J.~S., {Moore}, T.~J.~T., {Hoare}, M.~G., {et~al.} 2011{\natexlab{a}}, \mnras, 410, 1237, \dodoi{10.1111/j.1365-2966.2010.17514.x}

\bibitem[{{Urquhart} {et~al.}(2011{\natexlab{b}}){Urquhart}, {Morgan}, {Figura}, {Moore}, {Lumsden}, {Hoare}, {Oudmaijer}, {Mottram}, {Davies}, \& {Dunham}}]{Maser_Star_Formation_2}
{Urquhart}, J.~S., {Morgan}, L.~K., {Figura}, C.~C., {et~al.} 2011{\natexlab{b}}, \mnras, 418, 1689, \dodoi{10.1111/j.1365-2966.2011.19594.x}

\bibitem[{{Urquhart} {et~al.}(2018){Urquhart}, {K{\"o}nig}, {Giannetti}, {Leurini}, {Moore}, {Eden}, {Pillai}, {Thompson}, {Braiding}, {Burton}, {Csengeri}, {Dempsey}, {Figura}, {Froebrich}, {Menten}, {Schuller}, {Smith}, \& {Wyrowski}}]{ATLASGAL_2018}
{Urquhart}, J.~S., {K{\"o}nig}, C., {Giannetti}, A., {et~al.} 2018, \mnras, 473, 1059, \dodoi{10.1093/mnras/stx2258}

\bibitem[{{Urquhart} {et~al.}(2022){Urquhart}, {Wells}, {Pillai}, {Leurini}, {Giannetti}, {Moore}, {Thompson}, {Figura}, {Colombo}, {Yang}, {K{\"o}nig}, {Wyrowski}, {Menten}, {Rigby}, {Eden}, \& {Ragan}}]{ATLASGAL_2022}
{Urquhart}, J.~S., {Wells}, M.~R.~A., {Pillai}, T., {et~al.} 2022, \mnras, 510, 3389, \dodoi{10.1093/mnras/stab3511}

\bibitem[{{Urquhart} {et~al.}(2024){Urquhart}, {K{\"o}nig}, {Colombo}, {Karska}, {Wyrowski}, {Menten}, {Moore}, {Brand}, {Elia}, {Giannetti}, {Leurini}, {Figueira}, {Lee}, \& {Dumke}}]{OGHReS}
{Urquhart}, J.~S., {K{\"o}nig}, C., {Colombo}, D., {et~al.} 2024, \mnras, 528, 4746, \dodoi{10.1093/mnras/stad3983}

\bibitem[{van~der Walt {et~al.}(2014)van~der Walt, {S}ch\"onberger, {Nunez-Iglesias}, {B}oulogne, {W}arner, {Y}ager, {G}ouillart, {Y}u, \& the scikit-image contributors}]{scikit-image}
van~der Walt, S., {S}ch\"onberger, J.~L., {Nunez-Iglesias}, J., {et~al.} 2014, PeerJ, 2, e453, \dodoi{10.7717/peerj.453}

\bibitem[{{V{\'a}zquez-Semadeni} {et~al.}(2007){V{\'a}zquez-Semadeni}, {G{\'o}mez}, {Jappsen}, {Ballesteros-Paredes}, {Gonz{\'a}lez}, \& {Klessen}}]{Cloud_Evolution_1}
{V{\'a}zquez-Semadeni}, E., {G{\'o}mez}, G.~C., {Jappsen}, A.~K., {et~al.} 2007, \apj, 657, 870, \dodoi{10.1086/510771}

\bibitem[{{V{\'a}zquez-Semadeni} {et~al.}(2017){V{\'a}zquez-Semadeni}, {Gonz{\'a}lez-Samaniego}, \& {Col{\'\i}n}}]{Hierarchical_Collapse_2}
{V{\'a}zquez-Semadeni}, E., {Gonz{\'a}lez-Samaniego}, A., \& {Col{\'\i}n}, P. 2017, \mnras, 467, 1313, \dodoi{10.1093/mnras/stw3229}

\bibitem[{{V{\'a}zquez-Semadeni} {et~al.}(2019){V{\'a}zquez-Semadeni}, {Palau}, {Ballesteros-Paredes}, {G{\'o}mez}, \& {Zamora-Avil{\'e}s}}]{Hierarchical_Collapse_1}
{V{\'a}zquez-Semadeni}, E., {Palau}, A., {Ballesteros-Paredes}, J., {G{\'o}mez}, G.~C., \& {Zamora-Avil{\'e}s}, M. 2019, \mnras, 490, 3061, \dodoi{10.1093/mnras/stz2736}

\bibitem[{{VERA Collaboration} {et~al.}(2020){VERA Collaboration}, {Hirota}, {Nagayama}, {Honma}, {Adachi}, {Burns}, {Chibueze}, {Choi}, {Hachisuka}, {Hada}, {Hagiwara}, {Hamada}, {Handa}, {Hashimoto}, {Hirano}, {Hirata}, {Ichikawa}, {Imai}, {Inenaga}, {Ishikawa}, {Jike}, {Kameya}, {Kaseda}, {Kim}, {Kim}, {Kim}, {Kobayashi}, {Kono}, {Kurayama}, {Matsuno}, {Morita}, {Motogi}, {Murase}, {Nakagawa}, {Nakanishi}, {Niinuma}, {Nishi}, {Oh}, {Omodaka}, {Oyadomari}, {Oyama}, {Sakai}, {Sakai}, {Sawada-Satoh}, {Shibata}, {Shizugami}, {Sudo}, {Sugiyama}, {Sunada}, {Suzuki}, {Takahashi}, {Tamura}, {Tazaki}, {Ueno}, {Uno}, {Urago}, {Wada}, {Wu}, {Yamashita}, {Yamashita}, {Yamauchi}, \& {Yuda}}]{Maser_2}
{VERA Collaboration}, {Hirota}, T., {Nagayama}, T., {et~al.} 2020, \pasj, 72, 50, \dodoi{10.1093/pasj/psaa018}

\bibitem[{{Virtanen} {et~al.}(2020){Virtanen}, {Gommers}, {Oliphant}, {Haberland}, {Reddy}, {Cournapeau}, {Burovski}, {Peterson}, {Weckesser}, {Bright}, {van der Walt}, {Brett}, {Wilson}, {Millman}, {Mayorov}, {Nelson}, {Jones}, {Kern}, {Larson}, {Carey}, {Polat}, {Feng}, {Moore}, {VanderPlas}, {Laxalde}, {Perktold}, {Cimrman}, {Henriksen}, {Quintero}, {Harris}, {Archibald}, {Ribeiro}, {Pedregosa}, {van Mulbregt}, \& {SciPy 1. 0 Contributors}}]{SciPy}
{Virtanen}, P., {Gommers}, R., {Oliphant}, T.~E., {et~al.} 2020, Nature Methods, 17, 261, \dodoi{10.1038/s41592-019-0686-2}

\bibitem[{{Wang} {et~al.}(2016){Wang}, {Testi}, {Burkert}, {Walmsley}, {Beuther}, \& {Henning}}]{MST}
{Wang}, K., {Testi}, L., {Burkert}, A., {et~al.} 2016, \apjs, 226, 9, \dodoi{10.3847/0067-0049/226/1/9}

\bibitem[{{Wenger} {et~al.}(2018){Wenger}, {Balser}, {Anderson}, \& {Bania}}]{Distance_2}
{Wenger}, T.~V., {Balser}, D.~S., {Anderson}, L.~D., \& {Bania}, T.~M. 2018, \apj, 856, 52, \dodoi{10.3847/1538-4357/aaaec8}

\bibitem[{{Xu} {et~al.}(2021){Xu}, {Bian}, {Reid}, {Li}, {Menten}, {Dame}, {Zhang}, {Brunthaler}, {Wu}, {Moscadelli}, {Wu}, \& {Zheng}}]{Maser_3}
{Xu}, Y., {Bian}, S.~B., {Reid}, M.~J., {et~al.} 2021, \apjs, 253, 1, \dodoi{10.3847/1538-4365/abd8cf}

\bibitem[{{Yan} {et~al.}(2020){Yan}, {Yang}, {Su}, {Sun}, \& {Wang}}]{Emission_D5}
{Yan}, Q.-Z., {Yang}, J., {Su}, Y., {Sun}, Y., \& {Wang}, C. 2020, \apj, 898, 80, \dodoi{10.3847/1538-4357/ab9f9c}

\bibitem[{{Yan} {et~al.}(2021){Yan}, {Yang}, {Su}, {Sun}, {Xu}, {Wang}, {Zhou}, \& {Wang}}]{Emission_D1}
{Yan}, Q.-Z., {Yang}, J., {Su}, Y., {et~al.} 2021, \apj, 922, 8, \dodoi{10.3847/1538-4357/ac214f}

\bibitem[{{Yan} {et~al.}(2019){Yan}, {Yang}, {Sun}, {Su}, \& {Xu}}]{Emission_D6}
{Yan}, Q.-Z., {Yang}, J., {Sun}, Y., {Su}, Y., \& {Xu}, Y. 2019, \apj, 885, 19, \dodoi{10.3847/1538-4357/ab458e}

\bibitem[{{Yang} {et~al.}(2025){Yang}, {Chen}, {Xu}, {Yang}, {Chen}, {Li}, {Ju}, \& {Lu}}]{MWISP_Analysis_Yang_2025}
{Yang}, T., {Chen}, X., {Xu}, X.-Y., {et~al.} 2025, arXiv e-prints, arXiv:2503.16145, \dodoi{10.48550/arXiv.2503.16145}

\bibitem[{{Yuan} {et~al.}(2023){Yuan}, {Yang}, {Du}, {Su}, {Zhang}, {Yan}, {Sun}, {Zhou}, {Chen}, {Wang}, \& {Chen}}]{MWISP_Analysis_Yuan_2023_2}
{Yuan}, L., {Yang}, J., {Du}, F., {et~al.} 2023, \apj, 958, 7, \dodoi{10.3847/1538-4357/acf9ef}

\bibitem[{{Zhang} {et~al.}(2024){Zhang}, {Su}, {Chen}, {Fang}, {Yan}, {Zhang}, {Sun}, {Wang}, {Feng}, {Ma}, {Zhang}, {Zhuang}, {Zhou}, {Chen}, \& {Yang}}]{MWISP_Analysis_Zhang_2024}
{Zhang}, S., {Su}, Y., {Chen}, X., {et~al.} 2024, \aj, 167, 220, \dodoi{10.3847/1538-3881/ad2fcb}

\bibitem[{{Zhuang} {et~al.}(2024){Zhuang}, {Su}, {Zhang}, {Chen}, {Yan}, {Feng}, {Sun}, {Xu}, {Sun}, {Zhou}, {Wang}, \& {Yang}}]{MWISP_Analysis_Zhuang_2024}
{Zhuang}, Z., {Su}, Y., {Zhang}, S., {et~al.} 2024, \apj, 966, 202, \dodoi{10.3847/1538-4357/ad3552}

\end{thebibliography}
\bibliographystyle{aasjournal}

\appendix
\begin{appendices}
\renewcommand{\thetable}{\thesection.\arabic{table}}
\renewcommand{\thefigure}{\thesection.\arabic{figure}}

\section{Configuration Parameters}\label{Parameters_FacetClumps}

The parameters of FacetClumps used in this work are presented in Table \ref{InputPar}. We adopt a conservative threshold of 5$\times$RMS to ensure high reliability of our catalogs. The underestimation of parameters resulting from this threshold is recovered using the extrapolation method described in Appendix \ref{ExtMethod}. 

\setcounter{table}{0} 
\begin{table}
\centering
\caption{Parameters of FacetClumps along with their explanations and values.}
\begin{tabular}{p{4cm}p{10cm}p{2cm}}\hline\hline
    Parameters&Explanation&Value\\\hline
    FacetClumps.RMS&The noise rms or S/N of the data &1 (S/N data)\\\hline
    FacetClumps.Threshold&The minimum intensity or S/N used to truncate the signals&5$\times$RMS\\\hline
    FacetClumps.SWindow&The scale of the window function; in pixels&3\\\hline 
    FacetClumps.KBins&The coefficient used to calculate the number of eigenvalue bins&35\\\hline
    FacetClumps.FwhmBeam&The FWHM of the instrument beam; in pixels&2\\\hline
    FacetClumps.VeloRes&The velocity resolution of the instrument; in channels&2\\\hline 
    FacetClumps.SRecursionLBV&The minimum area of a region in the spatial direction (SRecursionLB) and the minimum length of a region in the velocity channels (SRecursionV) required for recursion termination. Identified clump regions must also satisfy these minimum size conditions. In pixels and channels, respectively.&[16,5]\\\hline
\end{tabular}
\begin{tablenotes}
    \item \textbf{Note.} A more detailed introduction to the parameters and related experiments can be found in \cite{FacetClumps}.
\end{tablenotes}
\label{InputPar}
\end{table}

\section{Catalogs of MWISP Clumps}\label{Tables}
The comprehensive catalogs of MWISP clumps are organized across three tables: detection parameters in Table \ref{Catalogue_Table_Detection}, distance measurements in Table \ref{Catalogue_Table_Distance}, and derived physical properties in Table \ref{Catalogue_Table_Physical}. Examples presented include representative clumps with various distance determinations \textit{KDA\_M} as shown in Table \ref{Catalogue_Table_Distance}, along with the selected clumps illustrated in Figure \ref{Imgs_ExClumps}. The complete catalogs, containing all identified clumps and their parameters, are available through the online repository at DOI: \href{10.57760/sciencedb.28751}{10.57760/sciencedb.28751}. 

\setcounter{table}{0} 
\begin{table*}[htbp]
\centering
\rotatebox{90}{
\begin{minipage}{\textheight}
\centering
\caption{\centering The Detection Catalog of Clumps.}
\begin{tabular}{lccccccccccccccc}
\toprule
MWISP CID& Peak1 & Peak2 & Peak3 & Cen1 & Cen2 & Cen3 & Size1 & Size2 & Size3 & Peak & Sum & Volume & Angle & Edge & SRID \\
 & (deg) & (deg) & (km s$^{-1}$) & (deg) & (deg) & (km s$^{-1}$) & (arcsec) & (arcsec) & (km s$^{-1}$) & (K) & (K km s$^{-1}$) & (pix) & (deg) & &\\
(1)&(2)&(3)&(4)&(5)&(6)&(7)&(8)&(9)&(10)&(11)&(12)&(13)&(14)&(15)&(16)\\\hline	
\midrule
MWISP00001 & 102.325 & 3.683 & -88.998 & 102.332 & 3.681 & -88.761 & 41.19 & 30.54 & 0.368 & 2.945 & 23.762 & 96 & 34.91 & 0 & 1 \\
MWISP03817 & 70.142 & 1.733 & -22.748 & 70.133 & 1.727 & -24.034 & 18.21 & 40.23 & 0.922 & 4.07 & 190.709 & 741 & 5.69 & 0 & 1652 \\
MWISP39790 & 16.967 & -0.617 & 18.929 & 16.993 & -0.615 & 17.905 & 162.24 & 97.29 & 0.702 & 4.152 & 257.379 & 1276 & 45.66 & 0 & 9206 \\
MWISP00011 & 84.417 & 1.675 & -85.013 & 84.42 & 1.679 & -85.138 & 75.15 & 29.1 & 0.281 & 2.233 & 19.006 & 96 & -13.38 & 0 & 9 \\
MWISP12372 & 80.775 & 0.092 & 5.479 & 80.78 & 0.095 & 5.631 & 112.71 & 32.49 & 0.242 & 2.007 & 17.836 & 91 & -29.41 & 0 & 7242 \\
MWISP11816 & 15.008 & -0.858 & 36.529 & 15.004 & -0.858 & 37.002 & 126.24 & 30.78 & 0.466 & 2.833 & 48.785 & 207 & -9.3 & 0 & 12416 \\
MWISP22478 & 43.292 & -0.15 & 9.298 & 43.29 & -0.158 & 9.701 & 189.54 & 31.86 & 0.302 & 2.437 & 18.43 & 78 & -70.69 & 0 & 8551 \\
MWISP28480 & 23.075 & 0.383 & 2.989 & 23.093 & 0.417 & 3.035 & 27.81 & 103.59 & 0.353 & 3.656 & 319.667 & 1035 & -25.41 & 0 & 6171 \\
MWISP26929 & 53.092 & 0.117 & 2.159 & 53.089 & 0.105 & 2.553 & 153.69 & 62.76 & 0.727 & 4.431 & 93.55 & 398 & -52.72 & 0 & 5776 \\
MWISP11006 & 15.017 & -0.667 & 19.261 & 15.003 & -0.657 & 19.964 & 85.44 & 117.12 & 1.886 & 30.663 & 10391.984 & 10198 & -74.88 & 0 & 2970 \\
MWISP57379 & 49.492 & -0.375 & 61.435 & 49.496 & -0.367 & 61.471 & 111.75 & 150.27 & 1.5 & 20.178 & 6228.878 & 9376 & -86.12 & 0 & 13207 \\
MWISP23367 & 202.025 & 1.383 & 0.166 & 202.016 & 1.398 & 0.092 & 41.67 & 31.68 & 0.19 & 1.608 & 8.03 & 53 & 56.38 & 0 & 5412 \\
MWISP57381 & 49.475 & -0.4 & 55.624 & 49.468 & -0.405 & 55.879 & 109.53 & 136.08 & 1.171 & 18.756 & 4753.458 & 7076 & -54.0 & 0 & 13207 \\
MWISP57616 & 13.717 & -0.067 & 43.171 & 13.715 & -0.067 & 43.474 & 166.41 & 30.78 & 0.436 & 2.088 & 31.98 & 157 & -1.19 & 0 & 13248 \\
\bottomrule
\end{tabular}
\begin{tablenotes}
    \item \textbf{Note.} Column (1): MWISP identification number. Columns (2)--(4): position of the peak intensity in Galactic coordinates and velocity (Peak1, Peak2, and Peak3). Columns (5)--(7): centroid position in Galactic coordinates and velocity (Cen1, Cen2, and Cen3). Columns (8)--(10): dispersion along the three axes (Size1, Size2, and Size3). Column (11): peak intensity in K. Column (12): Integrated intensity in K km s$^{-1}$. Column (13): volume of the clump in pixels. Column (14): position angle of the major axis. Column (15): edge flag (0: not at the edge of the survey; 1: at the edge). Column (16): ID of the SR to which the clump belongs.
\end{tablenotes}
\label{Catalogue_Table_Detection}
\end{minipage}
}
\end{table*}

\begin{table*}
\centering
\caption{\centering The Distance Catalog of Clumps.}
\begin{tabular}{lcccccccccc}
\toprule
MWISP CID& GLon & GLat & $V_{LSR}$ & Dist & Dist\_EL & Dist\_EU & KDA\_S & KDA\_M & KDA\_Ref&KDA\_Flag \\
 & (deg) & (deg) & (km s$^{-1}$) & (kpc) & (kpc) & (kpc) & & & &\\
(1)&(2)&(3)&(4)&(5)&(6)&(7)&(8)&(9)&(10)&(11)\\\hline	
\midrule
MWISP00001 & 102.33 & 3.683 & -88.748 & 8.64 & 0.0 & 0.49 & 0 & 0 & -1 & -1 \\
MWISP03817 & 70.128 & 1.726 & -23.956 & 6.41 & -0.6 & 0.73 & 2 & 1 & 1 & 1 \\
MWISP39790 & 16.982 & -0.603 & 17.976 & 1.74 & -0.34 & 0.27 & 0 & 1 & 1 & -1 \\
MWISP00011 & 84.419 & 1.679 & -85.138 & 10.52 & 0.0 & 0.43 & 2 & 2 & -1 & -1 \\
MWISP12372 & 80.779 & 0.095 & 5.628 & 0.13 & -0.12 & 0.84 & 0 & 3 & 3 & -1 \\
MWISP11816 & 15.004 & -0.858 & 36.998 & 3.33 & -0.26 & 0.25 & 0 & 4 & -1 & -1 \\
MWISP22478 & 43.291 & -0.158 & 9.703 & 11.42 & -0.33 & 0.31 & 2 & 5 & 5 & -1 \\
MWISP28480 & 23.083 & 0.414 & 2.975 & 0.21 & -0.18 & 0.33 & 0 & 6 & 1, 2 & -1 \\
MWISP26929 & 53.089 & 0.108 & 2.583 & 9.77 & -0.31 & 0.43 & 2 & 7 & 5 & -1 \\
MWISP11006 & 15.007 & -0.661 & 19.806 & 2.04 & -0.16 & 0.19 & 0 & 1 & 1, 2, 5 & 1 \\
MWISP57379 & 49.493 & -0.368 & 61.361 & 5.13 & -1.37 & 2.94 & 0 & 1 & 1, 2 & 2 \\
MWISP23367 & 202.016 & 1.399 & 0.084 & 0.11 & 0.0 & 0.32 & 0 & 0 & -1 & -1 \\
MWISP57381 & 49.471 & -0.408 & 55.913 & 5.13 & -1.37 & 2.94 & 0 & 1 & 1, 2 & 1 \\
MWISP57616 & 13.716 & -0.067 & 43.51 & 3.79 & -0.19 & 0.21 & 0 & 1 & 1 & 1 \\
\bottomrule
\end{tabular}
\begin{tablenotes}
    \item \textbf{Note.} Column (1): MWISP identification number. Column (2): Galactic longitude in degrees. Column (3): Galactic latitude in degrees. Column (4): LSR velocity in km s$^{-1}$. Column (5): heliocentric distance in kpc. Columns (6)--(7): lower and upper error bounds on the distance estimate in kpc. Column (8): KDA solution. Column (9): method used for distance determination. Column (10): references for distance determination. Column (11): KDA flag.
    
    a. Dist\_EL and Dist\_EU: -1 indicates no uncertainty estimate available.
    
    b. KDA\_S: 0--near distance, 1--unknown, 2--far distance.
    
    c. KDA\_M: 0--clumps in SGQ or TGQ, 1-7--methods numbered as Figure \ref{KDA_Flow}.
    
    d. When KDA\_M = 0 or 1, KDA\_Ref is the references of the maser: (1)\citep{Distance_1}, (2) \citep{Maser_2}, (3) \citep{Maser_3}, (4) \citep{Maser_4}, (5) \citep{Maser_5}, (6) \citep{Maser_6}, and (7) \citep{Maser_7}. When KDA\_M = 3, 5, or 7, KDA\_Ref is the references of the HI survey: (1) \citep{HI4PI}, (2) \citep{SGPS}, (3) \citep{CGPS}, (4) \citep{VGPS}, and (5) \citep{GalFa}. When KDA\_M = 6, KDA\_Ref is the references of the IRDC: (1) \citep{IRDC_1}, and (2) \citep{IRDC_2}.
    
    e. When KDA\_M = 0 or 1, KDA\_Flag is the flag of maser-associated clumps: (1) maser-hosting, (2) maser-closing, and (3) neighbor of maser-closing clump.
\end{tablenotes}
\label{Catalogue_Table_Distance}
\end{table*}

\begin{table*}[htbp]
\centering
\rotatebox{90}{
\begin{minipage}{\textheight}
\centering
\caption{The Physical Parameter Catalog of Clumps.}
\label{Catalogue_Table_Physical}
\small
\begin{tabular}{lccccccccccccccc}
\toprule
MWISP CID& GLon & GLat & $V_{LSR}$ & Dist & $V_{fwhm}$ & $T_{ex}$ & N($^{13}$CO) & $R_{eq}$ & Mass & $\alpha_{vir}$ &$\Sigma$& $n$(H$_2$) & $\Delta$ Flux & $\Delta \sigma_v$ & $\Delta R_{eq}$ \\
 & (deg) & (deg) & (km s$^{-1}$) & (kpc) & (km s$^{-1}$) & (K) & (cm$^{-2}$) & (pc) & ($M_{\odot}$) & &($M_{\odot}$ pc$^{-2}$)& (cm$^{-3}$) & & & \\
(1)&(2)&(3)&(4)&(5)&(6)&(7)&(8)&(9)&(10)&(11)&(12)&(13)&(14)&(15)&(16)\\
\midrule
MWISP00001 & 102.33 & 3.683 & -88.748 & 8.64 & 0.87 & 10.94 & 6.828e+14 & 3.373 & 1120 & 0.479 & 31.352 & 119.003 & -0.026 & -0.126 & -0.047 \\
MWISP03817 & 70.128 & 1.726 & -23.956 & 6.41 & 2.18 & 11.91 & 2.062e+15 & 4.514 & 3821 & 1.181 & 59.707 & 169.378 & -0.165 & -0.024 & -0.03 \\
MWISP39790 & 16.982 & -0.603 & 17.976 & 1.74 & 1.66 & 11.27 & 1.601e+15 & 1.758 & 281.381 & 3.63 & 28.973 & 210.999 & -0.221 & -0.167 & -0.076 \\
MWISP00011 & 84.419 & 1.679 & -85.138 & 10.52 & 0.66 & 8.09 & 8.056e+14 & 4.016 & 1235 & 0.301 & 24.39 & 77.772 & -0.126 & -0.044 & -0.077 \\
MWISP12372 & 80.779 & 0.095 & 5.628 & 0.13 & 0.57 & 8.27 & 5.983e+14 & 0.052 & 0.124 & 28.515 & 14.671 & 3621.948 & -0.082 & -0.039 & -0.116 \\
MWISP11816 & 15.004 & -0.858 & 36.998 & 3.33 & 1.1 & 8.94 & 1.644e+15 & 1.41 & 167.205 & 2.15 & 26.757 & 242.92 & -0.113 & -0.07 & -0.079 \\
MWISP22478 & 43.291 & -0.158 & 9.703 & 11.42 & 0.71 & 8.66 & 9.220e+14 & 3.823 & 985.108 & 0.409 & 21.458 & 71.879 & -0.374 & -0.005 & -0.38 \\
MWISP28480 & 23.083 & 0.414 & 2.975 & 0.21 & 0.84 & 9.07 & 2.360e+15 & 0.231 & 6.554 & 5.188 & 39.111 & 2168.326 & -0.083 & -0.075 & -0.016 \\
MWISP26929 & 53.089 & 0.108 & 2.583 & 9.77 & 1.72 & 10.04 & 1.391e+15 & 5.648 & 4018 & 0.872 & 40.091 & 90.884 & -0.089 & -0.017 & -0.051 \\
MWISP11006 & 15.007 & -0.661 & 19.806 & 2.04 & 4.47 & 30.32 & 1.884e+16 & 3.376 & 41,779 & 0.338 & 1167.056 & 4426.95 & -0.067 & -0.056 & -0.015 \\
MWISP57379 & 49.493 & -0.368 & 61.361 & 5.13 & 3.55 & 18.61 & 8.073e+15 & 9.596 & 102,011 & 0.249 & 352.627 & 470.539 & -0.109 & -0.065 & -0.074 \\
MWISP23367 & 202.016 & 1.399 & 0.084 & 0.11 & 0.44 & 6.79 & 3.427e+14 & 0.038 & 0.042 & 36.511 & 9.303 & 3142.02 & -0.118 & -0.009 & -0.064 \\
MWISP57381 & 49.471 & -0.408 & 55.913 & 5.13 & 2.77 & 18.91 & 6.122e+15 & 9.381 & 79,543 & 0.19 & 287.696 & 392.684 & -0.081 & -0.057 & -0.048 \\
MWISP57616 & 13.716 & -0.067 & 43.51 & 3.79 & 1.04 & 8.55 & 9.647e+14 & 1.605 & 130.575 & 2.765 & 16.131 & 128.674 & -0.079 & -0.062 & -0.009 \\
\bottomrule
\end{tabular}
\begin{tablenotes}
    \item \textbf{Note.} Column (1): MWISP identification number. Column (2): Galactic longitude in degrees. Column (3): Galactic latitude in degrees. Column (4): LSR velocity in km s$^{-1}$. Column (5): distance from the Sun in kpc. Column (6): FWHM velocity dispersion in km s$^{-1}$. Column (7): excitation temperature in kelvins. Column (8): $^{13}$CO column density in cm$^{-2}$. Column (9): equivalent radius in pc. Column (10): mass in solar masses; Column (11): virial parameter.  Column (12): H$_2$ surface density in M$_{\odot}$ pc$^{-2}$. Column (13): H$_2$ volume density in cm$^{-3}$. Column (14): flux fluctuation; Column (15): $\sigma_v$ fluctuation. Column (16): $R_{eq}$ fluctuation.
\end{tablenotes}
\end{minipage}%
}
\end{table*}

\section{Derivation Equation of Physical Properties}\label{Cal_Equations}

\subsection{Distance-independent Parameters}\label{Cal_Equations_1}
The physical properties of molecular gas based on CO observations have been extensively studied in the literatures \citep[e.g.,][]{CHIMPS_2,MWISP_Analysis_Li_2023,MWISP_Analysis_Dong_2023,MWISP_Analysis_Zhuang_2024,MWISP_Analysis_Zhang_2024,MWISP_Analysis_Feng_2024}. Following these studies, we assume that $^{12}$CO emission is optically thick \citep{MWISP_Analysis_Yuan_2023_2}. The excitation temperature ($T_{\mathrm{ex}}$) of $^{12}$CO can be derived as \citep{Tex}:

\begin{equation} 
  T_{\mathrm{ex}} = \frac{h\nu_{12}/k}{\mathrm{ln}\left [1+\frac{h\nu_{12}/k}{T_{\mathrm{MB,12_{peak}}+\frac{h\nu_{12}/k}{\mathrm{exp}(h\nu_{12}/kT_{\mathrm{bg}})-1}}}\right ]} = \frac{5.53}{\mathrm{ln}\left(1+\frac{5.53}{T_{\mathrm{MB,12_{peak}}}+0.819}\right)},
\end{equation}

\noindent where $\nu_{12}=\mathrm{115.27~GHz}$ is the frequency of $\rm ^{12}CO~(J=1\rightarrow 0)$ emission, $T_\mathrm{{MB,12_{peak}}}$ is the peak main-beam temperature of $^{12}$CO measured within the velocity range corresponding to each spatial position of a $^{13}$CO clump, and $T_{\mathrm{bg}}\approx2.7\rm ~K$ is the cosmic microwave background temperature.

The H$_2$ column density of molecular clumps can be derived using the LTE method, which assumes equal excitation temperatures for $^{12}$CO, $^{13}$CO, and C$^{18}$O. 

The optical depths of $^{13}$CO in each pixel of a clump are given by

\begin{align}
	\tau\!\left(\rm^{13}CO\right)=-\ln\!&\left[1-\frac{T{\rm_{MB,pk}\!\left(^{13}CO\right)}}{5.29}\right.\nonumber\left.\times\left(\frac{1}{e^{5.29/T_{\rm ex}}-1}-0.164\right)^{-1}\right]
\end{align}

The $^{13}$CO column densities in each spatial pixel of a clump can thus be expressed as \citep{Tex, Col_Density}

\begin{align}
N_{\rm^{13}CO}=2.&42\times{10}^{14}\times\frac{\tau\!\left(\rm^{13}CO\right)}{1-e^{-\tau\!\left(\rm^{13}CO\right)}}\nonumber\times\frac{1+0.88/T_{\rm ex}}{1-e^{-5.29/T_{\rm ex}}}	\int\! T_{\rm MB}\!\left(\rm^{13}CO\right)\,dv
\end{align}

The velocity dispersion ($\sigma_v$) of the clump is defined as the intensity-weighted standard deviation of velocities \citep{FacetClumps}, calculated as

\begin{equation}
\sigma_v = \sqrt{\frac{\sum I_i \cdot v_i^2}{\sum I_i} - \left(\frac{\sum I_i \cdot v_i}{\sum I_i}\right)^2}
\end{equation}

\noindent where $I_i$ is the intensity of voxel $i$ in the clump mask, and $v_i$ is the velocity coordinate of voxel $i$. The line width is estimated as

\begin{equation}
\Delta v=\sqrt{8\ln 2}\times \sigma_v
\end{equation}

\subsection{Distance-dependent Parameters}\label{Cal_Equations_2}
The Galactocentric radius ($R_{\rm GC}$) can be calculated from the heliocentric distance ($d$) and Galactic coordinates ($l, b$) using:

\begin{equation}
R_{\rm GC} = \sqrt{R_0^2 + d^2 - 2R_0d\cos l\cos b}
\end{equation}

\noindent where $R_0 \approx 8.15$ kpc is the distance from the Sun to the Galactic center \citep{Distance_1}. The relationship between $R_{\rm GC}$ and $d$ varies with Galactic position rather than following a simple proportionality.

To convert $^{13}$CO column density to H$_2$ column densities, we adopt the following abundance ratios: $\left[\rm^{12}C/^{13}C\right]=6.21\it R\rm_{GC}+18.71$ \citep[$R_{\rm GC}$ in kiloparsecs;][]{Trans_Factor_With_RGC}, and $\left[\rm H_2/^{12}C\right]=1.1\times10^4$ \citep{Ratio_CO12_H2}.

The H$_2$ column density \citep{Col_Density} is calculated as

\begin{equation}
\begin{aligned}
N_{\rm H_2} &= N_{\rm^{13}CO}\times \left[\rm^{12}C/^{13}C\right] \times \left[\rm H_2/^{12}C\right] \\
&= N_{\rm^{13}CO} \times (6.21\it R\rm_{GC}+18.71)\times 1.1\times10^4
\end{aligned}
\end{equation}

The mass of a clump is estimated by

\begin{equation}
M=2\mu m_{\scriptscriptstyle\rm H} a^2d^2\sum\nolimits_{i} N_{\rm H_2} \propto d^2
\end{equation}

\noindent where $\mu=1.36$ is the mean atomic weight accounting for the contribution of helium and metals, $m\scriptscriptstyle\rm_H$ is the mass of a hydrogen atom, $a=30\arcsec$ is the angular size of a pixel, and $d$ is the heliocentric distance. 

Assuming the clump is spherical, the equivalent radius is expressed as

\begin{equation}
R_{\rm eq} = d\,\sqrt{\frac{A}{\pi}-\frac{\theta^2_{\rm MB}}{4}} \propto d
\end{equation}

\noindent where $\theta_{\rm MB}\sim 50\arcsec$ is the beam size, $A = a^{\rm2}\times N_{\rm pixel}$ is the angular area (projected area) of the clump. 

The volume density, in cm$^{-3}$, is calculated as
\begin{equation}
n_{\rm H_2} = \frac{3M}{4\pi R_{\text{eq}}^3} \propto d^{-1}
\end{equation}

The mass surface density, in $M_\odot~\rm pc^{-2}$, is calculated as

\begin{equation}
\Sigma=\frac{M}{A\times d^2} = \frac{2\mu m_{\scriptscriptstyle\rm H} a^2d^2\sum\nolimits_{i} N_{\rm H_2}}{A\times d^2} = 2\mu m_{\scriptscriptstyle\rm H} \frac{\sum\nolimits_{i} N_{\rm H_2}}{N_{\rm pixel}}
\end{equation}

Notably, $N_{\rm H_2}$ depends on $R_{\rm GC}$ through the isotope ratio $\left[\rm^{12}C/^{13}C\right]$, and since $R_{\rm GC}$ depends on the heliocentric distance $d$, the surface density $\Sigma$ is indirectly distance dependent. As $R_{\rm GC}$ increases (which typically occurs with increasing $d$ for objects in similar Galactic directions), the $\left[\rm^{12}C/^{13}C\right]$ ratio increases, resulting in higher derived $N_{\rm H_2}$ values and consequently higher $\Sigma$.

The virial mass \citep{Virial_Analysis_4} can be calculated using:

\begin{equation}
M_{vir} = \frac{5\sigma_v^2 R_{eq}}{G}\approx210 \times R_{eq} \times (\Delta v)^2 \propto d
\end{equation}

\noindent where $R_{\rm eq}$ is in parsecs, $\Delta v$ is in kilometers per second, and the derived $M_{\rm vir}$ is in $M_\odot$. The virial parameter is defined as

\begin{equation}
\alpha_{vir} = \frac{M_{vir}}{M} = \frac{5\sigma_v^2 R_{eq}}{GM}\approx\frac{210 \times R_{eq} \times (\Delta v)^2}{M} \propto d^{-1}
\end{equation}

When $\alpha_{\rm vir} \approx 2$, the clump is in approximate virial equilibrium; when $\alpha_{\rm vir} > 2$, the clump is not gravitationally bound; and when $\alpha_{\rm vir} < 2$, the clump is gravitationally bound. Values of $\alpha_{\rm vir} \ll 1$ indicate strongly self-gravitating systems that may be undergoing collapse.

\section{Quantification and Correction of Parameter Biases Induced by Sensitivity Effects}\label{DPars}
\setcounter{figure}{0}

\begin{figure*}
\centering
\begin{minipage}[t]{0.4\textwidth}
    \centering
    \centerline{\includegraphics[width=7in]{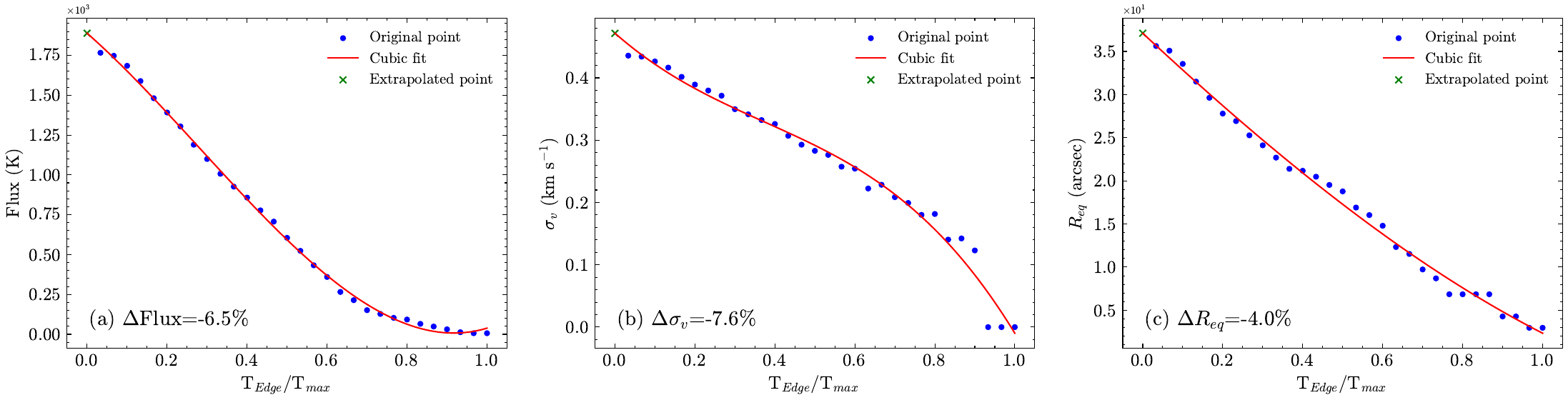}}
\end{minipage}
\caption{Parameter loss example for the clump located in the upper right corner of Figure~\ref{DSRegion_Flow}(d). The panels show how physical properties vary with normalized temperature gradient ($T_{\text{Edge}}/T_{\text{max}}$): (a) flux decreases by 6.5\%, (b) $\sigma_v$ decreases by 7.6\%, and (c) $R_{\text{eq}}$ decreases by 4.0\%. Blue dots represent observational measurements sampled at $T_{\text{Edge}}$ intervals of $\sim$ 0.22~K, corresponding to the typical instrumental noise level. Red curves show cubic polynomial fits to the data, while green crosses indicate extrapolated values at the theoretical zero-intensity boundary.}
\label{Delta_Flux_Imgs}
\end{figure*}

\begin{figure*}
\centering
\begin{minipage}[t]{0.32\textwidth}
    \centering
    \centerline{\includegraphics[width=2.2in]{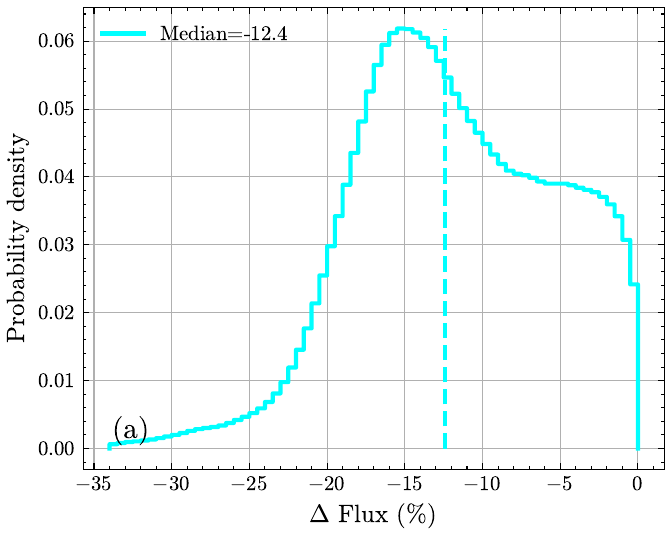}}
\end{minipage}
\begin{minipage}[t]{0.32\textwidth}
    \centering
    \centerline{\includegraphics[width=2.2in]{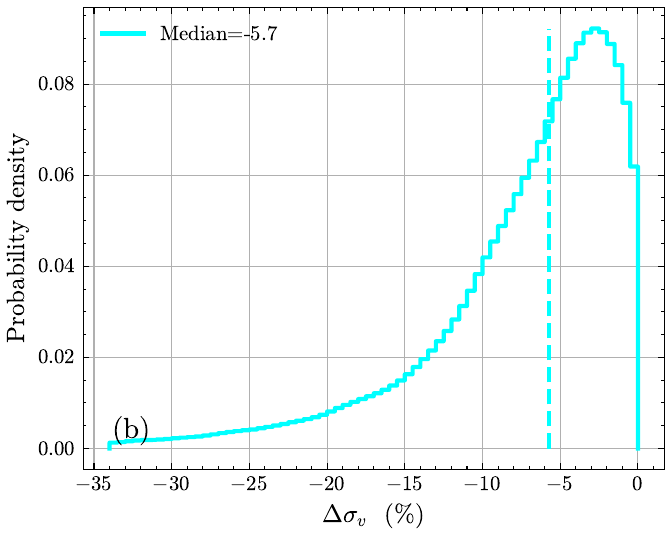}}
\end{minipage}
\begin{minipage}[t]{0.32\textwidth}
    \centering
    \centerline{\includegraphics[width=2.2in]{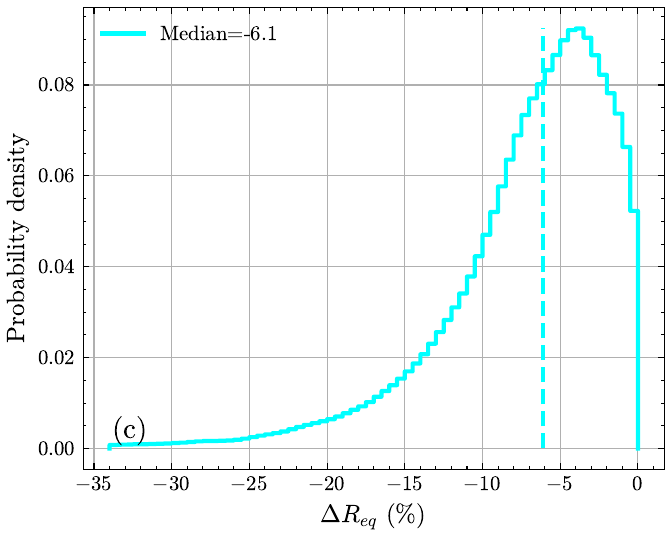}}
\end{minipage}
\caption{Probability density distributions of parameter loss rates for all MWISP clumps. The panels display: (a) the flux loss with a median value of $-12.4\%$, (b) the $\sigma_v$ loss with a median value of $-5.7\%$, and (c) the $R_{\rm eq}$ loss with a median value of $-6.1\%$. The dashed vertical lines indicate the median values. }
\label{Delta_Flux_Hists}
\end{figure*}

\subsection{Parameter Recovery through Extrapolation Method}\label{ExtMethod}
All clump detection algorithms inherently suffer from parameter underestimation biases. While the MWISP survey employs the FacetClumps algorithm, which significantly reduces flux loss \citep{FacetClumps}, systematic underestimation still occurs. Figure~\ref{Delta_Flux_Imgs} demonstrates how flux, velocity dispersion ($\sigma_v$), and equivalent radius ($R_{\text{eq}}$) vary as functions of normalized temperature gradient ($T_{\text{Edge}}/T_{\text{max}}$). These relationships usually exhibit nonlinear behavior that can be modeled using cubic or lower-order polynomial fits. 

We quantify this bias using a loss rate defined as:
\begin{equation}\label{DFlux_E}
\Delta Par = \frac{Par_{\text{obs}} - Par_{\text{ext}}}{Par_{\text{ext}}}
\end{equation}

\noindent where $Par_{\text{obs}}$ represents the observed parameter value at the detection threshold, and $Par_{\text{ext}}$ represents the extrapolated value at the zero-intensity boundary. 

Through the extrapolation method \citep{Extrapolation}, we can recover the true parameters of each clump. For the example shown in Figure~\ref{Delta_Flux_Imgs}, we measure losses of $-6.5\%$ for flux, $-7.6\%$ for $\sigma_v$, and $-4.0\%$ for $R_{\text{eq}}$. Figure~\ref{Delta_Flux_Hists} displays the probability density distributions of parameter loss rates for the entire MWISP sample as described in Section \ref{PProperties}. The asymmetric shapes of the distributions, particularly for flux, suggest that the underestimation bias varies among different clumps, likely depending on their density profiles and S/N. The median values indicate that flux measurements (-12.4\%) are most severely affected by the detection sensitivity limitations, followed by $R_{\text{eq}}$ (-6.1\%), while $\sigma_v$ (-5.7\%) shows the least bias. This is aligned with expectations, as flux measurements integrate over the entire clump volume and are thus more responsive to emission from diffuse outer regions that falls below detection thresholds.

\begin{figure*}
\centering
\vspace{0cm}
\begin{minipage}[t]{0.3\textwidth}
    \centering
    \centerline{\includegraphics[width=2.2in]{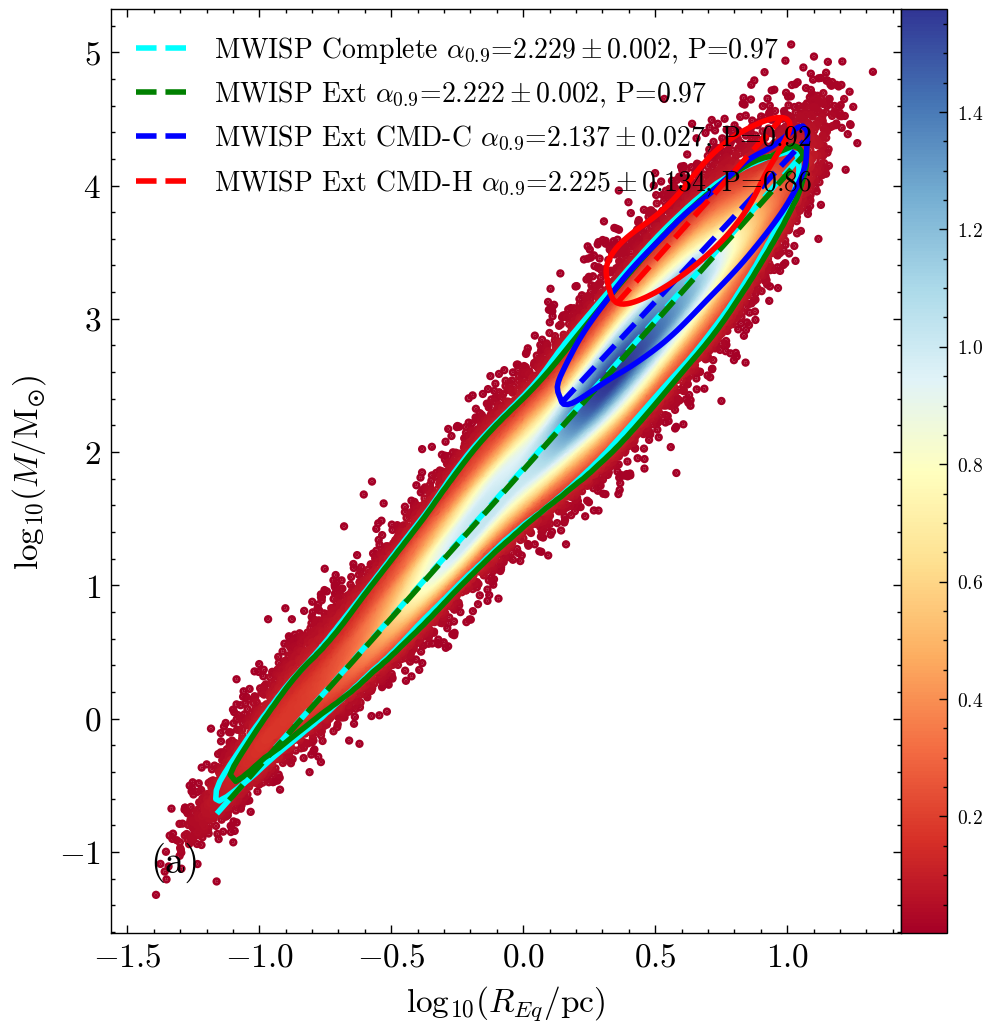}}
\end{minipage}
\begin{minipage}[t]{0.3\textwidth}
    \centering
    \centerline{\includegraphics[width=2.2in]{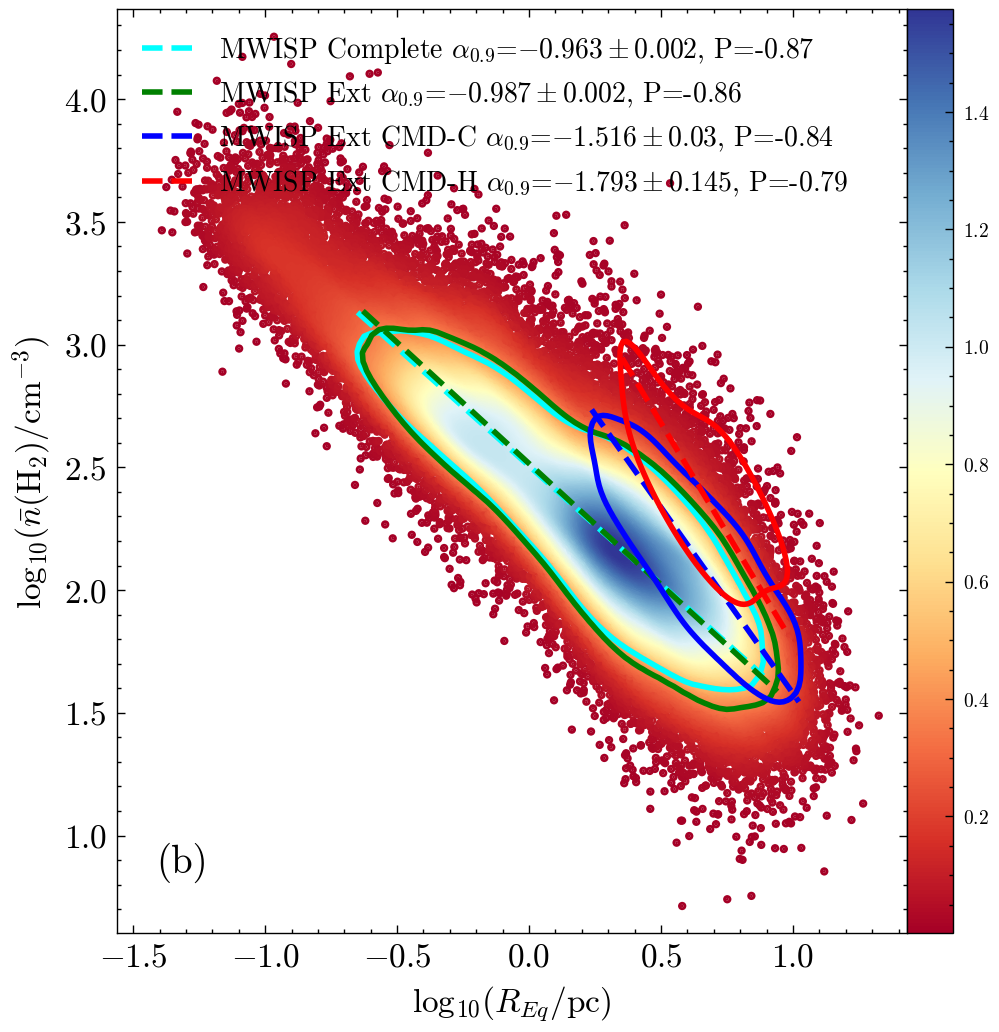}}
\end{minipage}
\begin{minipage}[t]{0.3\textwidth}
    \centering
    \centerline{\includegraphics[width=2.2in]{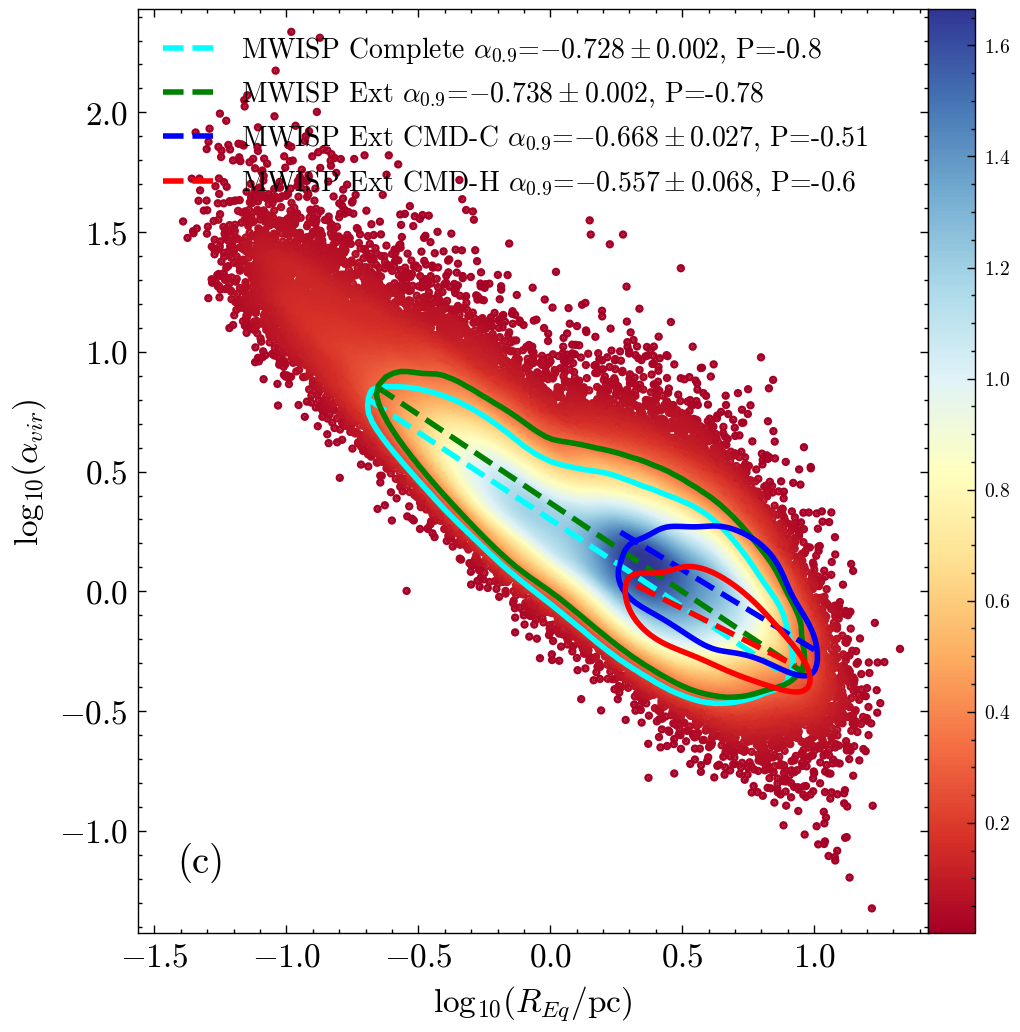}}
\end{minipage}

\begin{minipage}[t]{0.3\textwidth}
    \centering
    \centerline{\includegraphics[width=2.2in]{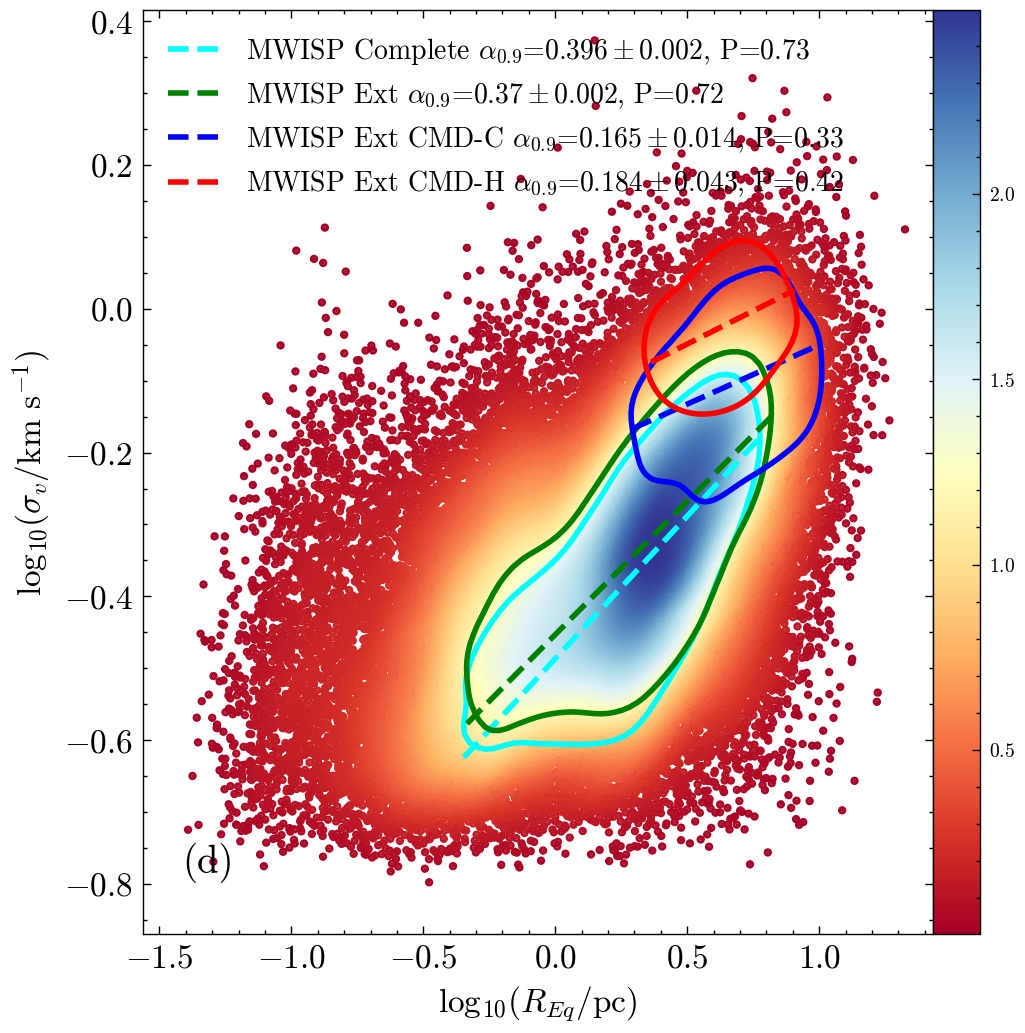}}
\end{minipage}
\begin{minipage}[t]{0.3\textwidth}
    \centering
    \centerline{\includegraphics[width=2.2in]{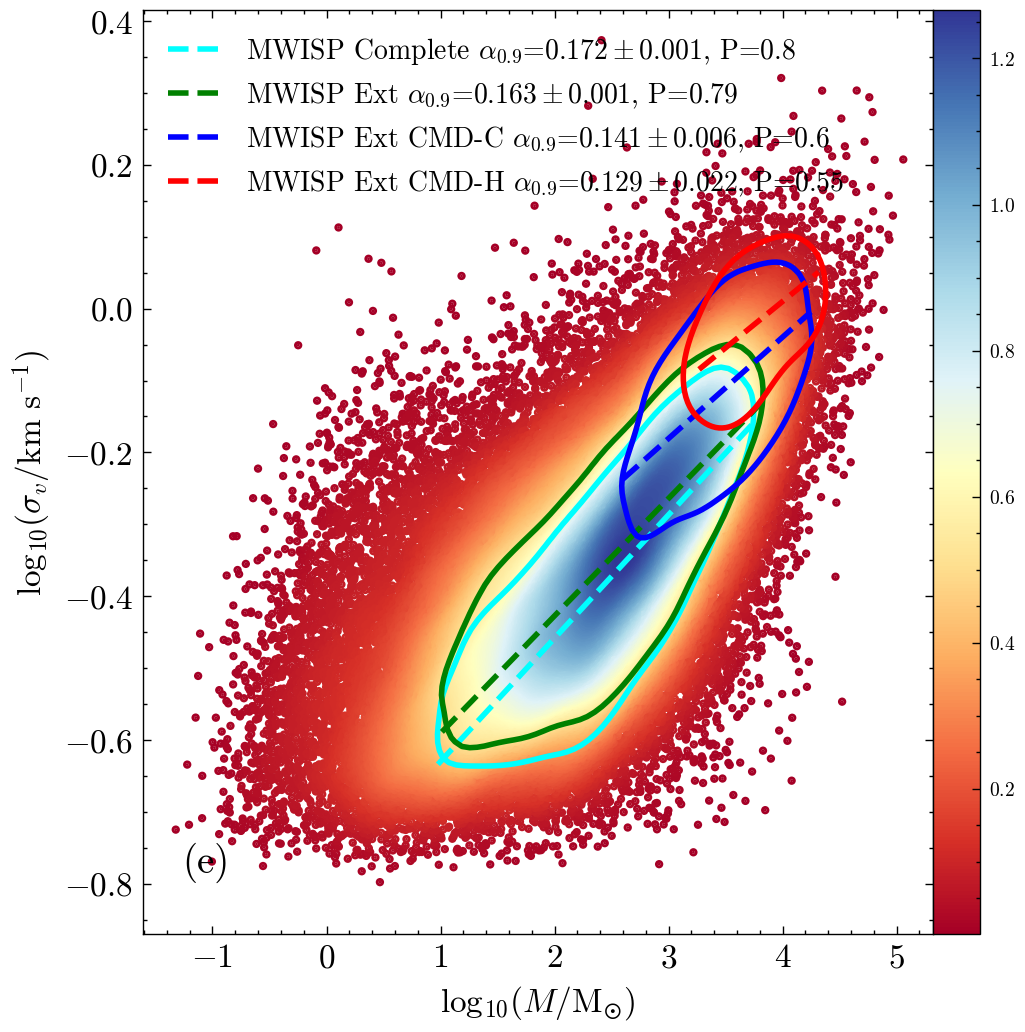}}
\end{minipage}
\begin{minipage}[t]{0.3\textwidth}
    \centering
    \centerline{\includegraphics[width=2.2in]{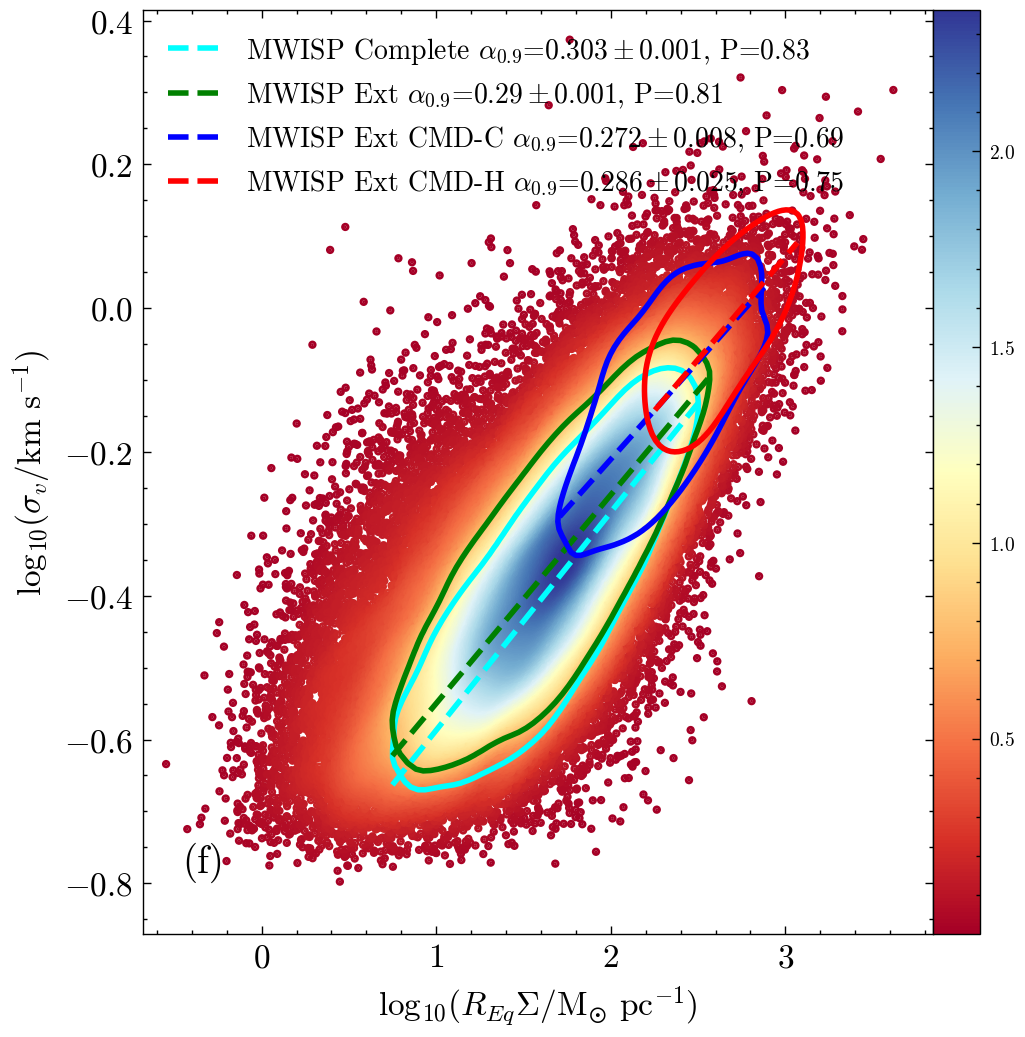}}
\end{minipage}

\begin{minipage}[t]{0.3\textwidth}
    \centering
    \centerline{\includegraphics[width=2.2in]{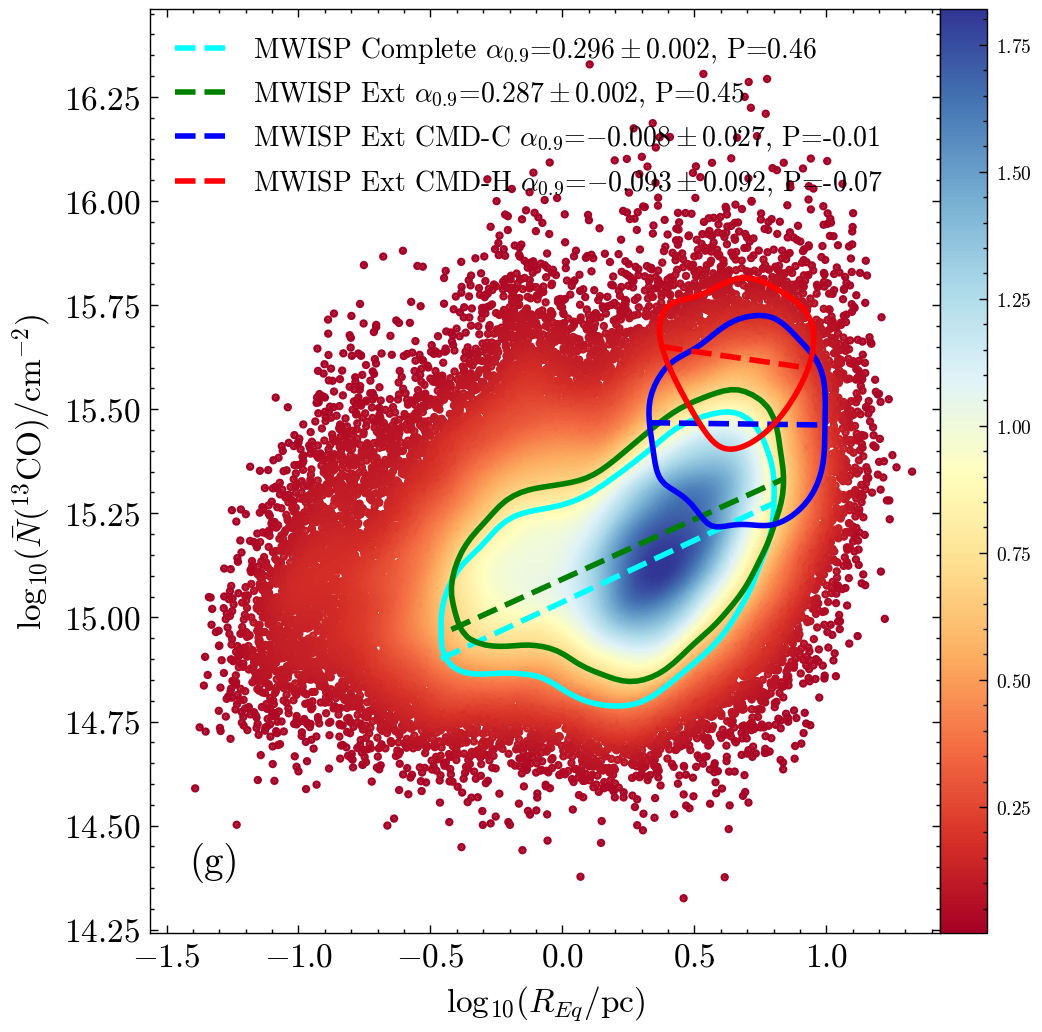}}
\end{minipage}
\begin{minipage}[t]{0.3\textwidth}
    \centering
    \centerline{\includegraphics[width=2.2in]{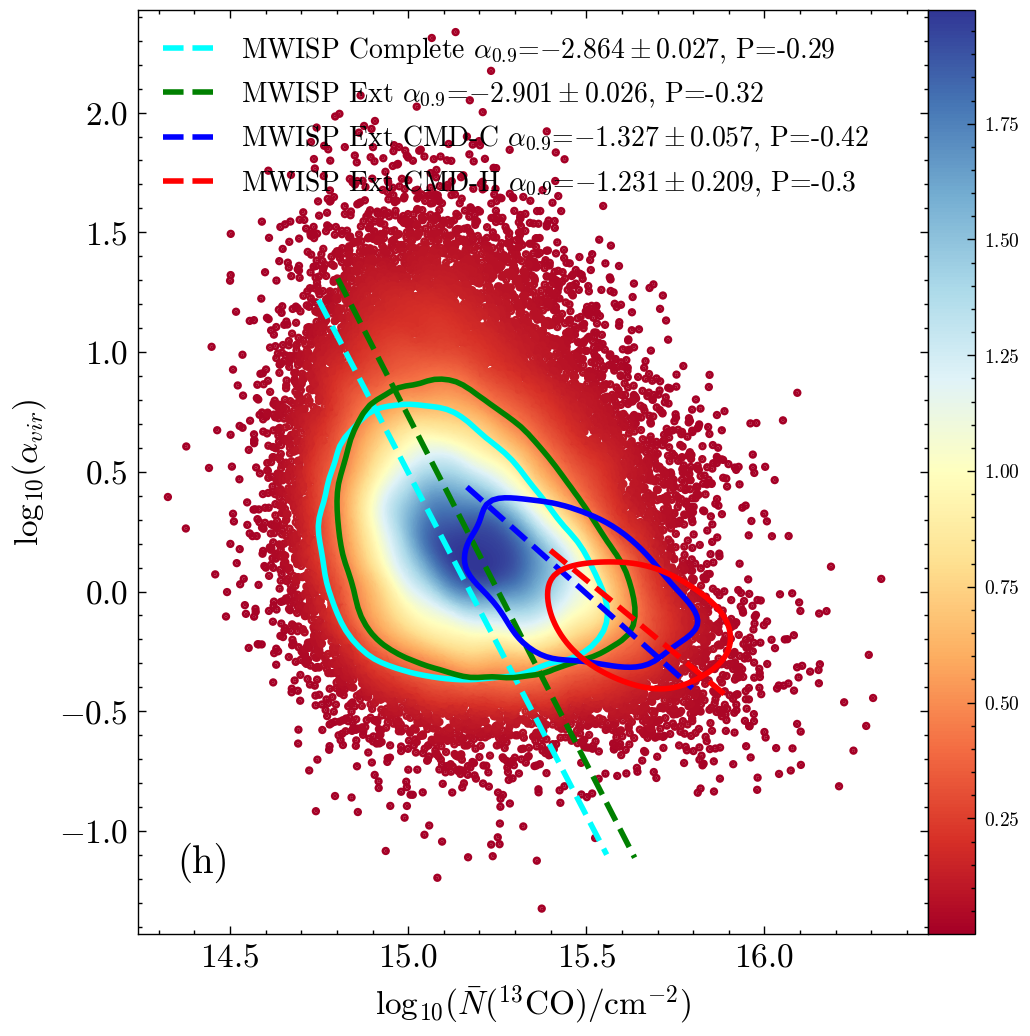}}
\end{minipage}
\begin{minipage}[t]{0.3\textwidth}
    \centering
    \centerline{\includegraphics[width=2.2in]{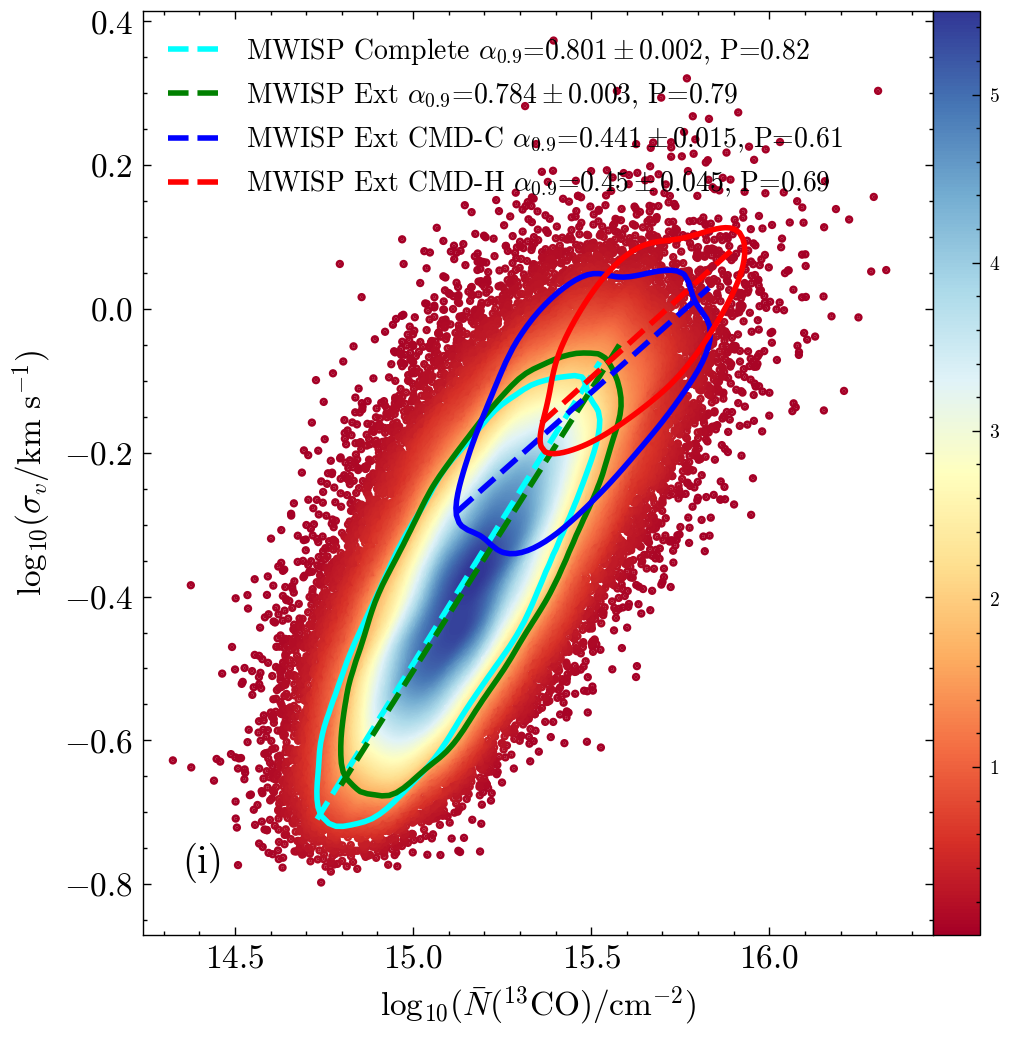}}
\end{minipage}
\caption{Scaling relations between extrapolated physical parameters of MWISP. Panels show (a) $M$-$R_{\rm eq}$, (b) $n({\rm H_2})$-$R_{\rm eq}$, (c)  $\alpha_{\rm vir}$-$R_{\rm eq}$, (d) $\sigma_v$-$R_{\rm eq}$, (e) $\sigma_v$-$M$, (f) $\sigma_v$-$R_{\rm eq}\Sigma$, (g) $N(^{13}\mathrm{CO})$-$R_{\rm eq}$, (h) $\alpha_{\rm vir}$-$N(^{13}\mathrm{CO})$, and (i) $\sigma_v$-$N(^{13}\mathrm{CO})$. The distributions are shown for four subsets: MWISP complete clumps (cyan contours), all extrapolated MWISP clumps (colored points with green contours), extrapolated CMD-C (blue contours), and extrapolated CMD-H (red contours). For each relation, the best-fit power-law index and corresponding $P$ are indicated in the legend.}
\label{Imgs_PR_Ext}
\end{figure*}

\subsection{Robustness of Scaling Relations Against Extrapolation Correction}\label{PRs_Ext}
After quantifying the biases in parameter measurements using the extrapolation method described in Appendix~\ref{ExtMethod}, we investigate whether these biases significantly affect the scaling relations established in the previous sections. Through extrapolation of flux, $\sigma_v$, and $R_{\text{eq}}$, we can derive corrections for other dependent parameters. For instance, mass is directly proportional to flux \citep{MWISP_Analysis_Sun_2021}, resulting in a calculated mass loss of -10.5\% from the flux loss. We estimate that the mass of 85.4\% of MWISP $^{13}$CO clumps is approximately $6.66\times10^7$ $M_{\odot}$, while the extrapolated total mass reaches $7.44\times10^7$ $M_{\odot}$. 

Figure~\ref{Imgs_PR_Ext} includes the distributions and scaling relations for both the original and extrapolated parameters across our sample categories. When comparing the complete non-extrapolated sample with its extrapolated counterpart, we observe only a minor shift in parameter space, with no significant changes in the fitting coefficients. Most parameters remain statistically equivalent within their respective error margins. The most substantial percentage change occurs in the $\sigma_v$-$R_{\rm eq}$ relation, where the $\Delta \alpha_{0.9}$ ratio is approximately 6.57\%. In conjunction with Figure~\ref{Imgs_PR_2}, Figure~\ref{Imgs_PR_Ext} further indicates that the fitting coefficients for extrapolated CMD-C and CMD-H samples maintain consistent trends relative to the complete sample, thereby preserving the differences documented in Section~\ref{PRs_Maser}. 

This consistency between extrapolated and non-extrapolated scaling relations demonstrates an important property of power-law relationships in molecular clumps: they are governed primarily by intrinsic physical processes rather than by limitations in detection sensitivity thresholds. While extrapolation yields more accurate absolute values for individual clump parameters, the proportional relationships between these parameters remain remarkably stable. This robustness validates our conclusions regarding systematic differences across different clump populations and their association with varying levels of star formation activity.

\section{Clump Examples}\label{Clump_Examples}
\setcounter{figure}{0}

\begin{figure*}
\centering
\vspace{0cm}
\begin{minipage}[t]{1\textwidth}
    \centering
    \centerline{\includegraphics[width=7in]{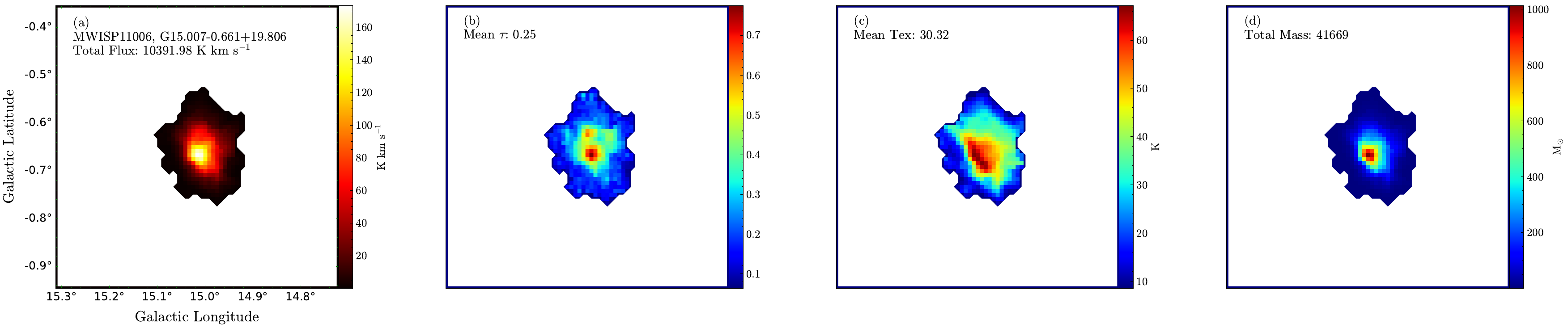}}
\end{minipage}

\begin{minipage}[t]{1\textwidth}
    \centering
    \centerline{\includegraphics[width=7in]{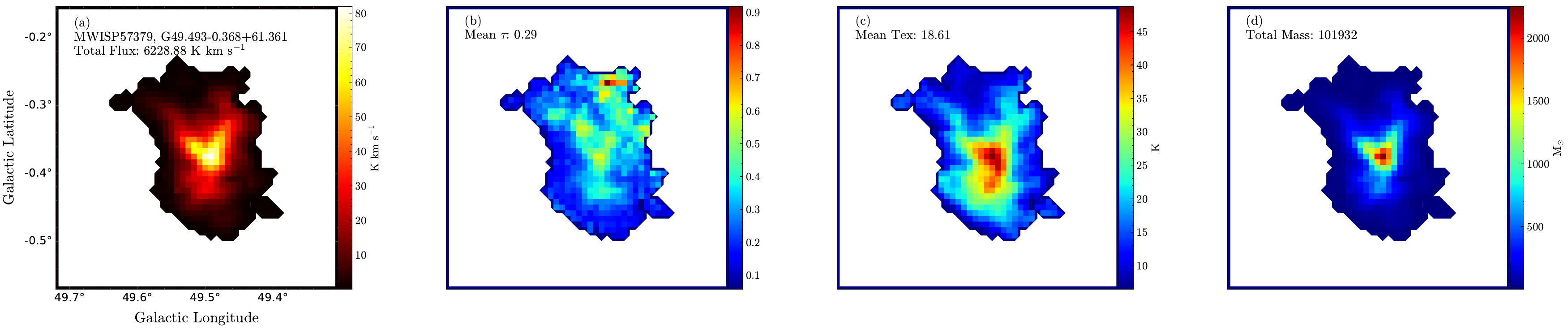}}
\end{minipage}

\begin{minipage}[t]{1\textwidth}
    \centering
    \centerline{\includegraphics[width=7in]{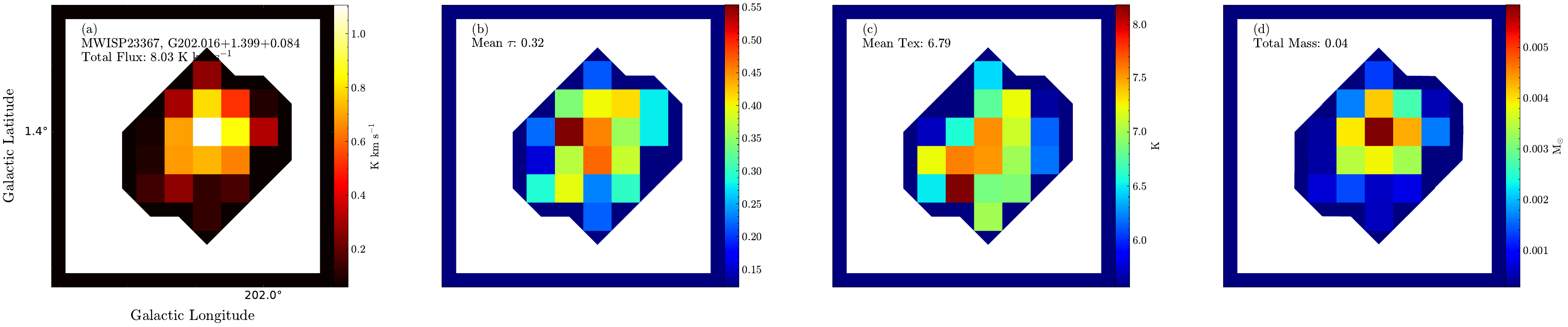}}
\end{minipage}

\begin{minipage}[t]{1\textwidth}
    \centering
    \centerline{\includegraphics[width=7in]{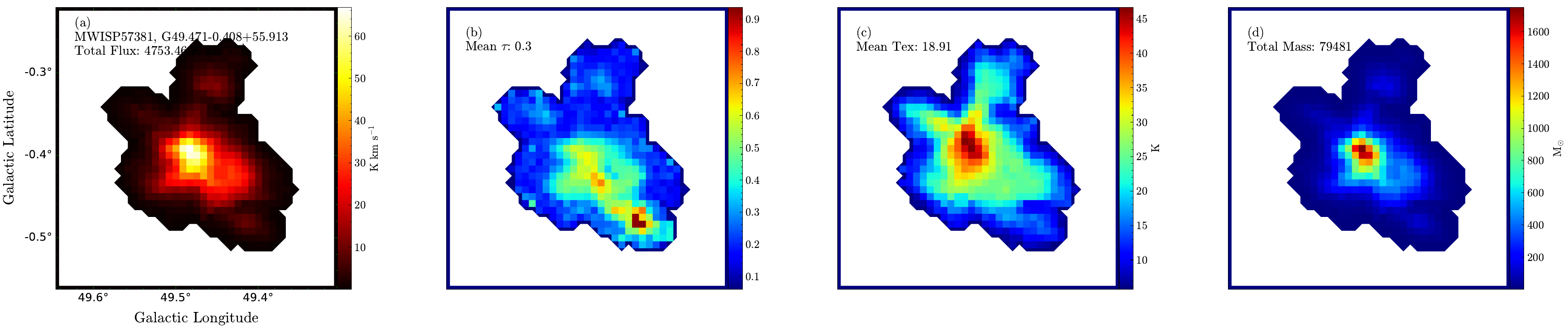}}
\end{minipage}

\begin{minipage}[t]{1\textwidth}
    \centering
    \centerline{\includegraphics[width=7in]{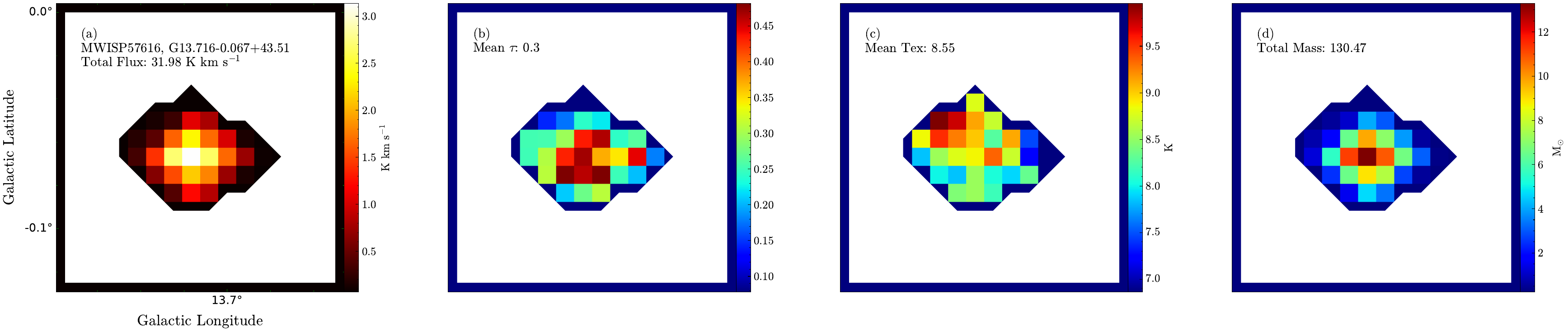}}
\end{minipage}
\caption{Examples of MWISP clumps. Row 1: clump corresponding to the maximum intensity (position: [15.017$^{\circ}$, -0.667$^{\circ}$, 19.261 km~s$^{-1}$]). Row 2: clump with the maximum mass. Row 3: clump with the minimum mass. Row 4: the most massive maser-hosting clump. Row 5: the least massive maser-hosting clump. Columns from left to right: (a) velocity-integrated intensity maps, (b) optical depth maps, (c) excitation temperature maps, and (d) mass maps. Labels show the catalog ID, Galactic coordinates, total flux, mean optical depth, mean excitation temperature, and total mass.}
\label{Imgs_ExClumps}
\end{figure*}

In Figure \ref{Imgs_ExClumps}, we present five exemplary clumps: the clump corresponding to the maximum intensity, the clump with the maximum mass, the clump with the minimum mass, the most massive maser-hosting clump, and the least massive maser-hosting clump. For each clump, we display the velocity-integrated intensity maps, optical depth maps, excitation temperature maps, and mass maps. The data blocks shown are slightly larger than the minimum bounding cube of the clumps. They represent the typical morphology of clumps in our catalogs. The parameters of these clumps are presented in the tables in Appendix \ref{Tables}.

\end{appendices}

\end{document}